\documentclass[10pt]{article}
\usepackage{amsfonts,color}
\usepackage{amsmath}
 \usepackage{epsfig}
\usepackage{lscape}
\usepackage{amssymb}
\usepackage{graphicx}
\usepackage{makeidx}
\usepackage{amssymb}
 \usepackage{footnote}

\DeclareMathOperator{\curl}{curl}
\DeclareMathOperator{\Div}{Div}

\DeclareMathOperator{\Curl}{Curl}

\newcommand{\Lp}[1]{L^{#1}(\Omega)}

\newcommand{\Hmp}[2]{H^{#1,#2}(\Omega)}
\newcommand{\Co}{C_0^{\infty}(\Omega)}

\newcommand{\norm}[1]{\|#1\|}

\newcommand{\R}{\mathbb{R}}

\newcommand{\C}{\mathbb{C}}

\renewcommand{\skew}{\mathop{\rm skew}}

\DeclareMathOperator{\sym}{sym}

\DeclareMathOperator{\axl}{axl}
\DeclareMathOperator{\anti}{anti}
\DeclareMathOperator{\dev}{dev}
\DeclareMathOperator{\sL}{\mathfrak{sl}}
\DeclareMathOperator{\so}{\mathfrak{so}}
\DeclareMathOperator{\gl}{\mathfrak{gl}}

\newcommand{\Sym}{ {\rm{Sym}} }

\newcommand{\Mprod}[2]{ {\langle #1 ,#2\rangle} }
\newcommand{\id}{ {1\!\!\!\:1 } }

\newcommand{\tr}[1]{ {\Tr \left[{#1}\right]} }

\allowdisplaybreaks[1]

\makeindex

\newtheorem{theorem}{Theorem}[section]

\newtheorem{corollary}[theorem]{Corollary}

\def\barr{\begin{array}}
\def\id{1\!\!1}
\def\tr{\textrm{tr}}
\def\dv{\textrm{div}}
\def\dvg{\textrm{Div}}
\def\crl{\textrm{Curl}}

\def\dd{\displaystyle}

\def\barr{\begin{array}}
\def\earr{\end{array}}
\def\bec#1{\begin{equation}\label{#1}}
\def\becn{\begin{equation*}}
\def\endec{\end{equation}}
\def\endecn{\end{equation*}}
 \def\C{\mathbb{C}}
 \def\H{\mathbb{H}}
  \def\L{\mathbb{L}_c}
  \def\cdp{\,\!\cdot\!\,}

\newtheorem{remark}{Remark}[section]
\begin{document}

\title{A unifying perspective: the relaxed linear micromorphic  continuum}
\author{
Patrizio Neff\thanks{Patrizio Neff,  \ \ Head of Lehrstuhl f\"{u}r Nichtlineare Analysis und Modellierung, Fakult\"{a}t f\"{u}r Mathematik, Universit\"{a}t Duisburg-Essen, Campus Essen, Thea-Leymann Str. 9, 45127 Essen, Germany, email: patrizio.neff@uni-due.de} \quad
and \quad
Ionel-Dumitrel Ghiba\footnote{Ionel-Dumitrel Ghiba, \ \ \ \ Lehrstuhl f\"{u}r Nichtlineare Analysis und Modellierung, Fakult\"{a}t f\"{u}r Mathematik, Universit\"{a}t Duisburg-Essen, Campus Essen, Thea-Leymann Str. 9, 45127 Essen, Germany;  Alexandru Ioan Cuza University of Ia\c si, Department of Mathematics,  Blvd. Carol I, no. 11, 700506 Ia\c si,
Romania;  Octav Mayer Institute of Mathematics of the
Romanian Academy, Ia\c si Branch,  700505 Ia\c si; and Institute of Solid Mechanics, Romanian Academy, 010141 Bucharest, Romania, email: dumitrel.ghiba@uaic.ro}
  \quad
and \quad\\
Angela Madeo\footnote{Angela Madeo, \ \  Laboratoire de G\'{e}nie Civil et Ing\'{e}nierie Environnementale,
Universit\'{e} de Lyon-INSA, B\^{a}timent Coulomb, 69621 Villeurbanne
Cedex, France; and International Center M\&MOCS ``Mathematics and Mechanics
of Complex Systems", Palazzo Caetani,
Cisterna di Latina, Italy,
 email: angela.madeo@insa-lyon.fr}
\quad
and \quad
Luca Placidi\footnote{Luca Placidi: Universit\`{a} Telematica Internazionale Uninettuno, Corso V. Emanuele II
39, 00186 Roma, Italy; and
International Center M\&MOCS ``Mathematics and Mechanics of Complex
Systems", Palazzo Caetani, Cisterna di Latina, Italy,
 email: luca.placidi@uninettunouniversity.net}
\quad
and \quad
Giuseppe Rosi\footnote{Giussepe Rosi: Laboratoire Mod\'{e}lisation Multi-Echelle, MSME UMR 8208 CNRS, Universit\'{e}
Paris-Est, 61 Avenue du General de Gaulle,
Creteil Cedex 94010, France; and  International Center M\&MOCS, University of
L'Aquila, Palazzo Caetani, Cisterna di Latina, Italy, email:
giuseppe.rosi@uniroma1.it}  }
\date{}
\maketitle

\begin{center}
``Das also war des Pudels Kern.", \emph{Faust I}, J.W. v. Goethe
\end{center}
\bigskip

\begin{abstract}
We formulate a relaxed linear elastic micromorphic continuum model with symmetric Cauchy force-stresses and curvature contribution depending only on the micro-dislocation tensor. Our relaxed model is still able to fully describe rotation of the microstructure and to predict non-polar size-effects. It is intended for the homogenized description of highly heterogeneous, but non polar materials with microstructure liable to slip and fracture. In contrast to classical linear micromorphic models our
free energy  is not uniformly pointwise positive definite in the
control of the independent constitutive variables.  The new relaxed micromorphic model supports  well-posedness results for the dynamic and static case. There, decisive use is made of  new coercive
inequalities recently proved by Neff, Pauly and  Witsch and by Bauer, Neff, Pauly and Starke.   The new relaxed micromorphic formulation can be related to dislocation dynamics, gradient plasticity and seismic processes of earthquakes. It unifies and simplifies the understanding of the linear micromorphic models.
\\
\vspace*{0.25cm}
\end{abstract}

\newpage

\tableofcontents

\section{Introduction}

\subsection{Motivation}

Microstructural motions are observed to produce new effects
that cannot be accounted for by classical translatory degrees of
freedom  (dof) used to formulate conventional theories. For instance, plane waves  in an unbounded elastic medium propagate without dispersion, i.e. the wave speed is independent of the frequency. However, experiments with  real solids disclose dispersive wave propagation. In order to incorporate the microstructure of the matter into the classical theory, generalized continuum models may be used. Among the various extended continuum theories we mention the higher gradient elasticity theories \cite{Mindlin65,Mindlin68,Neff_Jeong_IJSS09,Steinmann05,AskesAifantis} and  micromorphic models \cite{Eringen99,Neff_STAMM04,Mariano05,Mariano08a,Mariano08b,SteigmannIJNM12}.

General continuum models involving independent rotations have been introduced by the Cosserat brothers \cite{Cosserat09}
at the beginning of the last century.  A material point carrying three deformable directors  introduces nine extra degrees of freedom  besides the translational degrees of freedom from the classical theory.  Many developments have been reported since the
seminal work of the Cosserat brothers. The derived generalized theories   are called polar, micropolar, micro-elastic, micromorphic, Cosserat, multipolar, oriented, complex, etc., according to the  specifically considered kinematical variables and to the choice of the set of constitutive variables. All materials, whether natural or synthetic, possess microstructures if one considers sufficiently small scales.  Viewed first as a formal theoretical investigation, the micromorphic models (12 dof) derived by Eringen  and Suhubi \cite{Eringen64}, Mindlin \cite{Mindlin62,Mindlin63,Mindlin64,Mindlin68b} and Toupin \cite{Toupin62,Toupin64} are justified recently as more realistic continuum models based on molecular dynamics and ensemble averaging \cite{Chen03a,ChenLeeEskan2004,Chen02,WangLee,ZengChenLee2006}.

A large class of engineering materials, porous solids with deformable grains and pores, composites, polymers with deformable molecules, crystals, solids with microcracks, dislocations and disclinations \cite{Maugin09,Berdichevsky06}, and
biological tissues like ``bones and muscles"  may be modeled more realistically by means of the theory
of micromorphic materials. This is the reason why
micromorphic mechanics is a dynamic field of research both from a theoretical and practical
point of view.

 Considerations of the format of balance laws in geometrically nonlinear micromorphic
elasticity have been undertaken in \cite{Capriz89,Mariano08a,CaprizMariano,FabMariano,Yavari}. The only known
existence results for the static geometrically nonlinear formulation are due to Neff
\cite{Neff_micromorphic_rse_05} and to Mariano and Modica \cite{Mariano08a}.  In fact, Mariano and Modica [94] treat general microstructures
described by manifold-valued variables, even if they discuss
essentially what is called by Neff  in  \cite{Neff_micromorphic_rse_05}  macro-stability (two
other cases are treated  in  \cite{Neff_micromorphic_rse_05}, one leads to fractures - a
situation excluded in \cite{Mariano08a} - the other is left open). When the energy
analyzed by Mariano and Modica is reduced to micromorphic materials in
the splitted version considered by Neff \cite{Mariano08a}, their coercivity assumptions
result are more stringent than Neff's ones (the blow up of the determinant
of $\textrm{det}\, F$ a part), so they restrict the material response. However, the
direct comparison of the two existence results is not completely
straightforward. As for the numerical implementation, see  \cite{Mariano05} and the development in \cite{Klawonn_Neff_Rheinbach_Vanis09}. In \cite{Klawonn_Neff_Rheinbach_Vanis09}
the original problem  is decoupled into two separate problems. Corresponding domain-decomposition techniques for the subproblem related to balance of forces are investigated in \cite{Klawonn_Neff_Rheinbach_Vanis09}. The size effects involved in a natural way in the micromorphic models (see e.g. \cite{Sansour98b}) have recently received much attention in conjuction with nano-devices and foam-like structures. A geometrically nonlinear generalized continuum of micromorphic type in the sense of Eringen for the phenomenological description of metallic foams is given by Neff and Forest \cite{Neff_Forest_jel05}. Moreover, in  \cite{Neff_Forest_jel05} the authors proved the existence of  minimizers and they  identified the relevant effective material parameters.

A comparison of the geometrically nonlinear elastic micromorphic theories with affine microstructure with the intrinsically linear models of Mindlin and Eringen is given in \cite{Neff_STAMM04,Neff_Forest_jel05}.
In the present paper, a useful decomposition ({\it mixed variant}) of the constitutive choice for the strain energy density is presented for the {\it classical} linear-elastic micromorphic media of Mindlin-Eringen type. This decomposition allows to individuate, in the isotropic case, a unique parameter $\mu_c$ (called {\it Cosserat couple modulus}) which governs the asymmetry of the force stresses and which is strongly related to penalty formulations without intrinsic physical significance. This parameter is not included in the {\it relaxed model} we introduce in the second part of the present paper.
In the following, we refer to the classical micromorphic model, relaxed model etc.
according to the following understanding
\begin{itemize}
\item {\bf classical}: Dirichlet boundary condition  for $P$, the energy density is similar to\
$$\hspace{-0.5cm}
\mu_e\, \|\sym (\nabla u-P)\|^2+\framebox[3.3cm][l]{$\mu_c\,\|\skew(\nabla u-P)\|^2$}+\mu_h\, \|\sym  P\|^2 +\framebox[1.2cm][l]{$\| \nabla P\|^2$}\,\,;\notag
$$

\item {\bf relaxed}:  tangential boundary condition  for $P$, the energy density is similar to
$$\hspace{-3.6cm}\mu_e \|\sym (\nabla u-P)\|^2+\mu_h \|\sym  P\|^2\notag +\framebox[1.7cm][l]{$\|  \Curl P\|^2$}\,\,; \notag
  $$

\item {\bf mixed variant}: tangential boundary condition  for $P$, the energy density is similar to
$$\mu_e \|\sym (\nabla u-P)\|^2+\framebox[3.3cm][l]{$\mu_c\,\|\skew(\nabla u-P)\|^2$}+\mu_h \|\sym  P\|^2\notag  +\framebox[1.7cm][l]{$\|  \Curl P\|^2$}\,\,,$$
\end{itemize}
 where $u$ is the displacement and $P$ is the
micro-distortion. The precise definitions will be given in the Subsections \ref{Eringenmodel} and \ref{Sectrelaxmodel}.

In contrast to the Mindlin and Eringen models, we avoid the presence of the only one parameter in the force stress response which can not be directly related to simple experiments. However, the presence of the parameter $\mu_c$ may be necessary to completely describe the mechanical behavior of artificial metamaterials in which strong contrasts of the elastic properties are present at the microscopic level. This necessity is evident when studying e.g. phononic crystals which are especially designed to exhibit frequency band-gaps. This means that such metamaterials are conceived to block wave propagation in precise frequency ranges.
As far as standard heterogeneous materials (natural and artificial) are concerned, our reduced model with symmetric stress is  sufficient to fully describe their mechanical behavior. The new well-posedness results for the relaxed model include the well-posedness results for the classical model.

\subsection{Historical perspective}

The capability of continuum theories to describe the time evolution
and the deformation of the micro-structure of complex mechanical
systems was recognized in the very first formulations of continuum
mechanics (see the pioneering work by Piola \cite{Piola}). Piola
was led by stringent physical considerations to consider gradients
of displacement field higher than the first as needed independent
variables in the constitutive equation for the deformation energy
of continuous media (for a modern presentation of this subject see
e.g. \cite{FdISepp,SciarraFdICoussy,IsolaSciarraVidoliPRSA,Neff_Jeong_IJSS09,SteigmannIJNM12}).

However, more or less in the same period in which Piola was producing
his papers, Cauchy and Poisson managed to determine a very elegant
and effective format for continuum mechanics in which:

i) the only kinematical descriptor is the displacement from a reference
configuration,

ii) the crucial conceptual tool is the {\it symmetric Cauchy force stress tensor} $\sigma=\mathbb{C}.\varepsilon$ which is constitutively
related only to the  symmetrized gradient of displacement $\varepsilon=\sym \nabla u$,

iii) the crucial postulates are those concerning balance of mass,
linear and angular momentum and (eventually) energy.

The Cauchy and Poisson format is  very effective to describe the
mechanical behavior of a very wide class of natural and also artificial
materials. Nevertheless, when considering materials with  well-organized
microstructures subjected to particular loads and/or boundary conditions,
 a Cauchy continuum theory may fail to give accurate results. This is
the case for some engineering materials showing high contrast of material
properties (see e.g. \cite{PideriSeppecher,Boutin,FibrousComposites})
or for some natural materials which show highly heterogeneous hierarchical
microstructures (see e.g. \!\cite{BuechLakes}). In all these cases,
the introduction of more sophisticated models becomes mandatory if
one wants to catch all features of the mechanical behavior of such
complex materials.

About fifty years later, Piola's ideas were developed by the Cosserat
brothers, who were among the first authors who complemented the standard
kinematics constituted by a placement field with additional independent
kinematical fields. In their case, these suitable fields are given by
rigid rotations of the microstructure
with respect to the macroscopic continuum  displacement. This introduces three additional dof into the theory. Cosserat
contributions \cite{Cosserat09} were underestimated for another fifty
years and only starting from 1960 a group constituted by relevant
scientific personalities as Mindlin\footnote{R.D. Mindlin, 17.09.1906--22.11.1987, abandoned microstructure theory when he found that the systems did not mirror physical reality if compared with the ionic lattice theory.} \cite{Mindlin68,Mindlin64,Mindlin65},
Green and Rivlin \cite{GreenRivlinonCauchy,GreenRivlinFunctionalTheory,GreenRivlinMultipolar,GreenRivlinMultipoles},
Toupin \cite{Toupin62,Toupin64}, Eringen \cite{Eringen99,Eringen64,EringenSuhubi2}
and Germain \cite{Germain,Germain1} managed to establish (with still
some resistances) the formal validity of Cosserat's point of view.

Actually, the Cosserat's approach must be further generalized as not only
micro-rotations should be included in a macroscopic modelling picture,
but also micro-stretches, micro-strains, micro-shear or concentrated micro-distortions.
This can be done by introducing the so-called micro-structured or
micromorphic continuum models, which are suitably formulated by means
of a postulation process based on the principle of least action (see
e.g. \cite{Auffray}) or on the principle of virtual works (see the
beautiful works \cite{Kroner1,Maugin70,Maugin09,MauginVirtualPowers,Yavari,CaprizMariano})
even if later on the alternative postulation procedure based on ``generalized
balance laws'' has been also attempted (see \cite{Eringen99,Eringen64,EringenSuhubi2}).

Indeed, as already remarked in \cite{Piola}, when starting from a
discrete system characterized by a micro-structure spanning several
length scales and strong contrast of the elastic properties it is rather unlikely to get as a suitable macroscopic
model the simple standard Cauchy continuum. Whether the force stresses remain symmetric in such homogenization procedure is open \cite{Bigoni07,PideriSeppecher}.

In addition, the evolution of non purely mechanical phenomena can be described
within the framework of micromorphic continua theory: in this context
we refer for instance to those observed in nematic liquid crystals
(see e.g. \cite{VidoliIsola}) where also electromagnetic descriptors
need to be introduced to characterize completely the kinematics of the
system. In such a case we speak of polar materials in which the force stress may clearly become non-symmetric.

It has to be explicitly remarked that Piola's and Cosserat's models
can be reconciled by means of the introduction of suitable ``internal
constraints'' and ``Lagrange multipliers'' as clearly stated e.g.
in \cite{Bleustein}. Actually one can get Piola's deformation energies
depending on higher gradients of displacement as a limit of many different
(and physically non-equivalent) more detailed micromorphic models. If in the classical micromorphic model, in the formal limits, we assume  e.g. that the coefficients $\widehat{\mathbb{C}}\rightarrow\infty$  then this leads to the free energy \eqref{grpl} for the gradient elasticity model \cite{Neff_Jeong_ZAMP08,Neff_Jeong_IJSS09}. On the other hand, we can always relax Piola's gradient material into suitable micromorphic models.

\subsection{Approach in this work}

In the present paper we re-investigate the general micromorphic model with a focus on heterogeneous, but non-polar materials. More particularly, we start by recalling the classical Mindlin-Eringen model for micromorphic media with intrinsically non-symmetric force stresses.
As our contribution, we propose a relaxed linear micromorphic model with symmetric Cauchy force stresses and curvature response only due to dislocation energy, and we formulate the
 initial--boundary value problems.
 We prove that this new model is still well-posed \cite{GhibaNeffExistence}, i.e. we study the continuous dependence of solution with respect to the initial data and supply terms and existence and uniqueness of the solution. The main point in establishing the  existence, uniqueness and continuous dependence results \cite{GhibaNeffExistence} is represented by the new coercive inequalities recently proved by Neff, Pauly and  Witsch \cite{NeffPaulyWitsch,NPW2,NPW3} and by Bauer, Neff, Pauly and Starke \cite{BNPS1,BNPS2,BNPS3} (see also \cite{GhibaNeffExistence,LNPzamp2013}).

 This relaxed formulation of micromorphic elasticity has some similarities to recently studied models of gradient plasticity \cite{Ebobisse_Neff09,Neff_Chelminski07_disloc,NN-SIAM12,NN13,ForestJMPS10}. Indeed, in the static case, the micromorphic relaxed minimization problem has the same stored elastic energy. In gradient plasticity, however, the plastic distortion/micromorphic distortion is determined not by energy minimization, but instead by a flow rule. Our new approach, in sharp contrast to classical micromorphic models, features
a symmetric force stress tensor, which we  call Cauchy stress.  Teisseyre  \cite{Teisseyre73,Teisseyre74} in his model for the description of  seismic wave propagation phenomena \cite{Teisseyre84,Teisseyre00,Teisseyre01}  also used a symmetric force stress tensor. In fact, Teisseyre's model is a fully symmetric model and it is a particular case  of the dislocation dynamics theory proposed by Eringen and Claus \cite{Eringen_Claus69,EringenClaus,Eringen_Claus71}, where the relative force-stress is considered to be non-symmetric. The model of Eringen and Claus \cite{Eringen_Claus69,EringenClaus,Eringen_Claus71}  contains the linear Cosserat model \cite{Neff_ZAMM05,Neff_Jeong_Conformal_ZAMM08,Neff_JeongMMS08,Neff_Paris_Maugin09} with asymmetric force stresses upon suitable restriction. This is a situation we  avoid in the proposed relaxed micromorphic approach (see in Subsection \ref{Kroner} the motivation given by  Kr\"{o}ner for symmetric force stresses in dislocation dynamics). In fact, it turns out (to our surprise) that our relaxed model is the Eringen-Claus model \cite{Eringen_Claus69,EringenClaus,Eringen_Claus71}, albeit with symmetric Cauchy stresses and absent mixed coupling terms. In Section \ref{CompSect} we disclose the relation of our new relaxed model to the existing models in more detail.  Our critical remarks concerning the linear Cosserat model leave open the possible usefulness of a geometrically nonlinear Cosserat model \cite{Neff_ZAMM05,Neff_Muench_initial_plasticity_proc10} with symmetric Cauchy stresses.
In contrast with the models considered until now, our
free energy of the relaxed model is {\it not uniformly pointwise positive definite} in the
control of the constitutive variables.

The proposed relaxed  micromorphic model with symmetric force-stress may be thought to fully describe the mechanical behaviour of a great variety of natural and artificial microscopically heterogeneous materials. Granular assemblies are also a field of application of
micromorphic models. The possible non-symmetry of the force stress tensor in such
models has been discussed e.g. in  \cite{Goddard08} and it is proved that
in the absence of intergranular contact moments the grain rotation makes no direct contribution to quasi-static contact work, and that
the widely accepted formula based on volume averaging yields a symmetric
Cauchy stress. On the other hand, we are aware of the possible usefulness of Cosserat model for what concerns the modeling of artificial engineering metamaterials with strong contrast at the microscopic level. The results established in our paper can be extended to theories which include electromagnetic and thermal interactions \cite{GGI11,GalesEJMA12,Grekova05,Maugin13}. For isotropic materials, the models presented in this paper involve only a reduced number of constitutive parameters. This fact will allow us to find exact solutions for wave propagation problems using analog methods as in \cite{JanickeCMS12,ChiritaGhiba1,ChiritaGhiba2,ChiritaGhiba3} and we may also  compare the analytical solutions with experiments in order to identify the fewer relaxed constitutive coefficients.

It is known since the pionering works of Mindlin \cite{Mindlin64} that two
types of waves can propagate in a micromorphic continuum: {\it acoustic waves}, i.e. waves for which the frequency vanishes for vanishing
wavenumbers (wavelength which tends to infinity), and {\it optic waves}, i.e. waves which have non-vanishing, finite frequency
corresponding to vanishing wavenumbers (space independent oscillations). It can be shown that, for particular frequency ranges, also
a third type of waves may exist in our relaxed micromorphic media, namely so-called
{\it
standing waves}, i.e. waves which do not propagate inside the medium
but keep oscillating in a given region of space. These waves are impossible in the classical micromorphic model. Wave
propagation in the considered relaxed micromorphic model and the precise
effect of the considered elastic parameters will be carefully studied
in a forthcoming paper to show the interest of using our model to
proceed towards innovative technological applications. The paper \cite{Eremeyev5} gives a rich reference list on the wave propagation in second-gradient
materials and on generalized media, in general. Moreover, we will deal with the static model and consider the elliptic regularity question. The numerical treatment of our new model needs FEM-discretisations in ${\rm H}({\rm curl};\Omega)$, see \cite{Raviart79}. This will be left for future work.

\subsection{Notation}

For $a,b\in\R^3$ we let $\Mprod{a}{b}_{\R^3}$  denote the scalar product on $\R^3$ with
associated vector norm $\norm{a}_{\R^3}^2=\Mprod{a}{a}_{\R^3}$.
We denote by $\R^{3\times 3}$ the set of real $3\times 3$ second order tensors, written with
capital letters.
The standard Euclidean scalar product on $\R^{3\times 3}$ is given by
$\Mprod{X}{Y}_{\R^{3\times3}}=\tr({X Y^T})$, and thus the Frobenius tensor norm is
$\norm{X}^2=\Mprod{X}{X}_{\R^{3\times3}}$. In the following we omit the index
$\R^3,\R^{3\times3}$. The identity tensor on $\R^{3\times3}$ will be denoted by $\id$, so that
$\tr({X})=\Mprod{X}{\id}$.
We let $\Sym$  denote the set of symmetric tensors. We adopt the usual abbreviations of Lie-algebra theory, i.e.,
 $\so(3):=\{X\in\mathbb{R}^{3\times3}\;|X^T=-X\}$ is the Lie-algebra of  skew symmetric tensors
and $\sL(3):=\{X\in\mathbb{R}^{3\times3}\;| \tr({X})=0\}$ is the Lie-algebra of traceless tensors. For all  vectors $\xi,\eta\in\R^3$ we have the tensor product
$(\xi\otimes\eta)_{ij}=\xi_i\,\eta_j$  and $\epsilon_{ijk}$ is the  Levi-Civita symbol, also called the permutation symbol or antisymmetric symbol, given by
\begin{align}
\epsilon_{ijk}=\left\{
\begin{array}{ll}
1& \text{if} \quad (i,j,k) \quad \text{ is an even permutation of} \quad (1,2,3)\\
-1& \text{if} \quad (i,j,k) \quad \text{ is an odd permutation of} \quad (1,2,3)\\
0 & \text{otherwise}.
\end{array}\right.
\end{align}
 For all $X\in\mathbb{R}^{3\times3}$ we set $\sym X=\frac{1}{2}(X^T+X)\in\Sym$, $\skew X=\frac{1}{2}(X-X^T)\in \so(3)$ and the deviatoric part $\dev X=X-\frac{1}{3}\;\tr{X}\,\id\in \sL(3)$  and we have
the \emph{orthogonal Cartan-decomposition  of the Lie-algebra} $\gl(3)$
\begin{align}
\gl(3)&=\{\sL(3)\cap \Sym(3)\}\oplus\so(3) \oplus\mathbb{R}\!\cdot\! \id,\notag\\
X&=\dev \sym X+ \skew X+\frac{1}{3}\tr(X)\!\cdot\! \id\,.
\end{align}
By $\Co$ we denote infinitely
differentiable functions with compact support in $\Omega$. We employ the standard notation of Sobolev spaces, i.e.
$\Lp{2},\Hmp{1}{2},H_0^{1,2}(\Omega)$, which we use indifferently for
scalar-valued functions
as well as for vector-valued and tensor-valued functions. Throughout this paper (when we do not specify else) Latin subscripts take the values $1,2,3$. Typical conventions for differential
operations are implied such as comma followed
by a subscript to denote the partial derivative with respect to
 the corresponding cartesian coordinate, while $t$ after a comma denotes the partial derivative with respect to the time.
 The usual Lebesgue spaces of square integrable functions, vector or tensor fields on $\Omega$ with values in $\mathbb{R}$, $\mathbb{R}^3$ or $\mathbb{R}^{3\times 3}$, respectively will be denoted by $L^2(\Omega)$. Moreover, we introduce the standard Sobolev spaces \cite{Adams75,Raviart79,Leis86}
\begin{align}
&{\rm H}^1(\Omega)=\{u\in L^2(\Omega)\, |\, {\rm grad}\, u\in L^2(\Omega)\}, \ \ \ {\rm grad}=\nabla\, ,\notag\\\notag
&\ \ \ \ \ \ \ \ \ \ \ \ \ \ \ \|u\|^2_{{\rm H}^1(\Omega)}:=\|u\|^2_{L^2(\Omega)}+\|{\rm grad}\, u\|^2_{L^2(\Omega)}\, ,\\
&{\rm H}({\rm curl};\Omega)=\{v\in L^2(\Omega)\, |\, {\rm curl}\, v\in L^2(\Omega)\}, \ \ \ {\rm curl}=\nabla\times\, ,\\\notag
&\ \ \ \ \ \ \ \ \ \ \ \ \ \ \ \ \ \ \|v\|^2_{{\rm H}({\rm curl};\Omega)}:=\|v\|^2_{L^2(\Omega)}+\|{\rm curl}\, v\|^2_{L^2(\Omega)}\, ,\\\notag
&{\rm H}({\rm div};\Omega)=\{v\in L^2(\Omega)\, |\, {\rm div}\, v\in L^2(\Omega)\}, \ \ \ {\rm div}=\nabla\cdot\, ,\\\notag
&\ \ \ \ \ \ \ \ \ \ \ \ \ \ \ \ \ \ \|v\|^2_{{\rm H}({\rm div};\Omega)}:=\|v\|^2_{L^2(\Omega)}+\|{\rm div}\, v\|^2_{L^2(\Omega)}\, ,
\end{align}
of functions $u$ or vector fields $v$, respectively.

Furthermore, we introduce their closed subspaces $H_0^1(\Omega)$, and ${\rm H}_0({\rm curl};\Omega)$ as completion under the respective graph norms of the scalar valued space $C_0^\infty(\Omega)$, the set of smooth functions with compact support in $\Omega$. Roughly speaking, $H_0^1(\Omega)$ is the subspace of functions $u\in H^1(\Omega)$ which are
zero on $\partial \Omega$, while ${\rm H}_0({\rm curl};\Omega)$ is the subspace of vectors $v\in{\rm H}({\rm curl};\Omega)$ which are normal at $\partial \Omega$ (see \cite{NeffPaulyWitsch,NPW2,NPW3}). For vector fields $v$ with components in ${\rm H}^{1}(\Omega)$ and tensor fields $P$ with rows in ${\rm H}({\rm curl}\,; \Omega)$, resp. ${\rm H}({\rm div}\,; \Omega)$, i.e.,
\begin{align}
v=\left(
  \begin{array}{c}
    v_1 \\
    v_2 \\
    v_3 \\
  \end{array}
\right)\, , v_i\in {\rm H}^{1}(\Omega),
\ \quad
P=\left(
  \begin{array}{c}
    P_1^T \\
    P_2^T \\
    P_3^T \\
  \end{array}
\right)\, \quad P_i\in {\rm H}({\rm curl}\,; \Omega)\,  \quad \ \text{resp.} \quad P_i\in {\rm H}({\rm div}\,; \Omega)
\end{align}
we define
\begin{align}
 {\rm Grad}\,v=\left(
  \begin{array}{c}
   {\rm grad}^T\,  v_1 \\
    {\rm grad}^T\, v_2 \\
    {\rm grad}^T\, v_3 \\
  \end{array}
\right)\, ,
\ \ \ \ {\rm Curl}\,P=\left(
  \begin{array}{c}
   {\rm curl}^T\, P_1 \\
    {\rm curl}^T\,P_2 \\
    {\rm curl}^T\,P_3 \\
  \end{array}
\right),\,  \ \ \ \ {\rm Div}\,P=\left(
  \begin{array}{c}
   {\rm div}\, P_1 \\
    {\rm div}\,P_2 \\
    {\rm div}\,P_3 \\
  \end{array}
\right).
\end{align}

We note that $v$ is a vector field, whereas $P$, ${\rm Curl}\, P$ and ${\rm Grad}\, v$ are second order tensor fields. The corresponding Sobolev spaces will be denoted by
\begin{align}
{\rm H}({\rm Grad}\,; \Omega) \ \ \ \text{and}\ \ \ \ {\rm H}({\rm Curl}\,; \Omega)\, .
\end{align}

Furthermore, if ${T}$ is a third order tensor then we define
\begin{align}
\Div T:=(\Div T_1,\Div T_2, \Div T_3)^T\, ,
\end{align}
where $T_k=(T_{ijk})\in\mathbb{R}^{3\times3}$ are second order tensors. We recall that if $\mathbb{C}$ is a  fourth order tensor and $X\in \mathbb{R}^{3\times 3}$, then $\C. X\in \mathbb{R}^{3\times 3}$ with the components
\begin{align}
(\C.X)_{ij}=\sum\limits_{k=1}^{3}\sum\limits_{l=1}^{3}\mathbb{C}_{ijkl}X_{kl}\, ,
\end{align}
and
$\C^T. X\in \mathbb{R}^{3\times 3}$ with the components
\begin{align}
(\C^T.X)_{kl}=\sum\limits_{i=1}^{3}\sum\limits_{j=1}^{3}\mathbb{C}_{ijkl}X_{ij}\, .
\end{align}
If  $\mathbb{G}$ is a fifth  order tensor and $\mathbb{L}$ a sixth order tensor , then
\begin{align}
\mathbb{G}.\, Y\in \mathbb{R}^{3\times 3\times 3}\,\ \ \ \text{for all}\ \ Y\in \mathbb{R}^{3\times 3},\ \ \ (\mathbb{G}.\, Y)_{ijk}=\sum\limits_{m=1}^{3}\sum\limits_{n=1}^{3}\mathbb{G}_{mnijk}X_{mn},
\end{align}
and
\begin{align}
\mathbb{L}. Z\in \mathbb{R}^{3\times 3\times 3}\,\ \ \ \text{for all}\ \ Z\in \mathbb{R}^{3\times 3\times 3}, \ \ \ (\mathbb{L}. Z)_{ijk}=\sum\limits_{m=1}^{3}\sum\limits_{n=1}^{3}\sum\limits_{p=1}^{3}\mathbb{L}_{ijkmnp}Z_{mnp}\, .
\end{align}
\section{Formulation of the problem. Preliminaries}\setcounter{equation}{0}\label{Formulation}

We consider a micromorphic continuum which occupies a bounded domain $\Omega$ and is bounded by the piecewise smooth
surface $\partial \Omega$. Let $T>0$ be a given time. The motion of the body is referred to a fixed system of rectangular Cartesian axes $Ox_i$, $(i=1,2,3)$.

\subsection{Eringen's linear asymmetric micromorphic elastodynamics revisited}\label{Eringenmodel}

In this subsection, we present the initial-boundary value problem of the linear asymmetric micromorphic theory introduced by Eringen \cite{Eringen99}, which is basically identical to Mindlin's theory of elasticity with microstructure \cite{Mindlin64}. The micro-distortion (plastic distortion) $P=(P_{ij}):\Omega\times [0,T]\rightarrow \mathbb{R}^{3 \times 3}$  describes the substructure of the material which can rotate, stretch, shear and shrink, while $u=(u_i) :\Omega\times [0,T]\rightarrow  \mathbb{R}^3$  is the displacement of the macroscopic material points. In this dynamic micromorphic theory, the basic equations in strong form consist of the equations of motion
\begin{align}\label{EImotion}
\varrho\, {u}_{,tt}&=\Div \widehat{\sigma}+f,\\\notag
\varrho\, I.\,{P}_{,tt}&=\Div \widehat{m} +\widehat{\sigma}-s+M, \text{\ \ in \ \ } \Omega\times [0,T],
\end{align}
the constitutive equations
\begin{align}\label{EIconst}
&\widehat{\sigma}=\widehat{\mathbb{C}}.\,e+\widehat{\mathbb{E}}.\, \varepsilon_p+\widehat{\mathbb{F}}.\,\gamma,\notag\\
&s=\widehat{\mathbb{E}}^T.\,e+{\mathbb{H}}.\,\varepsilon_p+\widehat{\mathbb{G}}.\,\gamma,\\
&\widehat{m}=\widehat{\mathbb{F}}.\, e+\widehat{\mathbb{G}}.\,\varepsilon_p+\widehat{\mathbb{L}}.\,\gamma, \notag \text{\ \ in \ \ } \overline{\Omega}\times [0,T],
\end{align}
exclusively depending on the set of {\it independent  constitutive variables }
\begin{align}\label{EIkin}
e:=\nabla u-P, \quad \quad \quad \varepsilon_p:=\sym P, \quad \quad \quad \gamma:=\nabla P, \ \ \ \ \text{\ \ in \ \ } \overline{\Omega}\times [0,T].
\end{align}
The symmetric part of $e$ corresponds to the difference of material strain $\varepsilon$
and microstrain $\varepsilon_p$, whereas its skew-symmetric part accounts for the relative
rotation of the material with respect to the substructure. Various strain measures for the Cosserat continuum have been  extensively discussed in \cite{Eremeyev1,Eremeyev2}.
 Using the assumption of small strains and assuming skew-symmetry of $P$ the strain
measures \eqref{EIkin} coincide with the natural Cosserat strain measures which are
non-symmetric, in general.

The quantities involved in the above system of equations have the following physical signification:
\begin{itemize}
\item $(u,P)$ are the {\it kinematical variables},
\item $\varrho$ is the reference mass density,
\item $I$ is the microinertia tensor (second order),
\item $\widehat{\sigma}$ is   the  force-stress tensor (second order, in general non-symmetric),
\item $s$ is the microstress tensor (second order, symmetric),
\item $\widehat{m}$ is the moment stress  tensor (micro-hyperstress tensor, third order,  in general non-symmetric),
\item $u$ is the displacement vector (translational degrees of freedom),
\item $P$ is the micro-distortion tensor (``plastic distortion'', second order, non-symmetric),
\item $f$ is the body force,
\item $M$ is the body moment tensor (second order, non-symmetric),
\item $e:=\nabla u-P$ is the elastic distortion (relative distortion, second order, non-symmetric),
\item $\varepsilon_e:=\sym e=\sym(\nabla u-P)$ is the elastic strain tensor (second order, symmetric),
\item $\varepsilon:=\sym\nabla u$ is the total strain tensor (material strain tensor, second order, symmetric),
\item $\varepsilon_p:=\sym P $ is the micro-strain tensor (``plastic strain'', second order, symmetric),
\item  $\gamma:=\nabla P\in\mathbb{R}^{27}$ is the micro-curvature tensor (third order),
\item $\widehat{\mathbb{C}}=(\widehat{\mathbb{C}}_{ijmn})$, ${\mathbb{H}}=({\mathbb{H}}_{ijmn})$, $\widehat{\mathbb{E}}=(\widehat{\mathbb{E}}_{ijmn})$, $\widehat{\mathbb{F}}=(\widehat{\mathbb{F}}_{ijmnp})$ and $\widehat{\mathbb{G}}=(\widehat{\mathbb{G}}_{ijmnp})$ are tensors determining  the constitutive coefficients which satisfy the symmetry relations
    \begin{align}
    \widehat{\mathbb{C}}_{ijmn}&= \widehat{\mathbb{C}}_{mnij},\quad  \quad \quad \notag\\ {\mathbb{H}}_{ijmn}&={\mathbb{H}}_{mnij}={\mathbb{H}}_{jimn},\quad \quad \quad \widehat{\mathbb{E}}_{mnij}=\widehat{\mathbb{E}}_{mnji},\quad \quad\quad  \ \widehat{ \mathbb{G}}_{ijmnp}=\widehat{\mathbb{G}}_{jimnp},
    \end{align}
 \item The tensor $\widehat{\mathbb{L}}$ determines various {\it characteristic length scales} in the model, its unit is  [$\mathrm{MPa}\cdp \mathrm{m}^2$] and it satisfies the symmetry relations
    \begin{align}
    \widehat{\mathbb{L}}_{ijkmnp}=\widehat{\mathbb{L}}_{mnpijk}\, .
    \end{align}
\end{itemize}
The symmetries of $\widehat{\mathbb{E}}$ and $\widehat{\mathbb{G}}$ imply that $ \widehat{\mathbb{E}}, \widehat{\mathbb{G}}: \mathbb{R}^{3\times 3} \rightarrow\Sym(3)$. Thus, the microstress tensor $s$ is always symmetric. In contrast, the symmetries of $\widehat{\mathbb{C}}$ do not imply that $\widehat{\mathbb{C}}$ maps symmetric matrices into  symmetric matrices, while  $\mathbb{H}:\Sym(3)\rightarrow\Sym(3)$ has this property, as the classical elasticity tensor. For micro-isotropic materials the  microinertia tensor is given by $I=\frac{1}{3} J\cdot \id$, where $J$ is a known  scalar function on $\Omega$.

Using the micro-strain tensor $\varepsilon_p=\sym P$ instead of $P$ itself in the list of independent constitutive variables \eqref{EIkin} is mandatory for frame-indifference. The above equations lead to a system of 12 linear partial differential equations of Lam\'{e} type for the functions $u$ and $P$. In order to study the existence of solution of the resulting  system, \rm Hlav\'{a}\v{c}ek \cite{Hlavacek69}, Ie\c san and Nappa \cite{IesanNappa2001} and Ie\c san \cite{Iesan2002} considered null boundary conditions, i.e.
\begin{align}\label{EIbc}
u(x,t)=0 \quad \quad \text{and the {\it strong anchoring condition}}\quad \quad \ P(x,t)=0, \ \ \text{on} \ \ \partial \Omega\times(0,T).
\end{align}
We adjoin the initial conditions
\begin{align}
u(x,0)=u^0(x), \quad \ \ \ P(x,0)=P^0(x), \quad \ \ \ \dot{u}(x,0)=\dot{u}^0(x), \quad \ \ \ \dot{P}(x,0)=\dot{P}^0(x), \quad \ \ \text{on} \ \ \overline{\Omega},
\end{align}
where the quantities on the right-hand sides are prescribed, satisfying $u^0(x)=0$ and $P^0(x)=0$ on $\partial \Omega$.

 The  system of governing equations \eqref{EImotion}  is derived from the following elastic free energy
\begin{align}
2&\mathcal{E}(e,\varepsilon_p,\gamma)=\langle \widehat{\mathbb{C}}.\,(\nabla u-P),(\nabla u-P)\rangle
+\langle {\mathbb{H}}. \, \sym P,\sym P\rangle+
\langle \widehat{\mathbb{L}}.\,\nabla P,\nabla P\rangle\\\notag\ \ \ \ &\quad\quad\quad\quad\quad\quad+
2\langle \widehat{\mathbb{E}}.\,  \sym P ,(\nabla u-P)\rangle+
2\langle \widehat{\mathbb{F}}.\, \nabla P,(\nabla u-P)\rangle+
2\langle \widehat{\mathbb{G}}.\, \nabla P, \sym P \rangle\, ,\notag\\
\widehat{\sigma}&=D_e\, \mathcal{E}(e,\varepsilon_p,\gamma)\in\mathbb{R}^{3\times 3},\quad\quad
s=D_{\varepsilon_p}\, \mathcal{E}(e,\varepsilon_p,\gamma)\in \Sym(3),\quad\quad
\widehat{m}=D_\gamma\, \mathcal{E}(e,\varepsilon_p,\gamma)\in\mathbb{R}^{3\times 3\times 3}.\notag
\end{align}
Since the {\it elastic distortion}  $e:=\nabla u-P$ is in general {\it non-symmetric}, in this model the relative force-stress tensor $\widehat{\sigma}$ is also non-symmetric.
In case that $P$ is assumed to be  purely skew-symmetric, this model turns into the linear Cosserat model after orthogonal projection of the equation for the micro-distortion to the skew-symmetric subspace (see the Subsection \ref{cosapp}). The status  of the linear Cosserat model as a useful description of real material behaviour is still doubtful\footnote{for more details the reader may consult: http://www.uni-due.de/mathematik/ag\_neff/neff\_elastizitaetstheorie.} \cite{Neff_ZAMM05,Neff_Jeong_Conformal_ZAMM08} as far its application to classical heterogeneous materials is concerned even if the asymmetry of the stress tensor may be of use for some suitably conceived engineering metamaterials as e.g. phonon crystals. The existence results from  \cite{Hlavacek69,IesanNappa2001,Iesan2002} are established assuming that the  energy $\mathcal{E}$ is  a pointwise positive   definite quadratic form in terms of  the {\it independent constitutive variables}  $e$, $\varepsilon_p$ and $\gamma$, i.e. there is a positive constant $c^+$ such that
\begin{align}
\widehat{\mathcal{E}}(\nabla u-P,\sym P, \nabla P)\geq c^+\left(\|\nabla u-P\|^2+\|\sym P\|^2+\|\nabla P\|^2\right)\, .
\end{align}

A general feature of the asymmetric micromorphic model is its regularizing influence on the solution when coupled with other effects, e.g.  incompressible plasticity  is regularized by adding Cosserat effects, see  \cite{Neff_Chelminski03a,Neff_Chelminski05_dyn,Neff_Chelminski08,Neff_Muench_simple_shear09, Neff_Knees06,Neff_Muench_transverse_cosserat08,ForestJMPS10}.
When the body possesses a center of symmetry, the tensors $\widehat{\mathbb{F}}$ and $\widehat{\mathbb{G}}$ have to vanish. Thus, for  centro-symmetric elastic materials the two mixed terms $\langle \widehat{\mathbb{F}}.\, \nabla P,\nabla u-P\rangle$ and $\langle \widehat{\mathbb{G}}.\, \nabla P, \sym P\rangle$ are absent. The centro-symmetry of the material does not imply that $\widehat{\mathbb{E}}$ vanishes \cite{Eringen99}.  In the following we omit for simplicity the mixed term $\langle \widehat{\mathbb{E}}.\, \sym P,\nabla u-P\rangle$ in the energy since on the one hand its physical significance is unclear and it would induce nonzero relative stress $\widehat{\sigma}$ for zero elastic distortion $e=\nabla u-P=0$. Moreover, we show in the Subsections \ref{cosapp}, \ref{microstretchapp} and \ref{voidapp} how our energy without any mixed terms leads, in principle, to complete equations for the Cosserat model, the microstretch model and the microvoids model in dislocation format. This is a consequence of our choice of the {\it independent  constitutive variables}\footnote{For instance in the microvoids theory proposed by Cowin and Nunziato \cite{NunziatoCowin79,CowinNunziato83} and by Ie\c san \cite{IesanVoid} the presence of such mixed terms is mandatory because their absence leads to uncoupled equations, and thus to the incapability to take into account the microstructure effects. The same remark applies to the Eringen-Claus  isotropic model for dislocation dynamics \cite{EringenClaus}. This effect is due to an unfavourable  choice of the set of independent constitutive variables.}.  However, mixed terms may appear if
homogenization techniques are used, see \cite{Forest98b,Forest02,ForestTrinh11}. Our mathematical analysis can be extended in a straightforward manner to the case when the mixed terms are also present in the total energy.

If in the classical asymmetric micromorphic model, in the formal limits, we assume that the coefficients $\widehat{\mathbb{C}}\rightarrow\infty$  then this leads to the free energy from the gradient elasticity model \cite{Neff_Jeong_ZAMP08,Neff_Jeong_IJSS09} in which we do not have mixed terms either. Indeed, in this case $P=\nabla u$ and in consequence, for centro-symmetric materials, the free energy will reduce to
\begin{align}\label{grpl}
2\mathcal{E}(\sym\nabla u,\nabla (\nabla u))&=\langle {\mathbb{H}}. \, \sym\nabla u,\sym\nabla u\rangle+
\langle \widehat{\mathbb{L}}.\,\nabla (\nabla u),\nabla (\nabla u)\rangle\\
&=\langle {\mathbb{H}}. \, \sym\nabla u,\sym\nabla u\rangle+
\langle \widehat{\mathbb{L}}.\,D^2 u,D^2 u\rangle\, .\notag
\end{align}
Hitherto, the asymmetric micromorphic model has best been seen and motivated as a higher gradient elasticity model, in which the second derivatives have been replaced by a gradient of a new field \cite{SciarraFdICoussy,FdIGuarascio,FdIRosa,FdISepp,FdIPlacidiMadeo}. A distinctive feature of such a  second gradient elasticity model is that the {\it local force stresses} always remain {\it symmetric} \cite{IsolaSciarraVidoliPRSA,Neff_Svendsen08} and are characterized by the two classical Lam\'{e} constants $\mu, \lambda$ in the isotropic case. Therefore, the close connection between the classical micromorphic model and the gradient elasticity model is apparent. In many cases, the classical micromorphic model is thus used as a ``cheap" 2nd order numerical replacement for the ``expensive" 4th order model \cite{Neff_Jeong_ZAMP08,Neff_Jeong_IJSS09,Neff_Sydow_Wieners08,ForestJMPS10}. Let us remark that if in the free energy from isotropic strain gradient elasticity we replace the terms
\begin{equation}
\mathcal{E}(\nabla u,\nabla(\sym \nabla u))= \mu\,\|\sym \nabla u\|^2+\frac{\lambda}{2}\,[\tr(\nabla u)]^2+\mu L_c^2 \,\|\nabla(\sym \nabla u)\|^2,
\end{equation}
with
\begin{equation}
\mathcal{E}(\nabla u,P,\nabla P)=\mu\,\|\sym \nabla u\|^2+\frac{\lambda}{2}\,[\tr(\nabla u)]^2+\varkappa^+\mu \,  \|\sym \nabla u-\sym P\|^2+\mu L_c^2 \, \|\nabla(\sym P)\|^2,
\end{equation}
where $\varkappa^+$ is a dimensionless penalty coefficient, then Forest's microstrain theory (3+6 parameter theory) \cite{Forest06} (see Subsection \ref{forestapp}) is nothing else but a penalyzed strain gradient elasticity formulation. In the case of a strain gradient material both, the {\it local force  stresses and the  total force stresses} are {\it symmetric}, see the discussion from Subsection \ref{str-grad-sapp}.

Therefore, the asymmetry of the force stress tensor in a continuum theory is not a consequence of the presence of microstructure in the body, it is rather a constitutive assumption \cite{Bors}. Moreover, for an isotropic strain gradient material it is easy to see that both, the local force stresses and the nonlocal force stresses can be chosen symmetric, see Subsection \ref{str-grad-sapp}. We will deal with the complete modeling issue in another contribution.

\subsection{The relaxed micromorphic continuum  model}\label{Sectrelaxmodel}

The ultimate goal of  science is the reduction to a minimum of necessary complexity in the description of nature.  In the classical asymmetric micromorphic theory  there are involved more than 1000 constitutive coefficients in the general anisotropic case, and even for  isotropic materials the constitutive equations contain a great number of material constants (7+11 parameters in Mindlin's and Eringen's theory \cite{Eringen64,Mindlin64,EringenSuhubi2}). Unfortunately, this makes the general micromorphic model suitable for anything and nothing and has severely hindered the application of micromorphic models.

We consider  here a relaxed version of the classical micromorphic model with {\it symmetric Cauchy-stresses} $\sigma$ and drastically reduced numbers of constitutive coefficients. More precisely, our model is a subset of the classical model in which we allow the elasticity tensors $\widehat{\mathbb{C}}$ and $\widehat{\mathbb{L}}$ to become positive-semidefinite only. The proof of the well-posedness of this model \cite{GhibaNeffExistence} necessitates the {\it application of new mathematical tools}   \cite{NeffPaulyWitsch,NPW2,NPW3,BNPS1,BNPS2,BNPS3,LNPzamp2013}. The curvature dependence is reduced to a dependence only on the {\it micro-dislocation tensor} $\alpha:=\Curl e=-\Curl P\in\mathbb{R}^{3\times 3}$ instead of
$\gamma=\nabla P\in\mathbb{R}^{27}=\mathbb{R}^{3\times 3\times 3}$ and the local response is reduced to a dependence on the symmetric part of the elastic distortion (relative distortion) $\varepsilon_e=\sym e=\sym(\nabla u-P)$, while the {\it full kinematical degrees of freedom} for $u$ and $P$ are kept, notably {\it rotation of the microstructure remains possible}.

Our new set of {\it independent constitutive variables} for the relaxed micromorphic model is thus
\begin{align}\label{213}
\varepsilon_e=\sym(\nabla u-P), \quad \quad \quad \varepsilon_p=\sym P, \quad \quad \quad \alpha=-\Curl P.
\end{align}
 The stretch strain tensor defined in \eqref{213}$_1$ is
symmetric. For simplicity, the following systems of partial differential equations  are considered in a normalized form, i.e. the left hand sides of the equations are not multiplied with $\varrho $ or $\varrho\, I$, respectively.

We consider the following system of partial differential equations  which corresponds to this special linear anisotropic micromorphic continuum
\begin{align}\label{eqrelax}
u_{,tt}&=\dvg[ \C. \sym(\nabla u-P)]+f\, ,\\\notag
P_{,tt}&=- \crl[ \L.\crl\, P]+\C. \sym (\nabla u-P)-\H. \sym P+M\, \ \ \ \text {in}\ \ \  \Omega\times [0,T],
\end{align}
where $f :\Omega\times [0,T]\rightarrow  \mathbb{R}^3$ describes the body force and $M:\Omega\times [0,T]\rightarrow \mathbb{R}^{3 \times 3}$ describes the external body moment, $\C\!:\!\Omega\rightarrow L(\mathbb{R}^{3 \times 3},\mathbb{R}^{3 \times 3})$, $\L\!:\!\Omega\rightarrow L(\mathbb{R}^{3 \times 3},\mathbb{R}^{3 \times 3})$ and $\H\!:\!\Omega\rightarrow L(\mathbb{R}^{3 \times 3},\mathbb{R}^{3 \times 3})$
are fourth order elasticity tensors, positive definite and functions of class $C^1(\Omega)$.

For the rest of the paper we assume  that the constitutive coefficients have the following symmetries
\begin{align}\label{simetries}
&\C_{ijrs}=\C_{rsij}=\C_{jirs},\quad\quad\quad \quad \H_{ijrs}=\H_{rsij}=\H_{jirs}, \quad\quad\quad \quad {(\L)}_{ijrs}={(\L)}_{rsij}\,.
\end{align}

The system \eqref{eqrelax} is derived from the following free energy
\begin{align}\label{energyourrel}
\quad 2\,\mathcal{E}(\varepsilon_e,\varepsilon_p,\alpha)&=\langle \C.\, \varepsilon_e, \varepsilon_e\rangle
+ \langle \H.\,\varepsilon_p, \varepsilon_p\rangle+ \langle \L.\, \alpha, \alpha\rangle\\
&=\underbrace{\langle \C.\, \sym(\nabla u-P), \sym(\nabla u-P)\rangle}_{\text{elastic energy}}
+ \underbrace{\langle \H.\,\sym P, \sym P\rangle}_{\text{microstrain self-energy}\footnotemark}+ \underbrace{\langle \L.\, \Curl P, \Curl P\rangle}_{\text{dislocation energy}},\notag
\\\notag
&\hspace{-1.7cm}\sigma=D_{\varepsilon_e} \, \mathcal{E}(\varepsilon_e,\varepsilon_p,\alpha)\in \Sym(3),\quad \quad
s=D_{\varepsilon_p} \, \mathcal{E}(\varepsilon_e,\varepsilon_p,\alpha),\in \Sym(3),\quad \quad
m=D_{\alpha} \, \mathcal{E}(\varepsilon_e,\varepsilon_p,\alpha)\in \mathbb{R}^{3\times3}.
\end{align}
\footnotetext{In gradient plasticity theory, this term introduces {\it linear kinematic hardening} \cite{Neff_Svendsen08,Neff_Chelminski07_disloc,Neff_techmech07,Muehlhaus91,Forest02b,NN-SIAM12,NN13}, hence our notation $\H$. Kinematic hardening changes the state of the material, therefore, in a purely elastic setting, and in the interpretation provided by the elastic gauge theory of dislocations \cite{Popov92,Popov94a,Popov94b,Lazar2008,Lazar2009}, such a term may not appear.}

Note again that in this theory, the elastic distortion $e=\nabla u-P$ may still be non-symmetric but the possible asymmetry of $e$ does not produce a related asymmetric stress contribution. The comparison with the classical Eringen's equations \eqref{EImotion}--\eqref{EIkin} is achieved  through observing again that
\begin{align}\label{sCg}
&\langle \widehat{\mathbb{C}}.X,X\rangle_{\mathbb{R}^{3\times 3}}:=\langle \mathbb{C}.\sym X,\sym X\rangle_{\mathbb{R}^{3\times 3}},\\\notag
&\langle \widehat{\mathbb{L}}.\nabla P,\nabla P\rangle_{\mathbb{R}^{3\times 3\times 3}}:=\langle \mathbb{L}_c.\Curl P,\Curl P\rangle_{\mathbb{R}^{3\times 3}}
\end{align}
define only {\it positive semi-definite tensors $\widehat{\mathbb{C}}$ and $\widehat{\mathbb{L}}$} in terms of {\it positive definite tensors $\mathbb{C}$ and $\mathbb{L}_c$} acting on linear subspaces of $\gl(3)\cong \mathbb{R}^{3\times3}$. More precisely
\begin{align}
&\widehat{\mathbb{C}}:\mathbb{R}^{3\times 3}\rightarrow \mathbb{R}^{3\times 3},\quad\quad \quad
\widehat{\mathbb{L}}:\mathbb{R}^{3\times 3\times 3}\rightarrow \mathbb{R}^{3\times 3\times 3},
\end{align}
while
\begin{align}
&\mathbb{C}:\Sym(3)\rightarrow \Sym(3),\quad\quad \quad
\mathbb{L}_c:\mathbb{R}^{3\times 3}\rightarrow \mathbb{R}^{3\times 3}.
\end{align}
We assume that the new fourth order elasticity tensors $\C$, $\L$ and $\H$ are positive definite.
Then, there are  positive numbers ${c_M}$, ${c_m}$ (the maximum and minimum elastic moduli for $\C$), ${(L_c)}_M$ ,${(L_c)}_m$ (the maximum and minimum moduli for $\L$) and $h_M$, $h_m$ (the maximum and minimum moduli for $\H$) such that
\begin{align}\label{posdef}
{c_m}\|X \|^2\leq \langle\,\C. X,X\rangle\leq {c_M}\|X \|^2\,\quad  \ \ &\text{for all }\ \ X\in\Sym(3),\notag
\\
{(L_c)}_m\|X \|^2\leq \langle \L. X,X\rangle\leq {(L_c)}_M\| X\|^2\,\quad \ \ &\text{for all }\ \ X\in\mathbb{R}^{3\times 3},
\\
h_m\|X \|^2\leq \langle \H. X,X\rangle\leq h_M\| X\|^2\, \quad \ \ &\text{for all }\ \ X\in\Sym(3),\notag
\end{align}
Further we  assume, without loss of generality, that ${c_M}$, ${c_m}$, ${(L_c)}_M$, $h_M$, $h_m$ and ${(L_c)}_m$ are constants.

Our new approach, in marked contrast to classical asymmetric micromorphic models, features
a {\it symmetric Cauchy stress tensor} $\sigma=\C. \sym(\nabla u-P)$. Therefore, the linear Cosserat approach (\cite{Neff_ZAMM05}: $\mu_c>0$) is excluded here. Compared with Forest's microstrain theory \cite{Forest06}, the local force stress is similar, however, the micromorphic distortion  $P$ in our new model is not necessarily symmetric but endowed  with a weakest curvature response defined in terms of the {\it micro-dislocation} tensor $\alpha=-\Curl P$. The skew symmetric part is uniquely determined by the solution $P$ of the boundary value problem.

The relaxed formulation proposed  in the present  paper still shows size effects and smaller samples are relatively stiffer.  It is clear to us that for this reduced model of relaxed micromorphic elasticity {\it unphysical  effects of singular stiffening behaviour for small sample sizes} ("bounded stiffness", see \cite{Neff_JeongMMS08}) {\it cannot appear}. In case of the isotropic Cosserat model this is only true for a reduced curvature energy depending only on $\|\dev \sym \Curl P\|^2$, see the discussion in \cite{Neff_JeongMMS08,Neff_Paris_Maugin09}. Remarkably, the necessary property of bounded stiffness is impossible to obtain for the indeterminate couple stress model (elastic energy $\sim\|\sym \nabla u\|^2+\|\nabla\skew \nabla u\|^2$,  \cite{Neff_JeongMMS08}). Whether bounded stiffness is true for the general strain gradient model (elastic energy $\sim\|\sym \nabla u\|^2+\|\nabla\sym \nabla u\|^2$, \cite{Neff_JeongMMS08}) or the general gradient elasticity model (elastic energy $\sim\|\sym \nabla u\|^2+\| D^2 u\|^2$) is unclear.

 The model introduced by Teisseyre  \cite{Teisseyre74} for the study of seismic wave propagation due to earthquake processes \cite{Teisseyre84,Teisseyre00,Teisseyre01,Neff_Jeong_Conformal_ZAMM08} is also taking  a symmetric relative stress tensor (see \cite{Teisseyre74}, p. 204 and 208) and in fact it is a particular case  of the micromorphic approach to dislocation theory proposed by Eringen and Claus \cite{Eringen_Claus69,EringenClaus,Eringen_Claus71}. However, Teisseyre \cite{Teisseyre74,Teisseyre84,Teisseyre00,Teisseyre01} fails  in choosing a positive definite dislocation energy, see Subsection \ref{Teisapp}.

To our system of partial differential equations  we adjoin the weaker boundary conditions\footnote{Note that ${P}_i({x},t)\times n(x) =0, \ \ \ i=1,2,3$ is equivalent to ${P}_i({x},t)\cdot \tau(x) =0, \ \ \ i=1,2,3$  for all tangential vectors $\tau$ at $\partial \Omega$. The problem being posed for $P\in\textrm{H}(\textrm{Curl};\Omega)$, the variational setting only allows to prescribe tangential boundary conditions, i.e. $P_i\cdot \tau=0$ on $\partial \Omega$.} (compare with the conditions  \eqref{EIbc})
\begin{align} \label{bc}
{u}({x},t)=0, \ \ \  \text{and the {\it tangential condition}}  \quad {P}_i({x},t)\times \,n(x) =0, \ \ \ i=1,2,3, \ \ \
\ ({x},t)\in\partial \Omega\times [0,T],
\end{align}%
where $\times$  denotes the vector product, $n$ is the  unit outward normal vector at the surface  $\partial  \Omega$\,, $P_i$, $i=1,2,3$ are the rows of $P$. The model is driven by nonzero initial conditions
\begin{align}\label{ic}
&{u}({x},0)={u}_0(
x),\quad \quad \quad\dot{u}({x},0)=\dot{u}_0(
x),\quad \quad \quad
{P}({x},0)={P}_0(
x), \quad \quad \quad\dot{P}({x},0)=\dot{P}_0(
x),\ \ \text{\ \ }{x}\in \overline{\Omega},
\end{align}%
where   ${u}_0, \dot{u}_0, {P}_0$ and $\dot{P}_0$ are prescribed functions,  satisfying $u_0(x)=0$ and $P_{0i}(x)\times n(x)=0$ on $\partial \Omega$.

\begin{remark}Since $P$ is determined in $ {\rm H}({\rm Curl}\,; \Omega)$ in our relaxed model
the only possible description of boundary value is in terms of tangential traces
$P.\tau$. This follows from the standard theory of the  ${\rm H}({\rm Curl}\,; \Omega)$-space, see \cite{Raviart79}.
\end{remark}

 In contrast with  the 7+11 parameters isotropic Mindlin and Eringen model \cite{Mindlin64,Eringen64,EringenSuhubi2}, we have altogether only  seven parameters  $\mu_e,\lambda_e,  \mu_h, \lambda_h,\alpha_1, \alpha_2$, $\alpha_3$.  For isotropic materials, our system \label{eq} reads
\begin{align}\label{eqiso}
 u_{,tt}&=\dvg\,\sigma+f\, ,\\\notag
P_{,tt}&=-\Curl m+\sigma-s+M\,  \ \ \ \text {in}\ \ \  \Omega\times [0,T].
\end{align}
where
\begin{align}
\sigma&= 2\mu_e \sym(\nabla u-P)+\lambda_e \tr(\nabla u-P){\cdp} \id,\notag\\
m&=\alpha_1 \dev\sym \Curl P+\alpha_2 \skew \Curl P +\alpha_3\, \tr(\Curl P){\cdp} \id,\\
s&=2\mu_h \sym P+\lambda_h \tr (P){\cdp} \id\,. \notag
\end{align}
Thus, we obtain the complete system of linear partial differential equations in terms of the kinematical unknowns $u$ and $P$
\begin{align}\label{eqisoup}
 u_{,tt}&=\dvg[2\mu_e \sym(\nabla u-P)+\lambda_e \tr(\nabla u-P){\cdp} \id]+f\, ,\\\notag
P_{,tt}&=-\Curl [\alpha_1 \dev\sym \Curl P+\alpha_2 \skew \Curl P +\alpha_3\, \tr(\Curl P){\cdp} \id]\\&\quad\ +2\mu_e \sym(\nabla u-P)+\lambda_e \tr(\nabla u-P){\cdp} \id-2\mu_h \sym P-\lambda_h \tr (P){\cdp} \id+M\,  \ \ \ \text {in}\ \ \  \Omega\times [0,T].\notag
\end{align}

In this model, {\it the asymmetric parts of $P$} are entirely due only to {\it moment stresses} and {\it applied body moments}\,! In this sense, the {\it macroscopic} and {\it microscopic scales} are neatly {\it separated}.

The positive definiteness required for  the tensors $\mathbb{C}$, $\mathbb{H}$ and $\mathbb{L}_c$ implies for isotropic materials the following restriction upon the parameters $\mu_e,\lambda_e, \mu_h, \lambda_h, \alpha_1, \alpha_2$ and $\alpha_3$
\begin{align}\label{condpara}
\mu_e>0,\quad\quad  2\mu_e+3\lambda_e>0, \quad\quad  \mu_h>0, \quad\quad  2\mu_h+3\lambda_h>0, \quad\quad  \alpha_1>0,
\quad\quad  \alpha_2>0, \quad\quad  \alpha_3>0.
\end{align}
Therefore, positive definiteness for our isotropic model does not involve extra nonlinear side conditions \cite{Eringen99,Smith}. In our relaxed model, exclusively,  the material parameters $\mu_e,\lambda_e, \mu_h, \lambda_h$ can even be uniquely determined from homogenization theory, see \cite{Neff_STAMM04,Neff_Forest_jel05,Diebels09}: considering very large samples of an assumed heterogeneous  structure, i.e. the characteristic length tends to zero, we must have \cite{Neff_STAMM04,Neff_Forest_jel05}
\begin{align}\label{homfor}
\mu_e=\frac{\mu_h\,\mu}{\mu_h-\mu},\quad\quad\quad\quad 2\mu_e+3\lambda_e=\frac{(2\mu_h+3\lambda_h)\,(2\mu+3\lambda)}{(2\mu_h+3\lambda_h)-(2\mu+3\lambda)}\,,
\end{align}
where $\lambda$, $\mu$ are  the unique {\it macroscopic Lam\'{e} moduli} obtained in classical
experiments for large samples  and $\lambda_e,\mu_e$ are {\it isotropic scale transition parameters} that control the interaction between the macro and the micro deformation. Thus, the macroscopic Lam\'{e} moduli $\lambda$ and $ \mu$ should be always smaller than the {\it microstructural Lam\'{e} constants}  $\mu_h$ and $\lambda_h$ related to the response of a representative volume element of the substructure.

If, by  neglect of our guiding assumption, we add the anti-symmetric term $2\mu_c\skew(\nabla u-P)$ in the expression of the Cauchy stress tensor $\sigma$, where $\mu_c\geq0$ is the {\it Cosserat couple modulus}\footnote{A non symmetric local force stress tensor $\sigma$ deviates considerably from classical elasticity theory and indeed it does not appear in gradient elasticity, see the Subsection \ref{str-grad-sapp}. After more than half a century of intensive research there is no conclusive experimental evidence for the necessity of non-symmetric force stresses.  Therefore, in a purely mechanical (non-polar) context, we discard them in our model by choosing $\mu_c=0$ and this is mathematically sound! Nevertheless, some preliminary study on wave propagation show that a non-symmetric stress may be needed when dealing with very special metamaterials such as phononic crystals and lattice structures.},  then our analysis also works for $\mu_c\geq0$. The model in which $\mu_c>0$ is the isotropic Eringen-Claus  model for dislocation dynamics \cite{Eringen_Claus69,EringenClaus,Eringen_Claus71} (see also the Subsection \ref{cosapp} and \ref{ErClapp}) and it is derived from the following free energy
\begin{align}\label{XXXX}
\mathcal{E}(e,\varepsilon_p,\alpha)&=\mu_e \|\sym (\nabla u-P)\|^2+\mu_c\|\skew(\nabla u-P)\|^2+ \frac{\lambda_e}{2}\, [\tr(\nabla u-P)]^2+\mu_h \|\sym  P\|^2+ \frac{\lambda_h}{2} [\tr\,(P)]^2\notag\\&
 \quad \quad  +\frac{\alpha_1}{2}\| \dev\sym \Curl P\|^2 +\frac{\alpha_2}{2}\| \skew \Curl P\|+ \frac{\alpha_3}{2}\, \tr(\Curl P)^2.
 \end{align}
 For $\mu_c>0$ and if the other inequalities \eqref{condpara} are satisfied, the existence and uniqueness follow along the classical lines. There is no need for any new integral inequalities.

By means of a suitable decomposition \eqref{XXXX} of the Mindlin-Eringen strain energy density, we are able to attribute to the unique parameter $\mu_c $ the asymmetry of the stress tensor in the isotropic case \cite{Neff_ZAMM05}. We believe that this unique parameter plays a fundamental role in the description  of wave band-gaps in artificial metamaterials such as phononic crystals. Since the particular decomposition of the Mindlin-Eringen deformation
energy for isotropic micromorphic media which we introduce in Eq.
\eqref{XXXX} allows for isolating few additional constitutive parameters
with respect to standard Cauchy continuum theory, we may think to
associate to each of these additional parameters a particular effect
on wave propagation. Indeed, the search for wave solutions of the
set of governing equations associated to the introduced micromorphic
energy density may help to attribute a specific role to each of these
parameters. An exhaustive treatment of wave propagation in relaxed
Mindlin-Eringen media will be given in a forthcoming paper. Here,
we limit ourselves to show the most characteristic features of the
different elastic parameters introduced in this paper for the isotropic
case. To do so, we summarize the basic role of the most important
micromorphic parameters with respect to wave propagation:
\begin{itemize}
\item The parameter $\mu_{h}$ (associated to the microstrain energy $\left\Vert \mathrm{\text{\,}sym}\,{P}\,\right\Vert ^{2}$
in the energy density) regulates the propagation of acoustic waves
inside the considered medium. More particularly, when setting $\mu_{h}=0$
it can be observed that no acoustic waves can propagate in the considered
relaxed medium and hence only optic waves can propagate. This is sensible,
since when considering the limit case $\mathrm{\text{\,}sym}\,{P}=\mathrm{\text{\,}sym}\,\nabla {u}$
our relaxed model reduces to a second gradient continuum in which
$\mu_{h}$ is the only first gradient elastic parameter. It is indeed
known that only acoustic waves can propagate in second gradient continua
(see e.g. \cite{FdIPlacidiMadeo}).
\item When studying wave propagation phenomena in isotropic  micromorphic media, the fact of accounting for the curvature dependence only via the parameters $\alpha_1, \alpha_2, \alpha_3$ (the terms involved in the energy density, multiplying $\left\Vert \dev\sym\Curl{P}\,\right\Vert ^{2}$, $\left\Vert \skew\Curl{P}\,\right\Vert ^{2}$ and $[\tr (\Curl{P})]^2$, respectively) gives rise to dispersion curves (curves in the frequency/wavenumber plane) which have fixed concavity. This could, to some extent, make more difficult the fitting of the proposed relaxed model with some  very particular classes of possible material behaviours.
\item  It can be shown that the parameters $\alpha_1,\alpha_2,\alpha_3$ are related to the propagation of some particular optic waves. More
particularly, when setting  $\alpha_1=0,\,\alpha_2=0,\,\alpha_3=0$ in the considered relaxed model,
no propagation is associated to the microdisplacement field ${P}$,
which becomes an internal variable. Nevertheless, the global propagation
inside the considered relaxed medium is not affected by the presence
of  $\alpha_1,\alpha_2,\alpha_3$, since macroscopic optic and acoustic waves can always
propagate for all frequency ranges.
\item As far as  a nonvanishing Cosserat couple modulus $\mu_c>0$  (which is associated
to $\left\Vert \mathrm{\text{\,}skew}\,\left(\nabla {u}-{P}\right)\,\right\Vert ^{2}$
in the energy density) is considered in the presented relaxed model,
the micromorphic continuum starts exhibiting exotic properties which
may be of use to describe the mechanical behavior of very particular
metamaterials as lattice structures and phonon crystals. Indeed, when
setting $\mu_{c}\neq0$, the existence of frequency band-gaps is predicted by the considered micromorphic model. More particularly,
when switching on the parameter $\mu_{c}$, there exist some frequency
ranges in which neither acoustic nor optic waves can propagate. This means
that, in these frequency ranges, only standing waves can exist which
continue oscillating without propagating, thus keeping the energy trapped
in the same region. We can conclude that the modeling of such exotic
behavior is indeed directly related to the asymmetry of the stress
tensor, at least for what concerns the linearized case.
\end{itemize}
In the light of the aforementioned remarks, it is clear that the decomposition
\eqref{XXXX} of the strain energy density for the considered micromorphic media
allows for a very effective identification of the elastic parameters
and it may help in the identification of their physical meaning.

 Our model can also be  compared with the  model considered by Lazar and Anastassiadis \cite{Lazar2009}. In fact, in \cite{Lazar2009,Lazar-maugin}  a simplified static version of  the isotropic  Eringen-Claus  model for dislocation dynamics \cite{Eringen_Claus69}  has been investigated with $\mathbb{H}=0$ and $\mu_c>0$, with a focus on the gauge theory of dislocations (see Subsection \ref{Lazarapp}). However, the dynamical theory of Lazar \cite{Lazar2008,Lazar-MMS11} cannot be deduced from Mindlins dynamic theory,
since  in \cite{Lazar2008} there appears an additional gauge field which has no counterpart
in Mindlins model.

 The theory proposed by Teisseyre in \cite{Teisseyre74} is also using a symmetric force stress force and is a fully symmetric theory (see the assumption from \cite{Teisseyre74}, p. \!204) which means that $\mu_c=0$ \cite{Neff_ZAMM05,Neff_Jeong_Conformal_ZAMM08}, see Subsection \ref{Teisapp}. However, for the mathematical treatment there arises the need for new integral type inequalities which we present in the next section. In the energy density given by Teisseyre, there exists a dislocation  energy whose sign is not obvious. This is the reason why he did not take into account the influence of this energy.  Using the new results established by Neff, Pauly and  Witsch \cite{NeffPaulyWitsch,NPW2,NPW3} and by Bauer, Neff, Pauly and Starke \cite{BNPS1,BNPS2,BNPS3} we are now able to manage also  energies depending on the dislocation energy and  having symmetric Cauchy stresses \cite{GhibaNeffExistence}.

\subsection{Mathematical analysis}\label{Mathrelax}

In this subsection, for conciseness, we state only the obtained well-posedness results.  The full proof of these mathematical results are included in \cite{GhibaNeffExistence}. The boundary--initial value problem defined by the equations \eqref{eqrelax}, the boundary conditions \eqref{bc} and the initial conditions \eqref{ic} will be denoted by $(\mathcal{P})$.

In order to establish an existence
theorem for the solution of the problem $(\mathcal{P})$  we use the results of
the semigroup theory of linear operators. First, we will rewrite the initial boundary value problem
$({\mathcal{P}})$ as an abstract Cauchy problem in a Hilbert space \cite{Pazy,Vrabie}. Let us define the space
\begin{equation*}
\mathcal{X}\,{=}\,\big\{\,w=(u,v,P,K)\,|\,\ u{\in}{H}^1_0(\Omega),\quad v\in L^2(\Omega),\quad P \,{\in}\, H_0(\Curl; \Omega),\quad K \in L^2(\Omega)\big\}.
\end{equation*}

 Further, we introduce the operators
$
\dd A_1\, w=v,\quad
A_2\, {w}=\dvg[ \C. \sym(\nabla u-P)],\quad
A_3\, {w}=K,$ \break $
A_4\, {w}=- \crl[ \L.\crl\,P]+\C. \sym (\nabla u-P)-\H. \sym P\, ,$
where all the derivatives  of the functions are understood in the sense of distributions. Let $\mathcal{A}$ be the operator
$
\mathcal{A}=(A_1,A_2,A_3,A_4)
$
with domain
$$
\mathcal{D}(\mathcal{A})=\{{w}=(u,v,P,K)\in\mathcal{X}\ | \ \mathcal{A}{w}\in
\mathcal{X}\}.$$

With the above definitions, the problem $({\mathcal{P}})$
can be transformed into the following abstract equation in the Hilbert
space
$\mathcal{X}$
\begin{equation}\label{prcauchy1}
\frac{d{w}}{dt}(t)=\mathcal{A}{w}(t)+{\mathcal{F}}(t),\
\ {w}(0)={w}_0,
\end{equation}
where
$
{\mathcal{F}}(t)=\left({0},f,{0},M\right)
$ and
$
{w}_0=(u_0,\dot{u}_0, P_0, \dot{P}_0).
$

\begin{theorem}\label{teorexs1}{\rm (Existence and uniqueness  of the solution)} Assume that   $f,M\, \in C^1([0,t_1);L^2(\Omega))$,
${w}_0\in\mathcal{D}(\mathcal{A})$ and the fourth order elasticity tensors $\C$, $\L$ and $\H$ are symmetric and positive definite. Then, there exists a unique solution  ${w}\!\in\!
{C}^1((0,t_1);\mathcal{X})\cap
{C}^0([0,t_1);\mathcal{D}(\mathcal{A}))$ of the Cauchy problem
{\rm (\ref{prcauchy1}).} \hfill $\Box$
\end{theorem}

\begin{corollary}{\rm (Continuous dependence)} In the hypothesis of Theorem  {\rm \ref{teorexs1}} we have the following estimate
\begin{equation*}\barr{crl}
&&\|{w}(t)\|_\mathcal{X}\leq
\|{w}_0(t)\|_\mathcal{X}+C \dd\int
_0^{t}\left(\|f(s)\|_{{L}^2(\Omega)}+\|M(s)\|_{{L}^2(\Omega)}\right)ds ,\earr
\end{equation*}
where $C$ is a positive constant.
\hfill $\Box$
\end{corollary}

\section{Another further relaxed  problem}
\setcounter{equation}{0}

In this section, we weaken our energy expression further in the following model, where the corresponding elastic energy depends now only on the set of  {\it independent constitutive variables}
 \begin{align}
 \varepsilon_e=\sym(\nabla u-P),\quad\quad\quad \dev \varepsilon_p=\dev\sym P, \quad\quad\quad \dev\alpha=-\dev \Curl P.
 \end{align}
  In this model, it is neither implied that $P$ remains symmetric, nor that $P$ is trace-free, but only the trace free symmetric part of the micro-distortion $P$ and the trace-free part of the micro-dislocation tensor $\alpha$ contribute to the stored energy.

\subsection{Formulation of the problem}
The model in its general anisotropic form is:
\begin{align}\label{eqdev}
u_{,tt}&={\dvg}[ \C. \sym(\nabla u-P)]+f\, ,\\\notag
P_{,tt}&=- {\Curl}[ \dev [\L.\dev \Curl P]]+\C. \sym (\nabla u-P)-\H. \dev \sym P+M\, \ \ \ \text {in}\ \ \  \Omega\times [0,T].
\end{align}

In the isotropic case the model becomes
\begin{align}\label{eqisoup2}
 u_{,tt}&=\dvg[2\mu_e \sym(\nabla u-P)+\lambda_e \tr(\nabla u-P){\cdp} \id]+f\, ,\\\notag
P_{,tt}&=-\Curl [\alpha_1 \dev\sym \Curl P+\alpha_2 \skew \Curl P ]\\&\quad\ +2\mu_e \sym(\nabla u-P)+\lambda_e \tr(\nabla u-P){\cdp} \id-2\mu_h \dev\sym P+M\,  \ \ \ \text {in}\ \ \  \Omega\times [0,T].\notag
\end{align}

To the system of partial differential equations of this model we adjoin the weaker boundary conditions
\begin{align} \label{bcdev}
{u}({x},t)=0, \ \ \ \quad \quad {P}_i({x},t)\times n(x) =0, \ \ \ i=1,2,3, \ \ \
\ ({x},t)\in\partial \Omega\times [0,T],
\end{align}%
and the nonzero initial conditions
\begin{align}\label{icdev}
&{u}({x},0)={u}_0(
x), \quad\quad\quad \dot{u}({x},0)=\dot{u}_0(
x),\quad\quad\quad {P}({x},0)={P}_0(
x), \quad\quad\quad  \dot{P}({x},0)=\dot{P}_0(
x),\ \ \text{\ \ }{x}\in \bar{\Omega},
\end{align}%
where  ${u}_0, \dot{u}_0, {P}_0$ and $\dot{P}_0$ are prescribed functions,   satisfying $u_0(x)=0$ and $P_{0i}(x)\times n(x)=0$ on $\partial \Omega$.

We remark again that $P$ is not trace-free in this formulation and no projection is performed, compare with Subsection \ref{cosapp} and \ref{forestapp}. We denote the new problem defined by the above equations, the boundary conditions \eqref{bcdev} and the initial conditions \eqref{icdev} by $(\widetilde{\mathcal{P}})$.

\subsection{Mathematical analysis}

The study of problem $(\widetilde{\mathcal{P}})$ follows along the same lines as in Subsection \ref{Mathrelax}.
We consider  the operators
$
\dd \widetilde{A}_1\, w=v,\quad
\widetilde{A}_2\, {w}=\dvg[ \C. \sym(\nabla u-P)],\quad
\widetilde{A}_3\, {w}=K,\quad
\widetilde{A}_4\, {w}=- \crl[ \dev [\L.\dev\crl\, P]]+\C. \sym (\nabla u-P) $ $-\H. \dev \sym P\, ,
$
where all the derivatives  of the functions are understood in the sense of distributions, and the operator
$
\widetilde{\mathcal{A}}=(\widetilde{A}_1,\widetilde{A}_2,\widetilde{A}_3,\widetilde{A}_4)
$
with the domain
$
\mathcal{D}(\widetilde{\mathcal{A}})=\{{w}=(u,v,P,K)\in\mathcal{X} \ | \ \widetilde{\mathcal{A}}{w}\in
\mathcal{X}\}.$

\begin{theorem}\label{teorexs} Assume that   $f,M\, \in C^1([0,t_1);L^2(\Omega))$,
${w}_0\in\mathcal{D}(\widetilde{\mathcal{A}})$ and the fourth order elasticity tensors $\C$, $\L$ and $\H$ are symmetric and positive definite. Then, there exists a unique solution
${w}\!\in\!
{C}^1((0,t_1);\mathcal{X})\cap
{C}^0([0,t_1);\mathcal{D}(\widetilde{\mathcal{A}}))
$
 of the following Cauchy problem
$
\dd\frac{d{w}}{dt}(t)=\widetilde{\mathcal{A}}{w}(t)+{\mathcal{F}}(t),\
\ {w}(0)={w}_0,
$
where
$
{\mathcal{F}}(t)=\left({0},f,{0},M\right)
$ and
$
{w}_0=(u_0,\dot{u}_0, P_0, \dot{P}_0).
$
Moreover, we have the estimate
$$\|{w}(t)\|_\mathcal{X}\leq
\|{w}_0(t)\|_\mathcal{X}+C \dd\int
_0^{t}\left(\|f(s)\|_{{L}^2(\Omega)}+\|M(s)\|_{{L}^2(\Omega)}\right)ds ,$$
where $C$ is a positive constant.\hfill $\Box$
\end{theorem}

\section{New and/or existing relaxed models}\label{CompSect}
\setcounter{equation}{0}

In this section we propose a review of some existing relaxed models and we underline the possible connections between these models and the new relaxed models which we have proposed in this paper.

\subsection{Kr\"{o}ner's view}\label{Kroner}
\vspace*{-4mm}
\paragraph{\ref{Kroner}.1 \ Kr\"{o}ner's discussion of a dislocated body and the  Cosserat continuum: symmetric versus asymmetric force stresses\bigskip\\}

\bigskip

Beginning from mid 1950 Kr\"oner tried to link the theory of static
dislocations to the Cosserat model with asymmetric force stresses. However, since
1964 it was clear to Kr\"oner that the force stress $\sigma$ in the dislocation theory is
always symmetric\footnote{Kr\"{o}ner writes \cite[p.\,148]{Kroener64}: ``Im Gegensatz zu den
Momentenspannungen sind die Kraftspannungen stets symmetrisch. Dieser Befund ist
besonders wichtig, da seit Jahren bekannt ist, da{\ss} der geometrische Zustand
eines K\"orpers mit Versetzungen im allgemeinen Fall durch 15 Funktionen des Ortes
beschrieben wird...Im Gegensatz hierzu schienen 18 Funktionen des Ortes n\"otig, um
den statischen Zustand des K\"orpers vollst\"andig zu kennzeichnen, was eine
Inkonsistenz der Theorie andeutete. Der Befund, da{\ss} durch die besondere
Eigenschaft der Versetzungen, Tr\"ager der Gleitung zu sein, die drei
antisymmetrischen Freiheitsgrade des Kraftspannungstensors ausfallen, zerstreut
diese Sorgen in sehr
 befriedigender Weise."  and   ``...dabei stellte sich heraus, da{\ss} $\skew \sigma$ in der Feldtheorie der Versetzungen verschwindet, da die  $\skew \sigma$  zuzuordnende geometrische Gr\"{o}{\ss}e plastischer Natur ist..."

In our translation:

``Contrary to the moment-stresses the force stresses are always symmetric. This
statement is very important for it is known since
 many years, that the geometrical state of a dislocated body in the general case is
given through 15 functions of place... Contrary to this there seemed 18 functions of
place necessary in order to fully describe the static (equilibrium) of the body,
which seemed to indicate an inconsistency of the theory. The statement that, by the
very properties of dislocations to be carrier of slip, the three antisymmetric
degree of freedom of the force stress tensor are redundant, removes these concerns
in a elegant way." and ``...and it became obvious, that $\skew\sigma$ vanishes in the field theories of
dislocations, since the variable, which must be related to $\skew\sigma$ is of
plastic nature..."}.

\begin{figure}[htbp]
\centering
\includegraphics[width=0.45\textwidth]{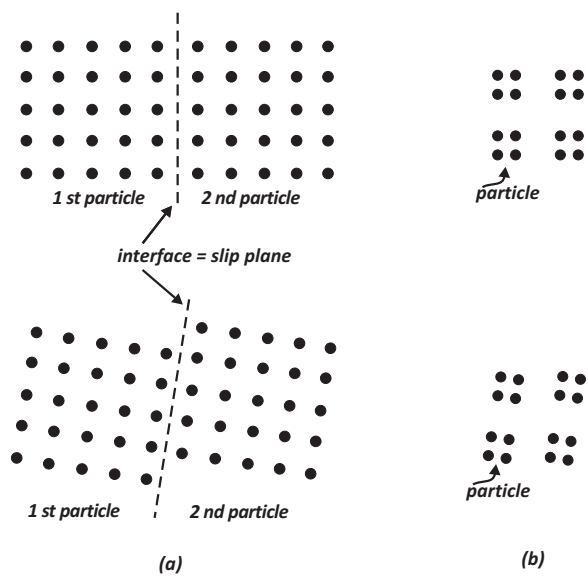}
\centering
\caption{\footnotesize{(a) Two adjacent "particles" of a crystalline body before and after a rotation through the same angle. This kind of rotation implies the slip of a dislocation along the interface. It {\it does not} change the {\it state} of the crystal. (b) Four adjacent ``particles" of a Cosserat type material before and after a rotation through the same angle. This kind of rotation {\it does} change the {\it state} of the body. }}
\label{fig}
\end{figure}

We reproduce here the old, but nevertheless refreshing and clear  comments of Kr\"{o}ner (\cite[p. 1059-1060]{Kroner}) regarding the papers by Eringen and Claus \cite{EringenClaus}, and Fox \cite{Fox70}. Kr\"{o}ner remarks: "I would like to make clear why the skew symmetric stress does not appear in dislocated bodies. Assume particles which are little crystalline domains, for instance little cubes which build up a perfect crystal. Now imagine two  of these particles to be isolated from the rest and be rotated through the same angle (Fig. \ref{fig}(a)). By this operation the atomic structure is not disturbed and the state of the crystal along the interface between particles is not changed. So there is no static response to this kind of deformation and that is why the skew symmetric part of the ordinary stresses vanishes in dislocation theory.

It does not vanish in Cosserat type theories where one considers oriented point particles which do not possess a crystalline structure (Fig. \ref{fig}(b)). Such bodies could be, for instance, non-primitive crystal lattices where atoms in a cell are so tightly bound that the deformation of a cell can be disregarded whereas the bonds between the cells are weak. In this example the cells are the particles of the Cosserat continuum; they possess the usual translational and rotational degrees of freedom. Now rotate these particles through the same angle and the body is in a different state. So you expect a response.

 I call the body described firstly a {\it dislocated body} and the other a {\it Cosserat continuum}. In the dislocated body one observes the occurrence of slip because the above described rotation of the two crystalline domains implies the slip of a dislocation along the interface between them. Slip has no meaning in the usual Cosserat continuum." The comments of Teodosiu\footnote{Teodosiu's comments: ``As long as we are concerned with developing a theory of continuum mechanics in which new invariant kinematical and dynamical quantities are involved, we can do it in a rather elegant way, and I think we have so far two beautiful examples. One is the theory of micropolar mechanics developed by Green and Rivlin, and the other that of micromorphic materials developed by Eringen and Suhubi. These two theories provide a good framework for developing other general theories of physical phenomena. But if we intend to describe new physical phenomena with these theories, we must be very careful when approaching physical objects to consider the descriptions by people studying such quantities. [...] In the meantime there is another point which is not quite clear to me: It is true that in previous developments of dislocation theory no asymmetric stress appears, and there is a good reason for this. We do not have in dislocation theory, in fact, a new independent degree of freedom such as a rotation, and as long as we don't have such independent rotations, we are always able to redefine the stress tensor in order to make it symmetrical. So there is no asymmetric part of the stress tensor in dislocation theory, as long as we don't introduce a new rotation. This question has also been discussed some time ago by Professor Kr\"{o}ner. "} \cite[p. 1053-1054]{Teodosiu} regarding the papers by Eringen and Claus \cite{EringenClaus}, and Fox \cite{Fox70} enforce the  Kr\"{o}ner's point of view.  In order to defend their theory \cite{EringenClaus}, Claus \cite[p. 1054-1055]{ClausIUTAM} gave the following answered to Teodosiu's comments: ``If one includes extra degrees of freedom into the angular momentum equation, we claim that the equation lead to a non-symmetric stress tensor, whether it is a couple stress or a stress moment tensor. [...] That is precisely the problem. Everybody these days is looking for situations in which the stress is non-symmetric. In continuum mechanics many people are trying  to think along these lines. Some of the areas of promise to be pointed out are liquid crystal experiments where inherently there is a structure to the liquid which could conceivably lead to a non-symmetric stress tensor. Another area is in a body which contains a polarization, and the behavior of that body in an internal field. Many people are trying to look for asymmetries there. But I cannot quote an experimental paper where it has been demonstrated.[...] Concerning the question about elastic and plastic distortion, the interpretation here is that we have a body with dislocations that are deforming elastically so there is no slip in a lattice sense. There is no plastic deformation taking place; you put loads on the body and get only elastic reactions. Obviously what we are trying to construct is a plasticity theory, and we think we have the beginning of a mechanism to do that." Moreover, Eringen said \cite[p. 1054-1055]{ClausIUTAM}: ``The ultimate goal of the present theory is to determine the motions and micromotions by solving an initial-boundary-value problem. Once they are determined, the dislocation density can be calculated in a straightforward manner. This point of view is, perhaps, in clash with the long-established traditions in other well-developed fields of continuum physics, I suggest that the continuum dislocation theory should offer a set of field equations subject to a set of well-posed initial and boundary conditions to predict the evolution of the motion and of the dislocations. Present practice in this field requires that the distribution of a second-oder tensor (the dislocation density) be given throughout the body at all times in order that we determine another second order tensor, namely, the stress tensor. This is not only unreasonable on logical grounds, but also not feasible experimentally. After all, why not ask for the stress tensor in the first place!".

\paragraph{\ref{Kroner}.2 \ The Popov-Kr\"{o}ner dislocation model\bigskip\\}\label{Popovapp}

\bigskip
If we combine  the dynamic model \cite{Popov92} of Popov with the Popov-Kr\"{o}ner static model of elastoplastic media with mesostructure \cite{Kroener01,Popov94b} we obtain the following equations
\begin{align}\label{ppeq}
 {u}_{,tt}&=\Div [{\mathbb{C}}.\,\sym(\nabla u- P)]+f,\\\notag
P_{,tt}&=-\Curl(\alpha_1\, \dev\sym\Curl P+\alpha_2\, \skew \Curl P) +{\mathbb{C}}.\,\sym(\nabla u-P)+ M, \text{\ \ in \ \ } \Omega\times [0,T]\, ,
\end{align}
where $\alpha_1, \alpha_2>0$. The Popov-Kr\"{o}ner model is derived from the  internal free energy
\begin{align}
2\mathcal{E}(\varepsilon_e, \alpha)=\langle \C.\, \sym(\nabla u-P), \sym(\nabla u-P)\rangle
 + W_{\Curl}(\alpha),
\end{align}
where
\begin{align}
W_{\Curl}(\alpha)=\mathfrak{a}_1\|\alpha \|^2+\mathfrak{a}_2\langle\alpha,\alpha^T\rangle
+\mathfrak{a}_3[\tr (\alpha)]^2,
\end{align}
and
\begin{align}
\mathfrak{a}_1&=\frac{\mu(2d)^2}{24}\left(3+\frac{2\nu}{1-\nu}\right),\quad\quad\quad
\mathfrak{a}_2=-\frac{\mu(2d)^2}{24}\frac{2\nu}{1-\nu},\quad\quad\quad
\mathfrak{a}_3=-\frac{\mu(2d)^2}{24},\\ \nu&=\frac{\lambda}{2(\mu+\lambda)}\quad\quad\quad \text{(the Poisson's ratio)},\quad\quad\quad  -1<\nu<\frac{1}{2}.\notag
\end{align}

The energy $W_{\Curl}$ can be expressed  in terms of $\dev\sym\Curl P,\skew\Curl P$ and $\tr(\Curl P)$ as in the following
\begin{align}
W_{\Curl}(\alpha)&=(\mathfrak{a}_1+\mathfrak{a}_2)\|\dev\sym\Curl P\|^2+(\mathfrak{a}_1-\mathfrak{a}_2)\|\skew\Curl P\|^2+\frac{\mathfrak{a}_1+\mathfrak{a}_2+3\mathfrak{a}_3}{3}[\tr (\Curl P)]^2\notag\\
&=\frac{3\mu(2d)^2}{24}\|\dev\sym\Curl P\|^2+\frac{\mu(2d)^2}{24}\left(3+\frac{4\nu}{1-\nu}\right)\|\skew\Curl P\|^2.
\end{align}

We remark that $\tr (\Curl P)$ is therefore, in fact, not present in the energy considered by Popov-Kr\"{o}ner \cite{Kroener01,Popov92,Popov94b} and in consequence it is also absent in the system of linear partial differential equations \eqref{ppeq}. Thus,  the Popov-Kr\"{o}ner equations \eqref{ppeq} coincide with our further relaxed model \eqref{eqisoup2} in which $\H=0$ and
\begin{align}\label{omPK}
\alpha_1=\frac{3\mu(2d)^2}{24},\quad\quad \ \alpha_2=\frac{\mu(2d)^2}{24}\left(3+\frac{4\nu}{1-\nu}\right),\quad\quad \alpha_3=0.
\end{align}
In gradient plasticity models, there is another modeling issue at work: the microstrain $\sym P$ is not a state variable, therefore it should not appear in the free energy, as such $\H=0$.  However, it may enter the equations through the notion of {\it equivalent plastic strain}, governing the isotropic linear hardening response \cite{Ebobisse_Neff09}.

Let us assume that $P$ is restricted to $\sL(3)$, which is the standard assumption in plasticity theory (plastic incompressibility, $\tr(P)=0$). By subsequent orthogonal projection of the second equation \eqref{ppeq}$_2$  to the space of trace-free matrices, the full system of equations for the Popov-Kr\"{o}ner model \cite{Kroener01,Popov92,Popov94b} become
\begin{align}\label{ppeqdev}
 {u}_{,tt}&=\Div [{\mathbb{C}}.\,\sym(\nabla u- \dev P)]+f,\\\notag
(\dev P)_{,tt}&=-\dev[\Curl(\alpha_1\, \dev\sym\Curl \dev P+\alpha_2\, \skew \Curl P)] +\dev[{\mathbb{C}}.\,\sym(\nabla u-\dev P)]+ \dev M.
\end{align}

The obtained model  \eqref{ppeqdev} is a 11 dof model, $(u,\dev P)$. In the isotropic case this is a 2+2 parameter model or a 2+1+2 parameter model if $\dev\sym P$ is taken into account as a constitutive variable. The constitutive variable $\sym P$ is not a state variable in this model. The model follows the line of argument given by Kr\"{o}ner\footnote{``In contrast to the moment stresses, the force stress tensor is always symmetric''} \cite[p. 148]{Kroener64}.

\subsection{Forest's  approach}\label{forestapp}

\paragraph{\ref{forestapp}.1 Forest's  dynamic microstrain model \bigskip\\}
\bigskip

In this Subsection we give a short description of the linear dynamic microstrain  model \cite{Forest06}. The basic system of partial differential equation of this model can be obtained assuming that the micro-distortion $P$ is restricted to $\Sym(3)$ and  by subsequent orthogonal projection of the second equation \eqref{EImotion}$_2$ (from the general Eringen's micromorphic dynamics) to the space of symmetric-matrices. In addition, the ordinary elasticity tensor   ${\mathbb{C}}:\Sym(3)\rightarrow\Sym(3)$  has to be taken, instead of Eringen's elasticity tensor  $\widehat{{\mathbb{C}}}$, since $\widehat{{\mathbb{C}}}$ does not map symmetric matrices into symmetric matrices. This leads to the system
\begin{align}\label{eqms}
 {u}_{,tt}&=\Div [{\mathbb{C}}.\,\sym(\nabla u- P)]+f,\\\notag
(\sym{P})_{,tt}&=\sym[\Div \widehat{\mathbb{L}}.\,\nabla( \sym P)] +{\mathbb{C}}.\,\sym(\nabla u-P)]+\mathbb{H}.\,\sym P+\sym M, \text{\ \ in \ \ } \Omega\times [0,T]\, .
\end{align}

The microstrain model is, however, incapable of describing rotation of the microstructure and features only $3+6$ degrees of freedom. For comparison we give the version without coupling  terms. In fact,  the mathematical problem for the microstrain  model is to find functions $u\in H^1(\Omega)$  and $\varepsilon_p=\sym P\in H^1(\Omega)$  which satisfy the partial differential equations \eqref{eqms}.  A noteworthy feature of this model is a symmetric Cauchy-stress tensor $\sigma=\C. \sym(\nabla u-P)$. The curvature is only active on the gradient of  microstrain, i.e.  curvature depends only on  $\nabla \varepsilon_p=\nabla(\sym P)$. Thus, the remaining set of {\it independent constitutive variables} in the microstrain theory is
\begin{align}
\varepsilon_e=\sym(\nabla u-P),\quad \quad \quad \varepsilon_p=\sym P, \quad \quad \quad \nabla \varepsilon_p=\nabla \sym P.
\end{align}

The total energy corresponding to the microstrain micromorphic model is given by
\begin{align}
2\,\widehat{E}(t):=\int_\Omega\bigg(\|u_{,t}\|^2&+ \|(\sym P)_{,t}\|^2+ \langle \C.\, \sym(\nabla u-P), \sym(\nabla u-P)\rangle
\vspace{2mm}\notag\\& \ \ \ \ \quad\  + \langle \H.\, \sym P,  \sym P\rangle + \langle \widehat{\mathbb{L}}.\, \nabla(\sym P), \nabla(\sym P)\rangle\bigg)dv\, .
\end{align}

  It is easy to obtain  qualitative properties (uniqueness, continuous dependence, existence) of the micro\-strain micromorphic model because we only have to use the well known Korn's inequality and the positive definiteness of $\C, \H, \widehat{\mathbb{L}}$ \cite{GhibaNeffExistence}, there is no need to specify Dirichlet boundary conditions on $\sym P$.

\paragraph{\ref{forestapp}.2 A microstrain-dislocation model without rotational degrees of freedom\bigskip\\}

 Let us consider now a new set of {\it independent constitutive variables}, i.e.
 \begin{align}
\varepsilon_e=\sym(\nabla u-P),\quad \quad \quad \varepsilon_p=\sym P, \quad \quad \quad \Curl \varepsilon_p=\Curl \sym P,
\end{align}
and the corresponding  total energy
\begin{align}
2\,\widehat{E}(t):=\int_\Omega\bigg(\|u_{,t}\|^2&+ \|(\sym P)_{,t}\|^2+ \langle \C.\, \sym(\nabla u-P), \sym(\nabla u-P)\rangle
\vspace{2mm}\notag\\& \ \ \ \ \quad\  + \langle \H.\, \sym P,  \sym P\rangle + \langle {\widehat{\mathbb{L}}}_c.\, \Curl\,\sym P, \Curl\, \sym P\rangle\bigg)dv\, .
\end{align}
Hence, the model equations are  given by
\begin{align}\label{eqms}
 {u}_{,tt}&=\Div [\,{\mathbb{C}}.\,\sym(\nabla u- P)]+f,\\\notag
(\sym{P})_{,tt}&=\sym[\Curl (\widehat{\mathbb{L}}_c.\,\Curl\, \sym P)] +{\mathbb{C}}.\,\sym(\nabla u-P)]+\mathbb{H}.\,\sym P+\sym M, \text{\ \ in \ \ } \Omega\times [0,T]\, .
\end{align}
Existence and uniqueness follows along the lines given by our relaxed model, without the need for new inequalities.

\paragraph{\ref{forestapp}.3 The microcurl model\bigskip\\}

The microcurl model is intended to furnish an approximation of a gradient plasticity model \cite{ForestJMPS10}.   The free energy of the original system reads
\begin{align}
2\mathcal{E}_\chi(\varepsilon_e, e_p,\Gamma_\chi)=\langle \C.\, \sym(\nabla u-P), \sym(\nabla u-P)\rangle
 + \langle \mathbb{L}_c.\, \Curl P, \Curl P\rangle
\end{align}
and leads to some difficulties when one implements nonlinear pde-systems due to couplings with plasticity theory \cite{Neff_Sydow_Wieners08,NN-SIAM12,NN13,Ebobisse_Neff09,Neff_Chelminski07_disloc,Reddy06b}.

 The idea then is to introduce a new micromorphic-type variable $\chi_p$ and to couple it to elasto-plasticity. The {\it independent constitutive variables} are the elastic strain tensor $\varepsilon_e=\sym(\nabla u-P)$,  the relative plastic strain $e_p=P-\chi_p$ measuring the difference between plastic deformation and the plastic microvariable, and the dislocation
density tensor $\Gamma_\chi=\Curl \chi_p$. The new free energy reads
\begin{align}
2\mathcal{E}_\chi(\varepsilon_e, e_p,\Gamma_\chi)=\langle \C.\, \sym(\nabla u-P), \sym(\nabla u-P)\rangle
 + \langle \H_\chi.\,(P-\chi_p), P-\chi_p\rangle+ \langle \mathbb{L}_\chi.\, \Curl \chi_p, \Curl \chi_p\rangle\,.
\end{align}
The quasistatic equations are
\begin{align}\label{eqrelax2}
0&=\dvg[ \C. \sym(\nabla u-P)]\, ,\\\notag
0&=- \crl[ \L.\crl\, \chi_p]+\H_\chi. (P-\chi_p)\notin\Sym(3)\,  \ \text {in general}\,,
\end{align}
together with flow rules for the plastic variable $P$ (these are missing here). Since $\chi_p\in\mathbb{R}^{3\times3}$, we have altogether 12 elastic  degrees of freedom.

Let us consider two alternative  energies (with different coupling of $P$ and $\chi_p$)
\begin{align}
 \mathcal{E}_{\chi}^{(1)} & =\langle \C.\, \sym(\nabla u-P), \sym(\nabla u-P)\rangle
 +\langle \mathbb{L}_\chi. \Curl \chi_p, \Curl \chi_p\rangle+
  \langle \H_\chi.\,(P-\chi_p), P-\chi_p\rangle,\\\notag
  \mathcal{E}_{\chi}^{(2)}
 & =\langle \C.\, \sym(\nabla u-P), \sym(\nabla u-P)\rangle
 +\langle \mathbb{L}_\chi.\Curl \chi_p, \Curl \chi_p\rangle+
    \langle \H_\chi.\,\sym(P-\chi_p), \sym(P-\chi_p)\rangle.
\end{align}
The corresponding minimization problem  in terms of the  energy
\begin{align}
\left\{\begin{array}{l}
\mathcal{E}_{\chi}^{(1)} \vspace{6mm}\\
 \mathcal{E}_{\chi}^{(2)}
\end{array}\right.=
\left\{\begin{array}{l}
\text{has a unique solution } \chi_p\in{\rm H}(\Curl,\Omega) \text{ for given }  P\in L^2(\Omega),\\
 \text{\quad \quad \quad \quad and there is no need for Dirichlet-boundary conditions (for uniqueness)},\\
\text{\quad \quad \quad \quad natural boundary conditions, determined by the variational formulation, suffice};\vspace{2mm}\\
 \text{has a solution } \chi_p\in{\rm H}(\Curl,\Omega) \text{ for given }  \sym P\in L^2(\Omega), \\
 \text{\quad \quad \quad \quad  uniqueness of } \chi_p \text{ requires tangential boundary conditions}.
\end{array}\right.
\end{align}
\subsection{The  asymmetric isotropic Eringen-Claus model for dislocation dynamics}\label{ErClapp}

This model is intended to describe a solid already containing dislocations undergoing elastic deformations: the dislocations bow out under the applied load, but do so reversibly.

The system of equations  derived by  Eringen and Claus  (\cite{EringenClaus}, Eq. (3.39)) consists, as  consequences of the balance laws of momentum and of the moment of momentum, of the following equations
\begin{align}\label{eqCE}
\varrho \,u_{l,tt}&=\widetilde{\sigma}_{kl,k}+f_l,\\
\varrho \,I{\cdp} P_{lm,tt}&=\epsilon_{kmn}m_{nl,k}+\widetilde{\sigma}_{ml}-\widetilde{s}_{ml}+M_{lm},\notag
\end{align}
where (see the constitutive equations (3.32), (3.33) and (3.41) from \cite{EringenClaus} and the equations (36) and (37) from \cite{Teisseyre73})
\begin{align}\label{cece}
\widetilde{\sigma}_{kl}&=(\overline{\lambda}+\tau)\,\varepsilon_{mm}\delta_{kl}+2(\overline{\mu}+\varsigma)\,\varepsilon_{kl}+\eta\, \overline{e}_{mm}\delta_{kl}+\overline{\nu}\, \overline{e}_{lk}+\kappa\,\overline{e}_{kl},\notag\\
\widetilde{s}_{kl}&=(\overline{\lambda}+2\tau)\,\varepsilon_{mm}\delta_{kl}
+2(\overline{\mu}+2\varsigma)\,\varepsilon_{kl}
+(2\eta-\tau)\, \overline{e}_{mm}\delta_{kl}+(\overline{\nu}+\kappa -\varsigma)\,(\overline{e}_{kl}+\overline{e}_{lk}),\\\notag
m_{kl}&=-a_3\,\alpha_{mm}\delta_{kl}-a_1\,\alpha_{kl}+(a_1-a_2+a_3)\,\alpha_{lk},
\end{align}
and the set of {\it  independent constitutive variables} (\cite{EringenClaus},  Eq. (1.7)) is
\begin{align}\label{ecicv}
\varepsilon=\sym \nabla u,\quad\quad\quad  \overline{e}=\nabla u^T+P,\quad\quad\quad
\alpha=-\Curl P.
\end{align}

The rest of the quantities have the same meaning as in Subsection \ref{Eringenmodel}. Let us remark that
\begin{align}
\varepsilon=\sym e+\varepsilon_p,\quad \quad\quad \text{and} \quad \quad\quad \overline{e}=\sym e+2 \varepsilon_p-\skew e\,
\end{align}
depend actually only on the {\it independent constitutive variables}\footnote{$\varepsilon$ and $\overline{e}$ are isomorphically equivalent with $e=\nabla u-P$ and $\varepsilon_p$.} $e, \varepsilon_p, \alpha$.

We also remark that
\begin{align}
\epsilon_{kmn}m_{nl,k}=-\epsilon_{mkn}m_{nl,k}.
\end{align}

According with the definition of the $\curl$ operator, we have
\begin{align}
(\curl v)_k=\epsilon_{klm}v_{m,l},\ \ \text{for any vector}\ \  v=(v_1,v_2,v_3)^T\in C^1(\Omega).
\end{align}
Hence, if we fix the indices $l$, then $\varepsilon_{ikn}m_{nl,k}$ gives the $i$-component of $\curl (m_{1l},m_{2l},m_{3l})$, i.e.
\begin{align}
(\epsilon_{kin}m_{nl,k})_{li}=-\big(\Curl (m^T)\big)_{li}.
\end{align}
Thus, written in terms of the operator $\Curl$, the constitutive equations \eqref{cece}$_3$ become
\begin{align}\label{ma1a2}
m=&a_3\,\tr(\Curl P){\cdp} \id+2a_1\,\skew \Curl P+(a_2-a_3)\,(\Curl P)^T.
\end{align}
In consequence, we deduce
\begin{align}
\Curl( m^T)&=\Curl \bigg[(a_2-a_3)\Curl P-2a_1\,\skew \Curl P+a_3\,\tr(\Curl P)\!\cdot\! \id\bigg]\\\notag
&=\Curl \bigg[(a_2-a_3)\,\dev\sym\Curl P+(a_2-a_3-2a_1)\,\skew \Curl P+\frac{2a_3+a_2}{3}\tr(\Curl P)\!\cdot\! \id\bigg].\notag
\end{align}
We are thus able to identify the constitutive coefficients of the dislocation energy in the Eringen-Claus model \cite{Eringen_Claus69,EringenClaus,Eringen_Claus71}  with the coefficients in our isotropic case, namely
\begin{align}\label{omE}
\alpha_1=a_2-a_3,\quad\quad \ \alpha_2=a_2-a_3-2a_1,\quad\quad\ \ \alpha_3=\frac{2a_3+a_2}{3}.
\end{align}

Regarding the  term $\widetilde{\sigma}_{ml}-\widetilde{s}_{ml}$ from the equations of motion  \eqref{eqCE}, if we take $\overline{\nu}=\kappa$, then we have only the elastic strain tensor $\varepsilon_e=\sym e=\sym(\nabla u-P)$  and the micro-strain tensor $\varepsilon_p=\sym P$ taken into account. The condition $\overline{\nu}=\kappa$  is necessary and sufficient in order to have a symmetric force-stress tensor $\widetilde{\sigma}$ (see the discussion from Subsection \ref{Kroner}), it corresponds to a  vanishing Cosserat couple modulus $\mu_c=0$.  Moreover, the force-stress tensor $\widetilde{\sigma}$ vanishes when $P=\nabla u$ if and only if $\overline{\mu}+\tau=-(\overline{\nu}+\kappa)$ and $ \overline{\lambda}+\tau=-2\overline{\nu}$. However, Eringen and Claus strictly considered $\mu_c>0$, i.e. the  asymmetry of the force stresses.

\subsection{The linear isotropic Cosserat model in terms of the dislocation density tensor}\label{cosapp}

In this subsection, we assume that the micro-distortion tensor is skew-symmetric, i.e.  $P\in \so(3)$. For isotropic materials and with the asymmetric term $2\mu_c\skew(\nabla u-P)$ incorporated \cite{BirsanEJMS09,BirsanIJES09}, by orthogonal projection of the second equation \eqref{eqiso}$_2$  to the space of skew-symmetric matrices, the full system of equations for our  reduced model is now
\begin{align}\label{eqisCos}
 u_{,tt}&=\dvg[2\mu_e \sym\nabla u+2\mu_c\skew(\nabla u-(\skew P))+ \lambda_e \tr(\nabla u){\cdp} \id]+f\, ,\\\notag
(\skew P)_{,tt}&=-\skew\Curl\bigg[\alpha_1 \dev\sym \Curl (\skew P)+\alpha_2 \skew \Curl (\skew P) +\alpha_3\, \tr(\Curl (\skew P)){\cdp} \id\bigg]\\
&\ \ \ \ \ \ +2\mu_c\skew(\nabla u-(\skew P))+\skew M\,  \ \ \ \text {in}\ \ \  \Omega\times [0,T].\notag
\end{align}
Then, switching to  $A:=\skew P$, the equations \eqref{eqisCos} become
\begin{align}\label{eqisCos1}
 u_{,tt}&=\dvg[2\mu_e \sym\nabla u+2\mu_c\skew(\nabla u-A)+ \lambda_e \tr(\nabla u){\cdp} \id]+f\, ,\\\notag
A_{,tt}&=-\skew\Curl\bigg[\alpha_1 \dev\sym \Curl A+\alpha_2 \skew \Curl A +\alpha_3\, \tr(\Curl A){\cdp} \id\bigg]\\
&\ \ \ \ \ \ +2\mu_c\skew(\nabla u-A)+\skew M\,  \ \ \ \text {in}\ \ \  \Omega\times [0,T].\notag
\end{align}
 Moreover, for antisymmetric $A\in\so(3)$ the  tangential boundary condition
\begin{align}\label{bcaxl}
{A}_i({x},t)\cdot\tau(x) =0, \ \ \ i=1,2,3\ \ \ \ \ \ \text{implies the strong anchoring condition }\ \ \ \ \ A=0\quad \text{on} \quad \partial \Omega.
\end{align}
We introduce the canonical identification of $\mathbb{R}^3$ with $\so(3)$. For
\begin{align}
A=\left(\begin{array}{ccc}
0 &-a_3&a_2\\
a_3&0& -a_1\\
-a_2& a_1&0
\end{array}\right)\in \so(3)
\end{align}
we introduce the operators $\axl:\so(3)\rightarrow\mathbb{R}^3$ and $\anti:\mathbb{R}^3\rightarrow \so(3)$ through
\begin{align}
\axl\left(\begin{array}{ccc}
0 &-a_3&a_2\\
a_3&0& -a_1\\
-a_2& a_1&0
\end{array}\right):=\left(\begin{array}{c}
a_1\\
a_2\\
a_3
\end{array}\right),\quad \quad A\cdot v=(\axl A)\times v, \quad \quad \forall v\in\mathbb{R}^3,\\\notag \quad A_{ij}=\sum\limits_{k=1}^3-\epsilon_{ijk}(\axl A)_k=:\anti(\axl A)_{ij}, \quad \quad (\axl A)_k=\sum\limits_{i,j=1}^3 -\frac{1}{2} \epsilon_{ijk}A_{ij}\,,
\end{align}
 where $\epsilon_{ijk}$ is the totally antisymmetric third order permutation tensor.
 We also have the following identities (see \cite{Neff_curl06}, Nye's formula \cite{Nye53} )
\begin{align}\label{curlaxl}
-\Curl A&=(\nabla \axl A)^T-\tr[(\nabla \axl A)^T]{\cdp} \id,\\
\nabla \axl A  &= -(\Curl A)^T+\frac{1}{2}\tr[(\Curl A)^T]{\cdp}\id,
\end{align}
 for all matrices $A\in \so(3)$. Using the above Curl-$\nabla\axl$ identity, it is simple to obtain
\begin{align}
\alpha_1 \dev\sym \Curl A&+\alpha_2 \skew \Curl A +\alpha_3\, \tr(\Curl A){\cdp} \id=\\\notag
&-\alpha_1 \dev\sym (\nabla \axl A)^T-\alpha_2 \skew (\nabla \axl A)^T -\alpha_3\, \tr(\nabla \axl A)^T{\cdp} \id+\\\notag
&+\alpha_1 \dev\sym \tr[(\nabla \axl A)^T]{\cdp} \id+\alpha_2 \skew \tr[(\nabla \axl A)^T]{\cdp} \id +\alpha_3\, \tr(\tr[(\nabla \axl A)^T]{\cdp} \id){\cdp} \id
\\\notag
&=-\alpha_1 \dev\sym (\nabla \axl A)+\alpha_2 \skew (\nabla \axl A) -\alpha_3\, \tr(\nabla \axl A){\cdp} \id +3\alpha_3\,\tr(\nabla \axl A){\cdp} \id
\\\notag
&=-\alpha_1 \dev\sym (\nabla \axl A)+\alpha_2 \skew (\nabla \axl A)+ 2\alpha_3\,\tr(\nabla \axl A){\cdp} \id.
\end{align}
Hence, we have after multiplication with $A_{,t}$
\begin{align}
\langle  \Curl[\alpha_1 \dev\sym \Curl A+\alpha_2&  \skew \Curl A +\alpha_3\, \tr(\Curl A){\cdp} \id],A_{,t}\rangle
\\\notag
=\langle \skew \Curl[\alpha_1 \dev\sym &\Curl A+\alpha_2 \skew \Curl A +\alpha_3\, \tr(\Curl A){\cdp} \id],A_{,t}\rangle
\\\notag
=\langle\skew \Curl[-\alpha_1 \dev&\sym (\nabla \axl A)+\alpha_2 \skew (\nabla \axl A)+ 2\alpha_3\,\tr(\nabla \axl A){\cdp} \id],A_{,t}\rangle
\\\notag
=\langle \Curl[-\alpha_1 \dev\sym& (\nabla \axl A)+\alpha_2 \skew (\nabla \axl A)+ 2\alpha_3\,\tr(\nabla \axl A){\cdp} \id],A_{,t}\rangle
\end{align}
which, using the strong anchoring  boundary conditions \eqref{bcaxl}, implies
\begin{align}
\int_\Omega&\langle \alpha_1 \dev\sym \Curl A+\alpha_2 \skew \Curl A +\alpha_3\, \tr(\Curl A){\cdp} \id,\notag\\&\quad \quad\quad\dev\sym\Curl A_{,t}+\skew\Curl A_{,t}+\frac{1}{3}\tr(\Curl A_{,t})\rangle\, dv
\\
&=\int_\Omega\langle-\alpha_1 \dev\sym (\nabla \axl A)+\alpha_2 \skew (\nabla \axl A)+ 2\alpha_3\,\tr(\nabla \axl A){\cdp} \id,\Curl A_{,t}\rangle\, dv.\notag
\end{align}
Moreover, we deduce that
\begin{align}
&\frac{1}{2}\frac{d}{dt}\int_\Omega\bigg( \alpha_1\| \dev\sym \Curl A\|^2 +\alpha_2\| \skew \Curl A\|+ {\alpha_3}\, \tr(\Curl A)^2\notag\bigg)\, dv\notag\\\notag
&=\int_\Omega\langle-\alpha_1 \dev\sym (\nabla \axl A)+\alpha_2 \skew (\nabla \axl A)+ 2\alpha_3\,\tr(\nabla \axl A){\cdp} \id,-(\nabla \axl A_{,t})^T+\tr[(\nabla \axl A_{,t})^T]{\cdp} \id\rangle\, dv
\\\notag
&=\int_\Omega\langle-\alpha_1 \dev\sym (\nabla \axl A)+\alpha_2 \skew (\nabla \axl A)+ 2\alpha_3\,\tr(\nabla \axl A){\cdp} \id,\\\notag
&\ \ \ \ \
-\dev\sym(\nabla \axl A_{,t})^T-\skew(\nabla \axl A_{,t})^T-\frac{1}{3}\tr[(\nabla \axl A_{,t})^T]{\cdp} \id+\tr[(\nabla \axl A_{,t})^T]{\cdp} \id\rangle\, dv
\\
&=\int_\Omega\langle-\alpha_1 \dev\sym (\nabla \axl A)+\alpha_2 \skew (\nabla \axl A)+ 2\alpha_3\,\tr(\nabla \axl A){\cdp} \id,\\\notag
&\ \ \ \ \
-\dev\sym(\nabla \axl A_{,t})^T-\skew(\nabla \axl A_{,t})^T+\frac{2}{3}\tr[(\nabla \axl A_{,t})^T]{\cdp} \id\rangle\, dv
\\\notag
&=\frac{1}{2}\frac{d}{dt}\int_\Omega\bigg(\alpha_1 \|\dev\sym (\nabla \axl A)\|^2+\alpha_2 \|\skew (\nabla \axl A)\|^2+ {4}\alpha_3\,[\tr(\nabla \axl A)]^2\bigg)\, dv.\notag
\end{align}
Because $A$ is skew-symmetric,  it is completely defined by its axial vector $\axl A$ and we have
\begin{align}\label{eqskew}
&\sym A=0,\quad\quad  \tr(A)=0,\quad\quad  \|A\|^2=2\|\axl A\|^2\notag,\quad\quad \tr(\Curl A)=2\,\tr(\nabla \axl A)\\
&\|\skew(\nabla u-A)\|^2=2\|\axl(\skew\nabla u)-\axl A\|^2=\frac{1}{2}\|\curl\, u-2\axl A\|^2.
\end{align}

 Then, the total energies
\begin{align}
\mathcal{L}_1&(u_{,t},A_{,t},\nabla u-A,\sym A,\Curl A)\notag\\
&=\int_\Omega\bigg(\frac{1}{2}\|u_{,t}\|^2+\frac{1}{2}\|A_{,t}\|^2+\mu_e \|\sym \nabla u\|^2+\mu_c\|\skew(\nabla u-A)\|^2+ \frac{\lambda_e}{2} \tr(\nabla u)^2\\\notag&
\ \ \ \ \ \ \ \ \ \ \quad \quad  +\frac{\alpha_1}{2}\| \dev\sym \Curl A\|^2 +\frac{\alpha_2}{2}\| \skew \Curl A\|+ \frac{\alpha_3}{2}\, [\tr(\Curl A)]^2\bigg)\, dv,
\end{align}
and
\begin{align}
\mathcal{L}_2&(u_{,t},(\axl A)_{,t},\nabla u-A,\axl A)\notag\\
&=\int_\Omega\bigg(\frac{1}{2}\|u_{,t}\|^2+\|(\axl A)_{,t}\|^2
+\mu_e \|\sym \nabla u \|^2+\mu_c\|\skew(\nabla u-A)\|^2+ \frac{\lambda_e}{2} [\tr(\nabla u)]^2
\\\notag&
\ \ \ \ \ \ \ \ \ \ \quad \quad  +\frac{\alpha_1}{2} \|\dev\sym (\nabla \axl A)\|^2+\frac{\alpha_2}{2} \|\skew (\nabla \axl A)\|^2+ {2}\alpha_3\,[\tr(\nabla \axl A)]^2\bigg)\,dv\,
\end{align}
are equivalent and lead to equivalent Euler-Lagrange equations. The power function is given by
\begin{align}
 \Pi(t)&=\int_\Omega (\dd\langle f,{u}_t\rangle +\langle M,A_t\rangle)\, dv\,=\dd\int_\Omega (\langle f,{u}_t\rangle +\langle \skew M,A_t\rangle)\, dv\, \\&=\int_\Omega (\dd\langle f,{u}_t\rangle +2\langle \axl\skew M,\axl A_t\rangle)\, dv\, .\notag
\end{align}

In conclusion, in view of \eqref{eqskew}, the Euler-Lagrange equation gives us the following system of partial differential equations for $u$ and $A$
\begin{align}\label{eqisaxl}
 u_{,tt}&=\dvg[2\mu_e \sym\nabla u+2\mu_c\skew(\nabla u-A)+ \lambda_e \tr(\nabla u){\cdp} \id]+f\, ,\notag\\\notag
\,(\axl A)_{,tt}&=\Div\bigg[\frac{\alpha_1}{2} \dev\sym (\nabla \axl A)+\frac{\alpha_2}{2}\skew (\nabla \axl A)+ {2}\alpha_3\,\tr(\nabla \axl A){\cdp} \id\bigg]\notag
\\&\ \ \ \ \ \ +2\mu_c\axl(\skew\nabla u-A)+\axl\skew M\,  \ \ \ \text {in}\ \ \  \Omega\times [0,T],\notag
\end{align}
which is completely equivalent with the system \eqref{eqisCos}.
In the case of the Cosserat theory  we must put $\mu_e=\mu$ and $\lambda_e=\lambda$, where $\mu$ and $\lambda$ are the Lam\'{e} constants from classical elasticity\footnote{In light of our ``homogenization formula" \eqref{homfor}, $\mu_e=\frac{\mu_h\,\mu}{\mu_h-\mu},\quad 2\mu_e+3\lambda_e=\frac{(2\mu_h+3\lambda_h)(2\mu+3\lambda)}{(2\mu_h+3\lambda_h)-(2\mu+3\lambda)}$  such a choice is inconsistent. The linear Cosserat model is physically doubtful \cite{Berglund82}. In \cite{Berglund77,Berglund82} the author
disproves micropolar effects to appear from the homogenization of heterogeneous
Cauchy material. In \cite{Neff_ZAMM05} a case is made, that the linear Cosserat model
leads to unphysical effects which are incompatible with a heterogeneous material. In
\cite{Bigoni07} the opposite claim is made, under the assumption that the inclusion in a Cauchy
matrix material is significantly less stiff than the matrix. These authors argue,
that a Cosserat model is not suitable for a stiffer inclusion.}. The set of {\it independent constitutive variables} for the Cosserat model is
\begin{align}
{e}=\nabla u-A,\quad\quad\quad {\alpha}=-\Curl A.
\end{align}
In terms of the microrotation vector $\vartheta=\axl A$, the above system turns into the classical format
\begin{align}
 u_{,tt}&=\dvg[2\mu_e \sym\nabla u+2\mu_c(\skew\nabla u-\anti(\vartheta))+ \lambda_e \tr(\nabla u){\cdp} \id]+f\, ,\notag\\\notag
\,\vartheta_{,tt}&=\Div\bigg[\frac{\alpha_1}{2} \dev\sym \nabla \vartheta+\frac{\alpha_2}{2}\skew \nabla \vartheta+ {2}\alpha_3\,\tr(\nabla \vartheta){\cdp} \id\bigg]\notag
\\&\ \ \ \ \ \ +2\mu_c\big[\axl(\skew\nabla u)-\vartheta\big]+\axl\skew M\,  \ \ \ \text {in}\ \ \  \Omega\times [0,T],\notag
\end{align}

We  point out that for the static case and for $\mu_c>0$ in this model, existence and uniqueness  can be shown for a very weak curvature energy, namely for $\alpha_1>0$, $\alpha_2,\alpha_3\geq0$, see \cite{Neff_JeongMMS08}. For $\mu_c=0$ in the linear Cosserat model, the system uncouples. This is another artefact of the linear Cosserat model.

Let us remark that if we relax the isotropic energy from the  gradient elasticity formulation \cite{IsolaSciarraVidoliPRSA,Neff_Svendsen08,MorroVianello13}
\begin{equation}
\mathcal{E}(\nabla u,\nabla(\skew \nabla u))= \mu\,\|\dev \sym \nabla u\|^2+\frac{2\mu+3\lambda}{6}\,[\tr(\nabla u)]^2+\mu L_c^2\, \|\nabla(\skew \nabla u)\|^2,
\end{equation}
corresponding to the {\it indeterminate couple stress problem},  such that
\begin{equation}
\mathcal{E}(\nabla u,A,\nabla A)= \mu\,\|\dev \sym \nabla u\|^2+\frac{2\mu+3\lambda}{6}\,[\tr(\nabla u)]^2+\mu L_c^2\, \|\nabla A\|^2
+\varkappa^+\mu \,\|\skew \nabla u-A\|^2,
\end{equation}
where $\varkappa^+$ is a dimensionless penalty coefficient, then we obtain the isotropic Cosserat model. The coefficient $\varkappa^+\mu=\mu_c$ is the Cosserat couple modulus. We  observe that no mixed terms appear.

\subsection{Lazar's translational gauge theory of dislocations}\label{Lazarapp}

The static equations used  by Lazar and Anastassiadis \cite{Lazar2009,Lazar2009b} in the isotropic gauge theory of dislocations  can be expressed  as
\begin{align}\label{L}
0=&\dd\Div [2{\mu}_e  \sym(\nabla u-{P})+2{\mu_c} \skew(\nabla u-{P})+{\lambda}_e\, \tr(\nabla u-{P}){\cdp} \id]+f,\notag\\
\sigma^0=&\dd-\Curl[\alpha_1\dev\sym(\Curl {P})+
\alpha_2\skew(\Curl P)
+{\alpha_3}\tr(\Curl{P}){\cdp} \id]\\\  &\dd+2{\mu}_e  \sym(\nabla u-{P})+2{\mu_c} \skew(\nabla u-{P})+{\lambda}_e\, \tr(\nabla u-{P}){\cdp} \id\, ,\notag
\end{align}
where the coefficients $\alpha_1,\,\alpha_2,\, \alpha_3$  correspond to $a_1,\,a_2,\, \dd\frac{a_3}{3}$ from the Lazar's notations, $\sigma^0$ is a statically admissible  background field (the body moment tensor $M$ in the Eringen-Claus model \eqref{eqCE}), i.e.
\begin{align}
\Div \sigma_0+f=0,\quad\quad\quad \sigma_0.\,n|_{\partial \Omega\setminus\Gamma}=N\,,
\end{align}
 with $N$ prescribed. Lazar and Anastassiadis have decomposed the dislocation tensor $\Curl P$ into its ${\rm SO}(3)$-irreducible pieces, ``the axitor", ``the tentor" and ``the trator" parts, i.e.
\begin{align}\label{tratoraxi}
\Curl P&=\underbrace{\dev\sym(\Curl {P})}_{\textrm{``tentor"}}+
\underbrace{\skew(\Curl P)}_{\textrm{``trator"}}
+\underbrace{\frac{1}{3}\tr(\Curl {P}){\cdp} \id}_{\textrm{``axitor"}}\, .
\end{align}
It is clear that the Lazar's model \cite{Lazar2009} is a simplified static version of the asymmetric isotropic  Eringen-Claus model for dislocation dynamics \cite{Eringen_Claus69} (see the Subsection \ref{ErClapp})  with $\mathbb{H}=0$ and $\mu_c>0$. The tensor $\mathbb{H}$ is absent since the term $\langle \mathbb{H}.\, \sym P,\sym P\rangle$ is not translation gauge invariant. In \cite{Lazar2009} various special solutions to \eqref{L}  for screw and edge dislocations are  constructed.

Abbreviating $\beta_e:=\nabla u-P\in\mathbb{R}^{3\times 3}$ the system is equivalent to the Euler-Lagrange equations of
\begin{align}\label{minimLazar}
\int_\Omega \bigg[&\mu_e\|\sym \beta_e\|^2+\mu_c\|\skew \beta_e\|^2+\frac{\lambda_e}{2}[\tr(\beta_e)]^2\notag\\
&+\frac{\alpha_1}{2}\|\dev \sym\Curl \beta_e\|^2+\frac{\alpha_2}{2}\|\dev \sym\Curl \beta_e\|^2+\frac{\alpha_3}{2}[\tr (\Curl \beta_e)]^2\\&+\langle \sigma_0,\beta_e\rangle\bigg]dv \quad\rightarrow \quad\min.\  \beta_e,\quad\quad\quad\quad \beta_e\cdot \tau=0 \quad \text{ on }\quad \Gamma\subset\partial \Omega.\notag
\end{align}
In the variational formulation, the dislocation model can be seen as an elastic (reversible) description of a material, which may respond to external loads by an elastic distortion field $\beta_e$  which is not anymore a gradient (incompatible). This is not yet an irreversible  plasticity formulation, since elasticity does not change the state of the body.

The  Euler-Lagrange equations turn out to be
\begin{align}\label{E-LL}
-f&=\dd\Div \sigma_0=\dd\Div [2{\mu}_e  \sym\beta_e+2{\mu_c} \skew\beta_e+{\lambda}_e\, \tr(\beta_e){\cdp} \id],\vspace{1mm}\notag\\
\sigma^0&=\dd\Curl[\alpha_1\dev\sym(\Curl {\beta_e})+
\alpha_2\skew(\Curl \beta_e)
+{\alpha_3}\tr(\Curl{\beta_e}){\cdp} \id]\vspace{1mm}\\\  &\dd\quad+2{\mu}_e  \sym\beta_e+2{\mu_c} \skew\beta_e+{\lambda}_e\, \tr(\beta_e){\cdp} \id\, .\notag
\end{align}
We will deal with the well-posedness for the minimization problem \eqref{minimLazar} in another work.

\subsection{The symmetric earthquake structure model of Teisseyre}\label{Teisapp}

Teisseyre \cite{Teisseyre73,Teisseyre74} followed closely the approach to micromorphic continuum theory developed by  Suhubi and Eringen \cite{Eringen64} and by Eringen and Claus \cite{EringenClaus}. In fact he used the equations of motion given by Eringen and Claus (\cite{EringenClaus}, Eq. (3.39)) written in terms of the divergence operator (see Eqs. (1)-(4) from \cite{Teisseyre73} and also \cite{EringenClaus})
\begin{align}\label{eqCEt}
\varrho \,u_{l,tt}&=\widetilde{\sigma}_{kl,k}+f_l,\\ P_{lm,tt}&=\Lambda_{plm,p}+\widetilde{\sigma}_{ml}-\widetilde{s}_{ml}+M_{lm},\notag
\end{align}
where  (see the constitutive equations (36)-(38)  from \cite{Teisseyre73})
\begin{align}\label{cecet}
\widetilde{\sigma}_{kl}&=(\overline{\lambda}+\tau)\,\varepsilon_{mm}\delta_{kl}+2(\overline{\mu}+\varsigma)\,\varepsilon_{kl}+\eta \overline{e}_{mm}\delta_{kl}+\overline{\nu}\, \overline{e}_{lk}+\kappa\,\overline{e}_{kl},\notag\\
\widetilde{s}_{kl}&=(\overline{\lambda}+2\tau)\,\varepsilon_{mm}\delta_{kl}
+2(\overline{\mu}+2\varsigma)\,\varepsilon_{kl}
+(2\eta-\tau)\, \overline{e}_{mm}\delta_{kl}+(\overline{\nu}+\kappa -\varsigma)\,(\overline{e}_{kl}+\overline{e}_{lk}),\\\notag
\Lambda_{plk}&=a_1\,\alpha_{rn}(\epsilon_{prn}\delta_{kl}-
\epsilon_{krn}\delta_{pl})+
a_2\, \epsilon_{pkn}\alpha_{ln}
+a_3\,(\epsilon_{pln}\alpha_{kn}-\epsilon_{kln}\alpha_{pn}),\end{align}
 the {\it constitutive variables}  $\varepsilon_{kl},\overline{e}_{kl}$ and $\alpha_{kl}$ are the same as in the Eringen-Claus theory (see Subsection \ref{ErClapp}) and the rest of the quantities have the same signification as in Subsection \ref{Eringenmodel}. For simplicity, the system \eqref{eqCEt} is considered in a normalized form.

 The constitutive equations and the equation of motion are the same as in the Eringen-Claus model \cite{EringenClaus}: in fact, using that $\Lambda_{klm}=-\Lambda_{mlk}$ from \eqref{cecet}$_3$, Eringen and Claus considered the tensor
\begin{align}
m_{kl}=\frac{1}{2}\epsilon_{kmn}\Lambda_{mln}
\end{align}
and rewrote the equation \eqref{eqCEt}$_3$ in the  following format
\begin{align}\label{formatcurl}
 P_{lm,tt}&=\epsilon_{kmn}m_{nl,k}
 +\widetilde{\sigma}_{ml}-\widetilde{s}_{ml}+M_{lm}\Leftrightarrow
 P_{,tt}=-\Curl(m^T)
 +\widetilde{\sigma}^T-\widetilde{s}^T+M\,.
\end{align}
 Moreover, because the tensor $m_{kl}$ is given by
 \begin{align}
 m_{kl}&=-a_3\alpha_{mm}\delta_{kl}-a_1\alpha_{kl}+(a_1-a_2+a_3)\alpha_{lk},
 \end{align}
 the equations of motion give us an energy whose form  can be found following the  Subsection \ref{ErClapp}. More precisely, the energy
 \begin{align}
 \mathcal{ K}_1(P)=\langle \Lambda,\nabla P\rangle,
 \end{align}
from \cite{Teisseyre74} is in fact
 the energy
 \begin{align}
 \mathcal{ K}_2(P)&=\langle m,\Curl P\rangle\notag\\&=(a_2-a_3)\|\dev\sym(\Curl {P})\|^2+
(a_2-a_3-2a_1)\|\skew(\Curl P)\|^2
+\frac{2a_3+a_2}{3}\,[\tr(\Curl{P})]^2
\notag\\&=\alpha_1\|\dev\sym(\Curl {P})\|^2+
\alpha_2\|\skew(\Curl P)\|^2
+\alpha_3\,[\tr(\Curl{P})]^2\, .
 \end{align}

The first assumption of Teisseyre is that $\kappa =\overline{\nu}$ which implies that the force stress tensor $\widetilde{\sigma}$ is symmetric. Imposing the additional assumption that the moments of rotations have to vanish he also requires  that  the corresponding differences between the stress moment components and body couples appearing in the equation vanish. This is the reason why he assumed that
\begin{align}\label{lambdaT}
\Lambda_{plk,p}=\Lambda_{pkl,p}, \quad\quad \quad M_{lk}=M_{kl}\, .
\end{align}
If \eqref{lambdaT} are satisfied, we see immediately that \eqref{eqCEt} determines $P_{,tt}$ to be symmetric.

Using the identity
 \begin{align}
\Lambda_{klm}=\epsilon_{kmn}m_{nl}
 \end{align}
the symmetry  constraint $\Lambda_{plk,p}=\Lambda_{pkl,p}$, can be rewritten in terms of  $m$, i.e.
\begin{align}\label{tconstrains}
(\Curl (m^T))_{ml}&=\epsilon_{lkn}m_{nm,k}=-\epsilon_{kln}m_{nm,k}=-\Lambda_{kml,k}\\
&=-\Lambda_{klm,k}=-\epsilon_{kmn}m_{nl,k}=\epsilon_{mkn}m_{nl,k}=(\Curl (m^T))_{lm}\,.\notag
\end{align}
In other words, \eqref{lambdaT}$_1$ demands that $m$ is such that
\begin{align}\label{tconstrains}
\Curl (m^T)\in \Sym(3).
\end{align}
 Obviously,  $P_{,tt}$  symmetric does not imply that $P$ must be symmetric.

Hence, in view of \eqref{ma1a2} the  constraint \eqref{tconstrains} means that
\begin{align}\label{ma1a3}
\Curl[\alpha_1\dev\sym\Curl P+\alpha_2\skew \Curl P+\alpha_3 \, \tr(\Curl P) {\cdp} \id]\in \Sym(3)\, ,
\end{align}
and further that
\begin{align}\label{ma1a4}
\Curl\{\alpha_1\dev\sym(\Curl P)^T-\alpha_2\skew (\Curl P)^T+\alpha_3 \,\tr[(\Curl P)^T]{\cdp} \id\}\in \Sym(3) .
\end{align}
In order to satisfy \eqref{lambdaT}$_1$,   Teisseyre considered the following sufficient condition\footnote{In fact the condition $a_2=a_1+a_3$ is necessary and sufficient to satisfy \eqref{lambdaT}$_1$ if $P\in\Sym(3)$. }
\begin{align}\label{Teiscond}
a_2=-a_3,\quad \quad\quad a_1=-2a_3\, .
\end{align}
In terms of our notations these imply that
\begin{align}\label{LTeiscond}
\alpha_1=-6\alpha_3,\quad \quad\quad \alpha_2=6\alpha_3\, .
\end{align}
The conditions \eqref{lambdaT} and \eqref{LTeiscond} are the so-called {\it Einstein choice} in three dimensions and they were used  by  Malyshev \cite{Malyshev} and Lazar \cite{LazarJPMG02,Lazar02} in order to investigate dislocations with symmetric force stress\footnote{In the Lazar's notations the conditions \eqref{LTeiscond} becomes  $a_2 = -a_1$ and $a_3 = -\frac{a_1}{2}$}.

In addition,  in another paper \cite{Teisseyre74},  Teisseyre assumed that $a_3=0$ which removes the effects of the {\it micro-dislocation} tensor $\alpha=-\Curl P$ completely.

In other words, the Einstein choice \eqref{LTeiscond}  leads to
\begin{align}\label{ma1a5}
\Curl\{\alpha_1&\dev\sym(\Curl P)^T-\alpha_2\skew (\Curl P)^T+\alpha_3 \tr[(\Curl P)^T]{\cdp} \id\}
\notag\\&=\Curl\{-6\alpha_3\dev\sym(\Curl P)^T-6\alpha_3\skew (\Curl P)^T+\alpha_3 \tr[(\Curl P)^T]{\cdp} \id\}
\\&=
-6\alpha_3\Curl\{\dev\sym(\Curl P)^T+\skew (\Curl P)^T-\frac{1}{6}\tr[(\Curl P)^T]{\cdp} \id\}\notag
\\&=
-6\alpha_3\Curl\{\dev\sym(\Curl P)^T+\skew (\Curl P)^T+\frac{1}{3}\tr[(\Curl P)^T]{\cdp} \id-\frac{1}{2}\tr[(\Curl P)^T]{\cdp} \id\}.\notag
\end{align}
Using the decomposition \eqref{tratoraxi} of the dislocation tensor $\Curl P$, \eqref{ma1a5} implies
\begin{align}\label{ma1a225}
\Curl\{\alpha_1&\dev\sym(\Curl P)^T-\alpha_2\skew (\Curl P)^T+\alpha_3 \tr[(\Curl P)^T]{\cdp} \id\}\\&=
-6\alpha_3\Curl[(\Curl P)^T]+3\alpha_3\Curl\{\tr[(\Curl P)^T]{\cdp} \id\}.\notag
\end{align}
Let us remark that for all differentiable functions $\zeta:\mathbb{R}\rightarrow\mathbb{R}$ on $\Omega$ we have
\begin{align}\label{diagcurl}
\Curl(\zeta{\cdp} \id)=\left(
\begin{array}{ccc}
0 &\zeta_{,3} &-\zeta_{,2}\\
-\zeta_{,3}& 0& \zeta_{,1}\\
\zeta_{,2} &-\zeta_{,1} &0
\end{array}\right)\in\so(3).
\end{align}
On the other hand we have
\begin{align}\label{idsymskew}
\Curl[(\Curl S)^T]&\in \Sym(3)\, ,\quad\quad\quad
\text{for all}\quad\quad S\in\Sym(3),\notag\\
\Curl[(\Curl A)^T]&\in \so(3)\,  ,\quad\quad\quad
\text{for all}\quad\quad A\in\so(3).\\\notag
\tr(\Curl S)&=0\, ,\quad\quad\quad
\text{for all}\quad\quad S\in\Sym(3).
\end{align}
Hence,  from \eqref{tconstrains}, \eqref{ma1a225}, \eqref{idsymskew} and \eqref{diagcurl} we obtain
\begin{align}\label{necesar}
\Curl (m^T)|_{\alpha_1=-6\alpha_3, \alpha_2=6\alpha_3}=&-6\alpha_3\underbrace{\Curl\{[\Curl(\sym P)]^T\}}_{\in\Sym(3)}\\\notag&-6\alpha_3\underbrace{\Curl\{[\Curl (\skew P)]^T\}}_{\in\so(3)}+3\alpha_3\underbrace{\Curl\{\tr[\big(\Curl( \skew P)\big)^T]{\cdp} \id\}}_{\in\so(3)}.
\end{align}

Let us also remark that if we consider
\begin{align}
\skew P=\left(\begin{array}{ccc}
0 &-p_3&p_2\\
p_3&0& -p_1\\
-p_2& p_1&0
\end{array}\right)
\end{align}
then we have
\begin{align}
\big(\Curl( \skew P)\big)^T=\left(
\begin{array}{ccc}
p_{3,3}+p_{2,2} & -p_{1,2} & -p_{1,3} \\
-p_{2,1} & p_{3,3}+p_{1,1}& -p_{2,3}\\
-p_{3,1} &-p_{3,2}& p_{2,2}+p_{1,1} \\
\end{array}
\right).
\end{align}
Thus, we deduce
\begin{align}\label{axlcurldiv}
\tr[\big(\Curl( \skew P)\big)^T]=2(p_{1,1}+p_{2,2}+p_{3,3})=2\,{\rm div}\, p\,,
\end{align}
where $p=\axl (\skew P)$.

Moreover, in view of  \eqref{diagcurl}, \eqref{axlcurldiv} implies
\begin{align}
\Curl\{\tr[\big(\Curl( \skew P)\big)^T]{\cdp} \id\}=2\left(
\begin{array}{ccc}
0 &\,{\rm div}\, p_{,3} &-{\rm div}\, p_{,2}\\
-{\rm div}\, p_{,3}& 0& {\rm div}\, p_{,1}\\
{\rm div}\, p_{,2} &-{\rm div}\, p_{,1} &0
\end{array}\right)=-2\anti\nabla({\rm div}\, p).
\end{align}

On the other hand, we have
\begin{align}
\Curl\{[\Curl (\skew P)]^T\}=\left(
\begin{array}{ccc}
0 &\,{\rm div}\, p_{,3} &-{\rm div}\, p_{,2}\\
-{\rm div}\, p_{,3}& 0& {\rm div}\, p_{,1}\\
{\rm div}\, p_{,2} &-{\rm div}\, p_{,1} &0
\end{array}\right)=-\anti\nabla({\rm div}\, p).
\end{align}
From the above two identities, we deduce
\begin{align}\label{happy}
-2\Curl\{[\Curl (\skew P)]^T\}+\Curl\{\tr[\big(\Curl( \skew P)\big)^T]{\cdp} \id\}=0, \quad\quad \text{for all}\quad \quad P\in\mathbb{R}^{3\times3}.
\end{align}

Thus, we obtain
\begin{align}
\Curl (m^T)|_{\alpha_1=-6\alpha_3, \alpha_2=6\alpha_3}=&-6\alpha_3\,\Curl\{[\Curl(\sym P)]^T\}\in\Sym(3),\quad\quad \text{for all}\quad \quad P\in\mathbb{R}^{3\times3}.
\end{align}

Summarizing,   we have the following result which gives information about the symmetry of the model.
\begin{remark}
\begin{itemize}\item[]
\item[i)] If  \ \ $\alpha_1=-6\alpha_3$ and \ \ $ \alpha_2=6\alpha_3$, then
    \begin{align}
\Curl \{\alpha_1 \dev \sym \Curl P+ \alpha_2 \skew \Curl P+ {\alpha_3}\ {\rm tr}(\Curl P)\!\cdot\!\id\} \in \Sym(3)\quad\text{ for all } \quad\quad  P\in\mathbb{R}^{3\times3}.
\end{align}
\item[ii)] Given $P\in\Sym(3)$, then we have
\begin{align}
\Curl \{\alpha_1 \dev \sym \Curl P+ \alpha_2 \skew \Curl P+ {\alpha_3}\ {\rm tr}(\Curl P)\!\cdot\!\id\} \in \Sym(3).
\end{align}
 if and only if $\ \alpha_1=-\alpha_2$.
\end{itemize}
\end{remark}

We conclude that, in view of \eqref{necesar} and \eqref{happy}, the Einstein choice \eqref{LTeiscond} implies that
\begin{align}\label{ma1a6}
\Curl[m^T]\in \Sym(3)\, , \quad\quad \text{for all}\quad \quad P\in\mathbb{R}^{3\times3}.
\end{align}
 Thus, the condition \eqref{Teiscond} is in concordance, without any restriction and projection of the equation, with the assumption that $P_{,tt}$ remains symmetric  since the right hand side of the equations \eqref{eqCE} is now symmetric. Therefore Teisseyre's model does have a symmetric stress tensor, it is based on the dislocation tensor $\alpha$, and determines nevertheless a symmetric micro-distortion such that $P_{,tt}$.

 In addition, if $P\in \Sym(3)$, then from \eqref{formatcurl} it follows that $\skew P$ is solution of the problem
 \begin{align}
 (\skew P)_{,tt}=0, \quad\quad\quad (\skew P)(x,0)=0,\quad\quad\quad (\skew P)_{,t}(x,0)=0.
 \end{align}
 The unique solution of the above problem is $\skew P=0$. Thus, we conclude that if $P(x,0)\in \Sym(3)$ and $P_{,t}(x,0)\in \Sym(3)$, then $P(x,t)\in \Sym(P)$ and in consequence the  Teisseyre's model is a fully ``symmetric" micromorphic model.

If we consider the supplementary conditions \eqref{Teiscond}, then the energy $\mathcal{K}_2$ becomes
 \begin{align}\label{a36}
 \mathcal{ K}_2(P)=-2a_3\,\|\dev\sym(\Curl {P})\|^2+4a_3\,\|\skew(\Curl P)\|^2
+\frac{a_3}{3}\,[\tr(\Curl{P})]^2\, ,
 \end{align}
 which has no sign!  The constraint \eqref{Teiscond}, introduced for having $P_{,tt}\in\Sym(3)$ destroys, therefore, the positive definiteness of the dislocation energy \eqref{a36}.   In fact, the most general form of the energy \eqref{a36} considered by Teisseyre is
\begin{align}
 \mathcal{ K}_2(\sym P)=\langle \widehat{\mathbb{L}}_c.\,\Curl \sym P, \Curl \sym P\rangle,
\end{align}
where $\widehat{\mathbb{L}}_c$ is a {\it non-positive definite} isotropic tensor. In view of \eqref{sCg}, this energy is equivalent with
\begin{align}
 \mathcal{ K}_T(\sym P)=\langle \widehat{\mathbb{L}}_T.\,\nabla \sym P, \nabla \sym P\rangle,
\end{align}
where
\begin{align}
\widehat{\mathbb{L}}_T:\mathbb{R}^{3\times 3\times 3}\rightarrow \mathbb{R}^{3\times 3\times 3}.
\end{align}
The energy $\mathcal{ K}_T(\sym P)$ is similar with the energy from the  gradient elasticity formulation \cite{IsolaSciarraVidoliPRSA}
\begin{align}
 \mathcal{ E}(\sym \nabla u)=\langle \widehat{\mathbb{L}}_T.\,\nabla (\sym \nabla u), \nabla (\sym \nabla u)\rangle.
\end{align}

If we extend the Teisseyre's model to the anisotropic case, then the total energy  is equivalent with the energy
\begin{align}
2\widehat{E}(t):=\int_\Omega\bigg(\|u_{,t}\|^2&+ \|(\sym P)_{,t}\|^2+ \langle \C.\, \sym(\nabla u-P), \sym(\nabla u-P)\rangle
\vspace{2mm}\notag\\& \ \ \ \ \quad\  + \langle \H.\, \sym P,  \sym P\rangle + \langle \widehat{\mathbb{L}}.\, \nabla(\sym P), \nabla(\sym P)\rangle\bigg)dv\, ,
\end{align}
from the microstrain model \cite{Forest06} (see Subsection \ref{forestapp}).  In conclusion, the Teisseyre's model is a special degenerate isotropic microstrain model  and it is therefore incapable of describing rotation of the microstructure, hence the name ``symmetric" micromorphic model.

\subsection{The asymmetric microstretch model in dislocation format}\label{microstretchapp}

It is well known that the theory of microstretch elastic materials is a special subclass of the class of micromorphic materials \cite{Eringen99,BirsanAltenbach13,BirsanZAMM08,BirsanJTS06}. In this Subsection we show that the microstretch model is already contained in our relaxed  micromorphic model in dislocation format.
To this aim, we assume that the micro-distortion tensor has the  form $P=\zeta{\cdp} \id+A$, where $A\in \so(3)$ and $\zeta$ is a scalar function. It is easy to check that
\begin{align}\Curl P=\Curl A+
\Curl(\zeta{\cdp} \id)=\Curl A+\left(
\begin{array}{ccc}
0 &\zeta_{,3} &-\zeta_{,2}\\
-\zeta_{,3}& 0& \zeta_{,1}\\
\zeta_{,2} &-\zeta_{,1} &0
\end{array}\right),
\end{align}
and
\begin{align}
\Curl(\Curl P)=\Curl(\Curl A)+\left(
\begin{array}{ccc}
-(\zeta_{,22}+\zeta_{,33}) &\zeta_{,12} &\zeta_{,13}\\
\zeta_{,12}& -(\zeta_{,11}+\zeta_{,33}) & \zeta_{,23}\\
\zeta_{,13} &\zeta_{,23} &-(\zeta_{,11}+\zeta_{,22})
\end{array}\right)\,.
\end{align}
As in the construction of the  linear Cosserat model in terms of the dislocation density tensor (Subsection \ref{cosapp}), for isotropic materials and with the asymmetric factor $2\mu_c\skew(\nabla u-P)$ incorporated,  the constitutive equations become
\begin{align}\label{consstr}
 \sigma&=2\mu_e\, \sym(\nabla u-\zeta{\cdp} \id)+2\mu_c\,\skew(\nabla u-A)+ \lambda_e \,\tr(\nabla u-\zeta{\cdp} \id){\cdp} \id\, ,\notag\\\notag
m&=\alpha_1 \dev\sym \Curl A+\alpha_2\, \skew \Curl  A +\alpha_3\, \tr(\Curl A){\cdp} \id\\
&\quad+\alpha_1\, \dev\underbrace{\sym \Curl (\zeta{\cdp} \id)}_{=0}+\alpha_2 \skew \Curl (\zeta{\cdp} \id) +\alpha_3\, \underbrace{\tr(\Curl (\zeta{\cdp} \id))}_{=0}{\cdp} \id\\
s&=2\mu_h \sym (\zeta{\cdp} \id)+\lambda_h \tr (\zeta{\cdp} \id){\cdp} \id\,,\quad \quad \quad\ \ \ \text {in}\ \quad \quad   \Omega\times [0,T].\notag
\end{align}
Observing that for all matrices $X\in\mathbb{R}^{3\times 3}$ we have the decomposition
\begin{align}
X-\dev\sym X=\skew X+\frac{1}{3}\tr (X){\cdp} \id,
\end{align}
we obtain by restriction and projection the equations
\begin{align}\label{eqistr-1}
 u_{,tt}&=\dvg[2\mu_e \sym(\nabla u-P)+2\mu_c\skew(\nabla u-P)+ \lambda_e \tr(\nabla u-P){\cdp} \id]+f\, ,\\\notag
P_{,tt}&-(\dev\sym P)_{,tt}=-\Curl\bigg[\alpha_1 \dev\sym \Curl P+\alpha_2 \skew \Curl  P+\alpha_3\, \tr(\Curl P){\cdp} \id\bigg]\notag\\
&\quad+\dev\sym\Curl\bigg[\alpha_1 \dev\sym \Curl P+\alpha_2 \skew \Curl  P +\alpha_3\, \tr(\Curl P){\cdp} \id\bigg]\notag\\
&\quad+2\mu_e \sym(\nabla u-P)+2\mu_c\skew(\nabla u-P)+\lambda_e \tr(\nabla u-P){\cdp} \id \notag\\
&\quad-2\mu_e \dev\sym(\nabla u-P)-2\mu_c\dev\skew(\nabla u-P)+\lambda_e \dev \tr(\nabla u-P){\cdp} \id \notag\\
&\quad-2\mu_h \sym P+2\mu_h \dev\sym P-\lambda_h \tr (P){\cdp} \id+\lambda_h \dev(\tr (P){\cdp} \id)\notag\\
&\quad+ M-\dev\sym M\,  \quad \quad \quad\ \ \ \text {in}\ \quad \quad  \Omega\times [0,T].\notag
\end{align}
and further
\begin{align}\label{eqistr0}
 u_{,tt}&=\dvg[2\mu_e \sym(\nabla u-\zeta{\cdp} \id)+2\mu_c\skew(\nabla u-A)+ \lambda_e \tr(\nabla u-\zeta{\cdp} \id){\cdp} \id]+f\, ,\\\notag
(\zeta{\cdp} \id+A)_{,tt}&-(\dev\sym (\zeta{\cdp} \id+A))_{,tt}=-\Curl\bigg[\alpha_1 \dev\sym \Curl A+\alpha_2 \skew \Curl  A +\alpha_3\, \tr(\Curl A){\cdp} \id\bigg]\notag\\
&\quad-\dev\sym\Curl\bigg[\alpha_1 \dev\sym \Curl A+\alpha_2 \skew \Curl  A +\alpha_3\, \tr(\Curl A){\cdp} \id\bigg]\notag\\&\quad -\alpha_2\Curl \Curl (\zeta{\cdp} \id)-\alpha_2\dev\sym\Curl \Curl (\zeta{\cdp} \id)\notag\\
&\quad+2\mu_e \sym(\nabla u-\zeta{\cdp} \id)-2\mu_e \dev\sym(\nabla u-\zeta{\cdp} \id)\notag\\
&\quad +2\mu_c\skew(\nabla u-A)-2\mu_c\dev\skew(\nabla u-A)\notag\\&\quad+ \lambda_e \tr(\nabla u-\zeta{\cdp} \id){\cdp} \id-\lambda_e \dev(\tr(\nabla u-\zeta{\cdp} \id){\cdp} \id) \notag\\&\quad-2\mu_h \sym (\zeta{\cdp} \id)+2\mu_h \dev\sym (\zeta{\cdp} \id)\notag\\&\quad-\lambda_h \tr (\zeta{\cdp} \id){\cdp} \id+\lambda_h \dev(\tr (\zeta{\cdp} \id){\cdp} \id)+ M-\dev\sym M\,  \quad \quad \quad\ \ \ \text {in}\ \quad \quad  \Omega\times [0,T].\notag
\end{align}
By orthogonal projection of the second equation \eqref{eqistr0}$_2$  to the space of skew-matrices and to the spherical part, respectively, and using the fact that $\tr(\Curl S)=0$\ \ for all $S\in\Sym(3)$ and $\tr[\Curl (\skew \Curl  A )]=0$ \ \ for all $A\in\so(3)$,
the full system of equations for our  reduced model is
\begin{align}\label{eqistr}
 u_{,tt}&=\dvg[2\mu_e \sym(\nabla u-\zeta{\cdp} \id)+2\mu_c\skew(\nabla u-A)+ \lambda_e \tr(\nabla u-\zeta{\cdp} \id){\cdp} \id]+f\,\notag ,\\\notag
\skew A_{,tt}&=-\skew\Curl\bigg[\alpha_1 \dev\sym \Curl A+\alpha_2 \skew \Curl  A +\alpha_3\, \tr(\Curl A){\cdp} \id\bigg]\\
&\ \ \ \ \ \ +2\mu_c\skew(\nabla u-A)+\skew M\, ,\ \\
\tr(\zeta_{,tt}{\cdp} \id)&=-\alpha_2\tr[\Curl \Curl (\zeta{\cdp} \id)]\notag\\\notag
&\quad\ +2\mu_e\tr[ \sym(\nabla u-\zeta{\cdp} \id)]+ 3\lambda_e \tr(\nabla u-\zeta{\cdp} \id)\notag\\&\quad\ -2\mu_h \tr\sym (\zeta{\cdp} \id)-3\lambda_h \tr (\zeta{\cdp} \id)+\frac{1}{3}\tr(M) \quad \quad \quad\ \ \ \text {in}\ \quad \quad  \Omega\times [0,T].\notag
\end{align}
But
$$
\Curl\Curl(\zeta{\cdp} \id)=\left(
\begin{array}{ccc}
-(\zeta_{,22}+\zeta_{,33}) &\zeta_{,12} &\zeta_{,13}\\
\zeta_{,12}& -(\zeta_{,11}+\zeta_{,33}) & \zeta_{,23}\\
\zeta_{,13} &\zeta_{,23} &-(\zeta_{,11}+\zeta_{,22})
\end{array}\right).
$$
Hence, we deduce
$
\tr[\Curl\Curl(\zeta{\cdp} \id)]=-2(\zeta_{,11}+\zeta_{,22} +\zeta_{,33})=-2\Delta  \zeta.
$
In terms of the microrotation vector $\vartheta=\axl A$, the above system becomes
\begin{align}\label{eqistraxl}
 u_{,tt}&=\dvg[2\mu_e\, \sym(\nabla u-\zeta{\cdp} \id)+2\mu_c\,(\skew\nabla u-\anti(\vartheta))+ \lambda_e\, \tr(\nabla u-\zeta{\cdp} \id){\cdp} \id]+f\,\notag ,\\\notag
\vartheta_{,tt}&=\Div\bigg[\frac{\alpha_1}{2} \,\dev\sym \nabla \vartheta+\frac{\alpha_2}{2}\,\skew \nabla \vartheta+ {2}\alpha_3\,\tr(\nabla \vartheta){\cdp} \id\bigg]\notag
\\&\ \ \ \ \ \ +2\mu_c\,\big[\axl(\skew\nabla u)-\vartheta\big]+\axl\skew M\\
\zeta_{,tt}&=\frac{2}{3}\alpha_2\,\Delta \zeta+\frac{2\mu_e+3\lambda_e}{3}\,\dv\,  u-(2\mu_e+3\lambda_e+2\mu_h+3\lambda_h)\,\zeta+\frac{1}{3}\,\tr(M) \quad \quad \quad\ \ \ \text {in}\ \quad \quad  \Omega\times [0,T].\notag
\end{align}

Using the micro-distortion tensor specific to the microstretch model,  the  tangential boundary condition \eqref{bc} implies the strong anchoring condition\footnote{$(\zeta{\cdp} \id+A)\tau=0$ for all $\tau$ tangent to the boundary implies $\zeta\langle \tau,\tau\rangle+\langle A\tau,\tau\rangle=0$. Since $\langle A\tau,\tau\rangle=0$ for skew-symmetric matrices $A$, we have $\zeta=0$ and further $A=0$.}
\begin{align}\label{bcaxlm}
A({x},t)=0 \ \quad \ \text{and}\ \quad  \ \ \ \ \zeta({x},t)=0\quad \quad \text{on} \quad \partial \Omega\times [0,T].
\end{align}
The above system is the microstretch model in dislocation format and the total energy for this model is given by
\begin{align}
\mathcal{L}_3=\int_\Omega\bigg(&\frac{1}{2}\|u_{,t}\|^2+ \|\vartheta_{,t}\|^2+\frac{3}{2}\|\zeta_{,t}\|^2\notag\\
&+\mu_e\,\| \sym(\nabla u-\zeta{\cdp} \id)\|^2+\mu_c\,\|\skew\nabla u-\anti(\vartheta)\|^2+ \frac{\lambda_e}{2}\, \tr(\nabla u-\zeta{\cdp} \id)^2\\\notag
&+\frac{\alpha_1}{2}\, \|\dev\sym \nabla \vartheta\|^2+\frac{\alpha_2}{2}\,\|\skew \nabla \vartheta\|^2+ {2}\alpha_3\,[\,\tr(\nabla \vartheta)]^2+\frac{3}{2}(2\mu_h+3\lambda_h)\,\zeta^2+\alpha_2\,\|\nabla\zeta\|^2\bigg)\, dv.\notag
\end{align}

We may also obtain the model of microstretch materials if we replace the energy from the  gradient elasticity formulation \cite{IsolaSciarraVidoliPRSA,Neff_Svendsen08,MorroVianello13}
\begin{align}
\mathcal{E}(\nabla u,\nabla(\skew \nabla u),\nabla(\tr(\nabla u)))=& \mu\,\|\dev \sym \nabla u\|^2+\frac{2\mu+3\lambda}{6}\,\,[\tr(\nabla u)]^2\\&
\quad+\mu L_{c_1}^2\, \|\nabla(\tr(\nabla u))\|^2
+\mu L_{c_2}^2\, \|\nabla(\skew \nabla u)\|^2,\notag
\end{align}
with
\begin{align}
\mathcal{E}(\nabla u,A,\nabla A,\zeta,\nabla \zeta)=&
 \mu\,\|\dev \sym \nabla u\|^2+\frac{2\mu+3\lambda}{6}\,[\tr(\nabla u)]^2\\&\quad +\mu\, L_{c_1}^2\, \|\nabla\zeta\|^2
+\mu\, L_{c_2}^2\, \|\nabla A\|^2+\varkappa_1^+ \mu\,[\tr(\nabla u-\zeta{\cdp} \id)]^2
+\varkappa_2^+\mu\, \|\skew \nabla u - A\|^2,\notag
\end{align}
where $\varkappa^+_1$ and $\varkappa^+_2$ are {\it dimensionless penalty coefficients}. The coefficient  $\varkappa^+_2\mu=\mu_c$  is the {\it  Cosserat couple modulus}\footnote{Certainly, penalty parameters do not have the status of material parameters: the doubtful role of the Cosserat couple modulus is again displayed.}. Again,  no mixed terms appear.

For comparison, the classical linear microstretch formulation has the energy  (see \cite{Eringen99}, p. 253)
\begin{align}
\mathcal{L}_4=\int_\Omega\bigg(&\frac{1}{2}\|u_{,t}\|^2+ \frac{1}{2}\|\vartheta_{,t}\|^2+\frac{1}{2}\|\zeta_{,t}\|^2\notag\\
&+\mu_e\,\| \sym(\nabla u-\zeta{\cdp} \id)\|^2+\mu_c\,\|\skew\nabla u-\anti(\vartheta)\|^2+ \frac{\lambda_e}{2}\, \tr(\nabla u-\zeta{\cdp} \id)^2\\\notag
&+\frac{\gamma_1}{2}\, \|\dev\sym \nabla \vartheta\|^2+\frac{\gamma_2}{2}\,\|\skew \nabla \vartheta\|^2+ \frac{\gamma_3}{2}\,[\,\tr(\nabla \vartheta)]^2+\frac{\lambda_1}{2}\,\zeta^2+\frac{\gamma_0}{2}\,\|\nabla\zeta\|^2\\&+\lambda_0\,\tr(\nabla u-\zeta{\cdp} \id)\,\zeta+b_0\,\langle \anti(\nabla\zeta),\nabla \vartheta\rangle\bigg)\, dv.\notag
\end{align}
The microstretch model in dislocation format involves only three curvature coefficients, instead of four considered in the classical model \cite{Eringen99}.

\subsection{The microvoids  model in dislocation format}\label{voidapp}

It is well known that the theory of elasticity with voids is a subset of the micromorphic model \cite{Eringen99,BulgariuGhibaDCDS2013,GhibaArchMech09,GhibaEJMS08,BirsanAltenbach11,BirsanIJSS05}. In this subsection we show  that the microvoid model is a special case of our relaxed  micromorphic model in dislocation format. Indeed this is a particular case of \eqref{eqisoup} in which we assume $P=\zeta{\cdp} \id$. Hence, using  \eqref{eqisoup} we obtain
\begin{align}\label{eqivoid}
 u_{,tt}&=\dvg[2\mu_e \sym(\nabla u-\zeta{\cdp} \id)+ \lambda_e \tr(\nabla u-\zeta{\cdp} \id){\cdp} \id]+f\,\notag ,\\\notag
\zeta_{,tt}&=\frac{2}{3}\alpha_2\Delta \zeta+\frac{2\mu_e+3\lambda_e}{3}\, \dv\,  u-(2\mu_e+3\lambda_e+2\mu_h+3\lambda_h)\zeta+\frac{1}{3}\tr(M) \quad \quad \quad\ \ \ \text {in}\ \quad \quad  \Omega\times [0,T].\notag
\end{align}

The boundary condition which follows from the  tangential boundary condition \eqref{bc} is the strong anchoring condition
\begin{align}
\zeta(x,t)=0,\quad \quad \text{on} \quad \partial \Omega\times [0,T],
\end{align}
which implies that on the boundary the volumes of the voids do not change.

This microvoids model in dislocation format is related to the theory of ``micro-voids" \cite{NunziatoCowin79,CowinNunziato83} and corresponds to the following choice of the total energy
 \begin{align}
\mathcal{L}_5=\int_\Omega\bigg(\frac{1}{2}\|u_{,t}\|^2+\frac{3}{2}\|\zeta_{,t}\|^2&+\mu_e\| \dev\sym(\nabla u-\zeta{\cdp} \id)\|^2+ \frac{2\mu_e+3\lambda_e}{6} [\tr(\nabla u-\zeta{\cdp} \id)]^2\\&
+\frac{3}{2}(2\mu_h+3\lambda_h)\,\zeta^2+\alpha_2\,\|\nabla\zeta\|^2\bigg)\, dv\notag.
\end{align}
 From the above equations we remark that the parameter $\alpha_2$ describes the creation of micro-voids.  This observation  suggests to skip $\alpha_2$ when $\tr(P)=0$.
 Cowin and Nunziato \cite{NunziatoCowin79,CowinNunziato83} introduced the following symmetric stress tensor
 \begin{align}
 {\sigma}_\mathrm{v}&=2\mu_\mathrm{v}\,\sym \nabla u+\lambda_{\mathrm{v}}\,\tr(\nabla u){\cdp} \id+b_\mathrm{v}\, \zeta{\cdp} \id,
 \end{align}
 while the {\it``balance of equilibrated  forces"} is given by
\begin{align}
{\zeta}_{,tt}=\alpha_\mathrm{v}\,\Delta\zeta-b_\mathrm{v}\,\dv\, u-\xi_\mathrm{v}\,\zeta+\ell,\ \
\end{align}
where $\lambda_\mathrm{v},\mu_\mathrm{v},b_\mathrm{v}, \alpha_\mathrm{v}$ and $\xi_\mathrm{v}$  are constitutive constants and $\ell$ is called {\it ``extrinsic equilibrated body force"}.

Our symmetric Cauchy stress tensor is given by
\begin{align}
\sigma=2\mu_e\, \sym\nabla u+ \lambda_e\, \tr(\nabla u){\cdp} \id
-(2\mu_e +3 \lambda_e )\,\zeta{\cdp} \id.
\end{align}
A direct identification of the coefficients gives us that the coefficient of the Cowin-Nunziato theory can be expressed in terms of our constitutive coefficients
\begin{align}
\mu_\mathrm{v}&=\mu_e,\quad \quad\quad \lambda_\mathrm{v}=\lambda_e,\quad\quad\quad \alpha_\mathrm{v}=\frac{2}{3} \alpha_2,\quad\quad\quad b_\mathrm{v}=-\frac{2\mu_e+3\lambda_e}{3},\\
 \xi_\mathrm{v}&=2\mu_e+3\lambda_e+2\mu_h+3\lambda_h=-3b_\mathrm{v}+2\mu_h+3\lambda_h.\notag
\end{align}
In our model we have only four parameters, because  $2\mu_h+3\lambda_h$ can be regarded as a single parameter, instead of five considered by Cowin and Nunziato \cite{CowinNunziato83}. Moreover, we have all the terms considered in the microvoids theory but without having any mixed terms involving two constitutive variables.

The positivity conditions for the Cowin-Nunziato theory with voids \cite{CowinNunziato83} are
\begin{equation}\label{positivedefinitvoid}
\mu_\mathrm{v} >0,\quad\quad\quad 2\mu_\mathrm{v}+3\lambda_\mathrm{v} >0,\quad\quad\quad\alpha_\mathrm{v} >0,\quad\quad\xi_\mathrm{v} >0,%
\quad\quad\quad(2\mu_\mathrm{v}+3\lambda_\mathrm{v} )\,\xi_\mathrm{v} >3\,b_\mathrm{v}^{2},
\end{equation}
while in our  microvoids model in dislocation format the positivity conditions are obvious
 \begin{align}
\mu_e>0,\quad\quad\quad 2\mu_e+3\lambda_e>0,\quad\quad\quad2\mu_h+3\lambda_h>0,\quad\quad\quad \alpha_2>0\,.
\end{align}

Let us consider the energy from the isotropic second gradient poromechanics model  \cite{FdIGuarascio,IsolaSciarraVidoliPRSA}
\begin{equation}
\mathcal{E}(\nabla u,\nabla(\tr( \sym \nabla u)))= \mu\,\|\dev \sym \nabla u\|^2+\frac{2\mu+3\lambda}{6}\,[\tr(\sym\nabla u)]^2+\mu L_c^2\, \|\nabla(\tr(\sym\nabla u)\|^2.
\end{equation}
If we rewrite this energy into a two-field formulation for $u$ and $\zeta$
\begin{align}
\mathcal{E}(\nabla u,\zeta,\nabla \zeta)=&
 \mu\,\|\dev \sym \nabla u\|^2+\frac{2\mu+3\lambda}{6}\,[\tr(\sym \nabla u)]^2\\&\quad +\mu L_c^2\, \|\nabla\zeta\|^2
+\varkappa^+\mu\, [\tr(\nabla u-\zeta{\cdp} \id)]^2,\notag
\end{align}
where $\varkappa^+$ is a dimensionless penalty coefficient,  we obtain a 4-parameter microvoids model in which no mixed terms appear.

We give below, only for comparison, the total energy of the  classical linear elastic microvoid model (see \cite{IesanCarteElast,GhibaZAMM13})
\begin{align}
\mathcal{L}_6=\int_\Omega\bigg(&\frac{1}{2}\|u_{,t}\|^2+\frac{1}{2}\|\zeta_{,t}\|^2\notag+\mu_{\mathrm{v}}\,\| \sym \nabla u \|^2+ \frac{\lambda_{\mathrm{v}}}{2}\, \tr(\nabla u)^2+\frac{\xi_{\mathrm{v}}}{2}\,\zeta^2+\frac{\alpha_{\mathrm{v}}}{2}\,\|\nabla\zeta\|^2+b_{\mathrm{v}}\,\tr(\nabla u)\,\zeta\bigg)\, dv.\notag
\end{align}
According to Lakes \cite{Lakes85b}, the Cowin-Nunziato theory of porous materials  predicts that size effects will occur in bending of bars but not in torsion and not in tension in an isotropic material. However, size effects occur always both in bending and in torsion, which means that the void theory cannot adequately describe materials with microvoids.

\subsection{A glimpse on the isotropic strain gradient model}\label{str-grad-sapp}

Let us consider the general energy from the isotropic strain gradient  model  \cite{Germain,Germain1,FdIGuarascio,IsolaSciarraVidoliPRSA,Neff_Svendsen08,AskesAifantis,MorroVianello13}
\begin{equation}\label{germaneq}
\mathcal{E}(\nabla u,D^2u)= \mu\,\|\sym \nabla u\|^2+\frac{\lambda}{2}\,[\tr(\sym\nabla u)]^2+W_{\rm curv}(\nabla\sym\nabla u).
\end{equation}
In general, the strain gradient models have the great advantage of simplicity and physical transparency.

Due to isotropy, the curvature energy $W_{\rm curv}(\nabla\sym\nabla u)$ involves 5 additional constitutive constants. Taking free variations in the energy \eqref{germaneq}, we obtain
 \begin{align}\label{germaneq2}
\int_\Omega\bigg[ 2\mu\,\langle\sym \nabla u, \sym \nabla \delta u \rangle_{\mathbb{R}^{3\times3}}+\lambda \tr(\nabla u)&\tr(\nabla \delta u)\\&+\langle D\,W_{\rm curv}(\nabla\sym\nabla u), \nabla\sym\nabla \delta u\rangle_{\mathbb{R}^{27}}\bigg]dv =0, \quad\quad\quad \forall\, \delta u\in C_0^\infty(\Omega).\notag
\end{align}
But for all $\delta u\in C^\infty(\Omega)$, we have
 \begin{align}\label{curvatureg}
\int_\Omega\bigg[ \langle D&\,W_{\rm curv}(\nabla\sym\nabla u), \nabla\sym\nabla \delta u\rangle_{\mathbb{R}^{27}}\bigg]dv \\ &\notag=\int_\Omega\bigg[\dvg\big[\big( D\,W_{\rm curv}(\nabla\sym\nabla u)\big)^T.\, (\sym\nabla \delta u)\big]-\langle \dvg (D\,W_{\rm curv}(\nabla\sym\nabla u)), \sym\nabla \delta u\rangle_{\mathbb{R}^{3\times 3}}\bigg] dv\notag
  \\ &\notag=\int_\Omega\bigg[\dvg\big[\big( D\,W_{\rm curv}(\nabla\sym\nabla u)\big)^T.\, (\sym\nabla \delta u)\big]-\langle \sym \dvg (D\,W_{\rm curv}(\nabla\sym\nabla u)), \nabla \delta u\rangle_{\mathbb{R}^{3\times 3}}\bigg] dv\notag
  \\ &\notag=\int_\Omega\bigg[\dvg\big[\big( D\,W_{\rm curv}(\nabla\sym\nabla u)\big)^T.\, (\sym\nabla \delta u)\big]
  -\dvg\big[ \big(\sym \dvg\, D\,W_{\rm curv}(\nabla\sym\nabla u)\big)^T .\,\delta u\big]\notag\\\notag&\quad\quad \quad \quad+\langle \dvg[ \sym \dvg (D\,W_{\rm curv}(\nabla\sym\nabla u))],  \delta u\rangle_{\mathbb{R}^{3\times 3}}\bigg] dv\notag.\notag
\end{align}
Hence, the relation \eqref{germaneq2} leads to
\begin{align}\label{germaneq3}
\int_\Omega&\bigg\{\langle
\dvg\big\{\underbrace{\underbrace{ 2\mu\,\sym \nabla u+\lambda \tr(\nabla u)\!\cdot\! \id}_{\text{``local force stress"}} -\underbrace{\sym \dvg (\underbrace{D\,W_{\rm curl}(\nabla\sym\nabla u)}_{\text{``hyperstress"}})}_{\text{``nonlocal force stress"}}}_{\text{``total force stress"}}\big\},\delta u\rangle_{\mathbb{R}^{3}}\bigg\}\,dv\\
&\quad-\int_{\partial\Omega}\bigg\{\langle\big( D\,W_{\rm curv}(\nabla\sym\nabla u)\big).\,n, \sym\nabla \delta u\rangle
  +\langle \big(\sym \dvg\, D\,W_{\rm curv}(\nabla\sym\nabla u)\big).\, n,\delta u\rangle\bigg\}\,da=0, \notag
\end{align}
for all $ \delta u\in C^\infty(\Omega)$.
In this representation, the local and nonlocal parts of the force stress tensor are both seen to be symmetric. This is in line with the observation that the generalized
Cauchy stresses in a second grade elastic material can always be assumed in
symmetric form if frame-indifference is satisfied \cite{Neff_Svendsen08,MorroVianello13}, see the footnote 6.

\section{Conclusion}

Let us summarize the main thrust of the paper regarding the new relaxed micromorphic model. We
\begin{itemize}
\item reconcile Kr\"{o}ner's rejection of antisymmetric force stresses in dislocation theory with the dislocation model of Eringen and Claus and  show that the concept of asymmetric force stress is not needed in order to describe rotations of the microstructure in non-polar materials.
\item preserve full kinematical freedom (12 degree of freedom) by reducing the model in order to obtain symmetric Cauchy stresses. The proposed relaxed model is still able to fully describe rotations of the microstructure and to fit a huge class of mechanical behaviours of materials with microstructure.
\item note that the possible non-symmetry of the micro-distortion $P$ is governed solely by moment stresses and applied body moments. The macroscopic and microscopic scales are separated, in this sense.
\item define a dependence of the free energy  only on the elastic strain, microstrain and dislocation density tensor.
\item provide a standard set   of tangential boundary conditions for the micro-distortion, i.e. $P.\,\tau=0$ ($P_i\times\,n=0$) on $\partial \Omega$ and not the usual strong anchoring condition $P=0$ on $\partial \Omega$.
    \item obtain  well-posedness results for the relaxed formulation regarding: existence, uniqueness and continuous dependence for tangential
boundary conditions.
\item disclose as unnecessary the concept of asymmetric force stresses for a wide class of microstructured materials.
 \item conclude that the linear Cosserat theory is a redundant model for dislocated bodies and for the description of a huge class of material behaviours.
\item remark that for the isotropic relaxed micromorphic model only 3 curvature parameters remain to be determined, which may eventually be reduced to 2 parameters, which are needed for fitting bending and torsion experiments.
\item allow in principle for non-smooth solutions and the possibility of fracture. The solution space for the elastic distortion and micro-distortion is only ${\rm H}(\Curl;\Omega)$ and for the macroscopic displacement $u\in {\rm H}^1(\Omega)$. For non-smooth external data we expect slip lines.
\item introduce a suitable decomposition of the Mindlin-Eringen strain energy density for micromorphic media (see Eq. \eqref{XXXX}) which allows to determine a unique parameter $\mu_c$ which is responsible for eventual asymmetry of the stress tensor.
\item observe that the Cosserat couple modulus $\mu_c\geq0$ can be set to zero, the relaxed micromorphic model is still capable to describe rotations of the microstructure and to fit a large class of microstructured material behaviours.
\item understand that for $\mu_c=0$, the seemingly   absent (local) control of the antisymmetric part of the elastic distortion is provided by the dislocation energy, the microstrain energy and the  tangential boundary conditions. Thus, the skew-symmetric part of the distortion is fully determined by the boundary value problem.
\item note that the asymmetry of the Cauchy stress tensor may arise in theories where there are couple stresses due to a non-mechanical nature, e.g. in models with electromagnetic interactions and in polarizable media (piezoelectricity, ferroelectricity). As far as purely mechanical models are considered in the framework of linear elasticity, the need of introducing asymmetric stresses becomes rarer. Indeed, only some very special engineering materials like lattice structures and phononic crystals may be seen to need asymmetric stresses to disclose their complete mechanical behavior.
\end{itemize}

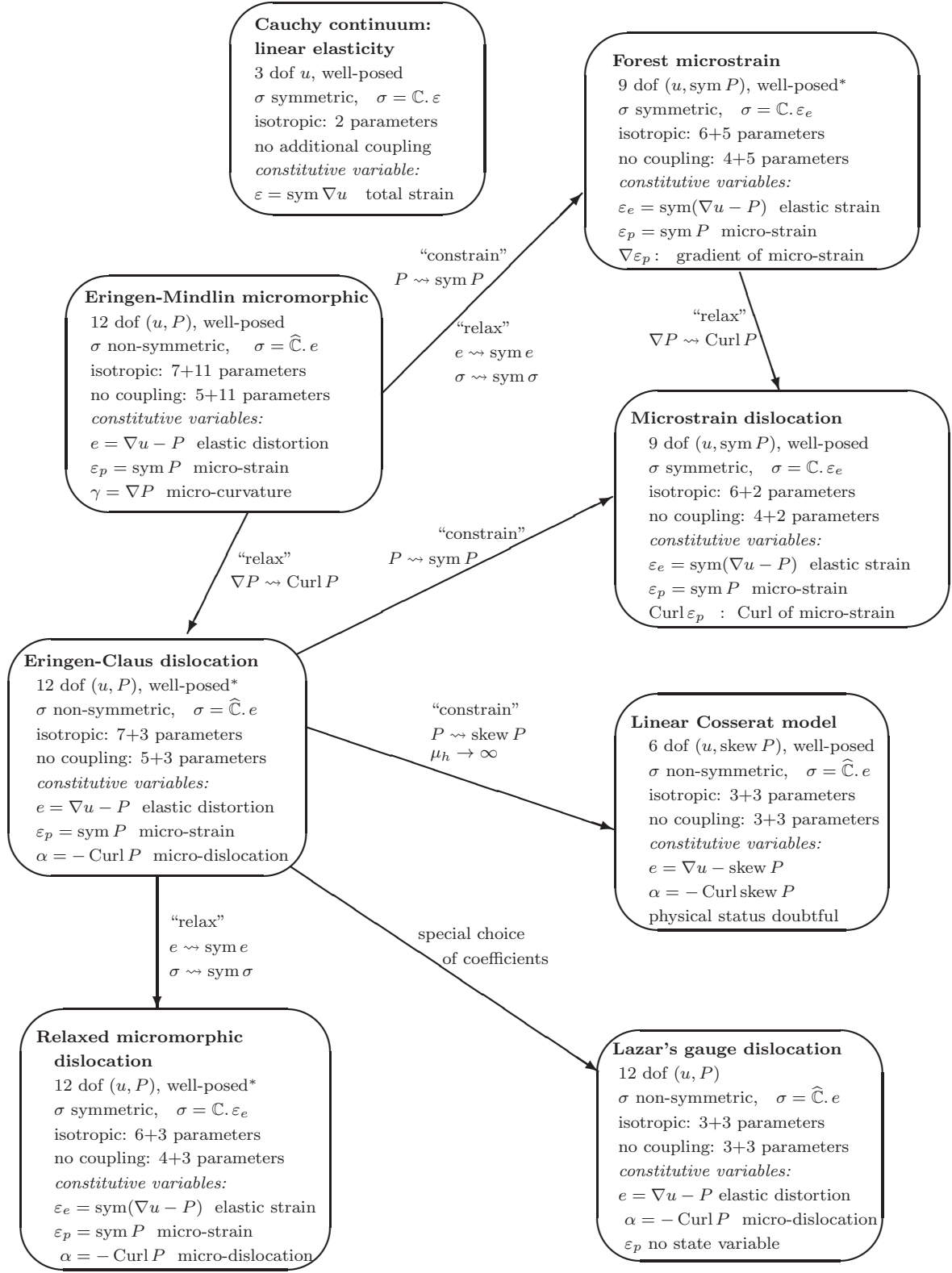
\begin{figure}
\setlength{\unitlength}{1mm}
\begin{center}
\begin{picture}(5,200)
\thicklines
\put(-20,210){\small{\bf A  family of generalized continua}}
\put(-18,187){\oval(42,35)}
\put(-35,200){\footnotesize{\bf Cauchy continuum:}}
\put(-35,196){\footnotesize{\bf linear elasticity}}
\put(-35,192){\footnotesize{3 dof $u$, well-posed}}
\put(-35,188){\footnotesize{$\sigma$ symmetric,\quad $\sigma=\mathbb{C}.\, \varepsilon$}}
\put(-35,184){\footnotesize{isotropic: 2 parameters}}
\put(-35,180){\footnotesize{no additional coupling}}
\put(-35,176){\footnotesize{\it constitutive variable:}}
\put(-35,172){\footnotesize{$\varepsilon=\sym\nabla u$ {\ \ total strain}}}
\put(-40,140){\oval(52,39)}
\put(-63,155){\footnotesize{\bf Eringen-Mindlin micromorphic}}
\put(-62,151){\footnotesize{12 dof $(u,P)$, well-posed}}
\put(-62,147){\footnotesize{$\sigma$ non-symmetric}, \quad $\sigma=\widehat{\mathbb{C}}.\, e$}
\put(-62,143){\footnotesize{isotropic: 7+11 parameters}}
\put(-62,139){\footnotesize{no coupling: 5+11 parameters}}
\put(-62,135){\footnotesize{\it constitutive variables:}}
\put(-62,131){\footnotesize{$e=\nabla u-P$ {\ elastic distortion}}}
\put(-62,127){\footnotesize{$\varepsilon_p=\sym P$  {\ micro-strain}}}
\put(-62,123){\footnotesize{$\gamma=\nabla P$ {\ micro-curvature}}}
\put(-51,80){\oval(49,39)}
\put(-73,95){\footnotesize{\bf Eringen-Claus dislocation}}
\put(-71,91){\footnotesize{12 dof $(u,P)$, well-posed$^*$}}
\put(-71,87){\footnotesize{$\sigma$ non-symmetric,\quad $\sigma=\widehat{\mathbb{C}}.\, e$}}
\put(-71,83){\footnotesize{isotropic: 7+3 parameters}}
\put(-71,79){\footnotesize{no coupling: 5+3 parameters}}
\put(-71,75){\footnotesize{\it constitutive variables:}}
\put(-71,71){\footnotesize{${e}=\nabla u-P$ {\ elastic distortion}}}
\put(-71,67){\footnotesize{$\varepsilon_p=\sym P$  {\ micro-strain}}}
\put(-71,63){\footnotesize{$\alpha=-\Curl P$ {\ micro-dislocation}}}
\put(52,121){\oval(55,39)}
\put(27,135){\footnotesize{\bf Microstrain dislocation}}
\put(30,131){\footnotesize{9 dof $(u,\sym P)$, well-posed}}
\put(30,127){\footnotesize{$\sigma$ symmetric,\quad $\sigma=\mathbb{C}.\, \varepsilon_e$}}
\put(30,123){\footnotesize{isotropic: 6+2 parameters}}
\put(30,119){\footnotesize{no coupling: 4+2 parameters}}
\put(30,115){\footnotesize{\it constitutive variables:}}
\put(30,111){\footnotesize{$\varepsilon_e=\sym(\nabla u-P)$ {\ elastic strain}}}
\put(30,107){\footnotesize{$\varepsilon_p=\sym P$  {\ micro-strain}}}
\put(30,103){\footnotesize{$\Curl \varepsilon_p$ \, :\ {\  Curl of micro-strain}}}
\put(47,71){\oval(45,39)}
\put(27,85){\footnotesize{\bf Linear Cosserat model}}
\put(30,81){\footnotesize{6 dof $(u,\skew P)$, well-posed}}
\put(30,77){\footnotesize{$\sigma$ non-symmetric,\quad $\sigma=\widehat{\mathbb{C}}.\, e$}}
\put(30,73){\footnotesize{isotropic: 3+3 parameters}}
\put(30,69){\footnotesize{no coupling: 3+3 parameters}}
\put(30,65){\footnotesize{\it constitutive variables:}}
\put(30,61){\footnotesize{$e=\nabla u-\skew P$ }}
\put(30,57){\footnotesize{$\alpha=-\Curl \skew P$ }}
\put(30,53){\footnotesize{physical status doubtful}}
\put(-48,17){\oval(51,43)}
\put(-71,33){\footnotesize{\bf Relaxed micromorphic }}
\put(-68,29){\footnotesize{\bf dislocation}}
\put(-68,25){\footnotesize{12 dof $(u,P)$, well-posed$^*$}}
\put(-68,21){\footnotesize{$\sigma$ symmetric,\quad $\sigma={\mathbb{C}}.\, \varepsilon_e$}}
\put(-68,17){\footnotesize{isotropic: 6+3 parameters}}
\put(-68,13){\footnotesize{no coupling: 4+3 parameters}}
\put(-68,9){\footnotesize{\it constitutive variables:}}
\put(-68,5){\footnotesize{$\varepsilon_e=\sym(\nabla u-P)$ {\ elastic strain}}}
\put(-68,1){\footnotesize{$\varepsilon_p=\sym P$  {\ micro-strain}}}
\put(-68,-3){\footnotesize{ $\alpha=-\Curl P$ {\ micro-dislocation}}}
\thicklines
\put(47,180){\oval(55,39)}
\put(24,194){\footnotesize{\bf Forest microstrain}}
\put(25,190){\footnotesize{9 dof $(u,\sym P)$, well-posed$^*$}}
\put(25,186){\footnotesize{$\sigma$ symmetric,\quad $\sigma=\mathbb{C}.\, \varepsilon_e$}}
\put(25,182){\footnotesize{isotropic: 6+5 parameters}}
\put(25,178){\footnotesize{no coupling: 4+5 parameters}}
\put(25,174){\footnotesize{\it constitutive variables:}}
\put(25,170){\footnotesize{$\varepsilon_e=\sym(\nabla u-P)$ {\ elastic strain}}}
\put(25,166){\footnotesize{$\varepsilon_p=\sym P$  {\ micro-strain}}}
\put(25,162){\footnotesize{$\nabla \varepsilon_p$\,:\, {\ gradient of micro-strain}}}
\put(45,160){\vector(1,-3){6.5}}
\put(37,152){\footnotesize{``relax"}}

\put(30,148){\footnotesize{$\nabla P\rightsquigarrow\Curl P$}}
 \put(45,16){\oval(47,38)}
 \put(24,31){\footnotesize{\bf Lazar's gauge dislocation}}
\put(25,27){\footnotesize{12 dof $(u, P)$}}
\put(25,23){\footnotesize{$\sigma$ non-symmetric,\quad $\sigma=\widehat{\mathbb{C}}.\, e$}}
\put(25,19){\footnotesize{isotropic: 3+3 parameters}}
\put(25,15){\footnotesize{no coupling: 3+3 parameters}}
\put(25,11){\footnotesize{\it constitutive variables:}}
\put(25,7){\footnotesize{$e=\nabla u- P$ {elastic distortion} }}
\put(25,3){\footnotesize{ $\alpha=-\Curl  P$  { micro-dislocation}}}
\put(25,-1){\footnotesize{ $\varepsilon_p$  {no state variable}}}
\thicklines
\put(-8,162){\footnotesize{``constrain"}}
\put(-12,158){\footnotesize{$P\rightsquigarrow\sym P$}}
\put(-2,150){\footnotesize{``relax"}}
\put(-2,146){\footnotesize{$e\rightsquigarrow\sym e$}}
\put(-2,142){\footnotesize{$\sigma\rightsquigarrow\sym \sigma$}}
\put(-14,140){\vector(1,1){33}}
\put(-26.5,85){\vector(3,-1){50.7}}
\put(-6,87){\footnotesize{``constrain"}}
\put(-6,83){\footnotesize{$P\rightsquigarrow\skew P$}}
\put(-6,80){\footnotesize{$\mu_h\rightarrow\infty$}}
\put(-28,97){\vector(2,1){52}}
\put(-29,62){\vector(3,-2){50.7}}
\put(-9,50){\footnotesize{ special choice}}
\put(-4,46){\footnotesize{of  coefficients}}
\put(-36,120.5){\vector(-1,-2){10}}
\put(-38,112){\footnotesize{``relax"}}
\put(-39,108){\footnotesize{$\nabla P\rightsquigarrow\Curl P$}}
\put(-51,60.5){\vector(0,-1){22}}
\put(-49,52){\footnotesize{``relax"}}
\put(-49,48){\footnotesize{$e\rightsquigarrow\sym e$}}
\put(-49,44){\footnotesize{$\sigma\rightsquigarrow\sym \sigma$}}
\put(-5,116){\footnotesize{``constrain"}}
\put(-13,112){\footnotesize{$P\rightsquigarrow\sym P$}}      \thicklines
\end{picture}
\end{center}
\caption{Situation for centro-symmetric materials.  All models are defined by a positive definite quadratic form in the given set of independent constitutive variables,
$\mathbb{C}$ being a symmetric fourth order tensor such that $\mathbb{C}:\Sym(3)\rightarrow\Sym(3)$, while
$\widehat{\mathbb{C}}:\mathbb{R}^{3\times 3}\rightarrow\mathbb{R}^{3\times 3}$  is a fourth order tensor which does not map symmetric matrices into symmetric metrices. For the isotropic case we add the number of constitutive parameters $\#1+\#2$, where $\#1$ represents the number of constitutive parameters in the force-stress response and  $\#2$ is the number of constitutive parameters in the curvature energy. By $^*$ we specify that the
well-posedness is discussed in this work. }\label{diagram}
\end{figure}

In Figure \ref{diagram} we indicate the place in the literature of our relaxed model and we point out the relations between the existing models.
Moreover,  Figure \ref{RMV} gives the relations between our relaxed micromorphic,  microstretch model, Cosserat model, microstrain model and microvoid models in
 dislocation format.
\begin{figure}
\setlength{\unitlength}{1mm}
\begin{center}
\begin{picture}(10,200)
\thicklines
\put(-34,193){\oval(51,42)}
\put(-55,210){\footnotesize{\bf Relaxed micromorphic }}
\put(-55,206){\footnotesize{\bf dislocation}}
\put(-55,202){\footnotesize{12 dof $(u,P)$, well-posed$^*$}}
\put(-55,198){\footnotesize{$\sigma$ symmetric,\quad $\sigma={\mathbb{C}}.\, \varepsilon_e$}}
\put(-55,194){\footnotesize{isotropic: 6+3 parameters}}
\put(-55,190){\footnotesize{no coupling: 4+3 parameters}}
\put(-55,186){\footnotesize{\it constitutive variables:}}
\put(-55,182){\footnotesize{$\varepsilon_e=\sym(\nabla u-P)$ {elastic strain}}}
\put(-55,178){\footnotesize{$\varepsilon_p=\sym P$  { micro-strain}}}
\put(-55,174){\footnotesize{ $\alpha=-\Curl P$ { micro-dislocation}}}
\put(-9.5,175){\vector(1,-1){35.5}}
\put(-13,162){\footnotesize{``constrain"}}
\put(-14,158){\footnotesize{$P\rightsquigarrow\zeta{\cdp} \id+A$}}
\put(-3,154){\footnotesize{$A\in\so(3)$}}
\put(-5,150){\footnotesize{$\skew(\nabla u-P)$}}
\put(-3,146){\footnotesize{incorporated\&}}
\put(6,142){\footnotesize{orthogonal}}
\put(8,138){\footnotesize{ projections}}
\put(-35,172){\vector(-1,-2){11}}
\put(-56,166){\footnotesize{non-positive}}
\put(-64,162){\footnotesize{curvature energy}}
\put(-60,158){\footnotesize{$P\rightsquigarrow\sym P$}}
\put(-36,164){\footnotesize{no coupling  }}
\put(-38,160){\footnotesize{terms for }}
\put(-39,156){\footnotesize{simplicity }}
\put(-16,172){\vector(0,-1){84}}
\put(-32,104){\footnotesize{``constrain"}}
\put(-32,100){\footnotesize{$P\rightsquigarrow\zeta\!\cdot\! \id$}}
\put(-32,96){\footnotesize{orthogonal}}
\put(-32,92){\footnotesize{ projections}}
\put(51,195){\oval(49,39)}
\put(30,210){\footnotesize{\bf Eringen-Claus dislocation}}
\put(30,206){\footnotesize{12 dof $(u,P)$, well-posed$^*$}}
\put(30,202){\footnotesize{$\sigma$ non-symmetric\quad $\sigma=\widehat{\mathbb{C}}.\, e$}}
\put(30,198){\footnotesize{isotropic: 7+3 parameters}}
\put(30,194){\footnotesize{no coupling: 5+3 parameters}}
\put(30,190){\footnotesize{\it constitutive variables:}}
\put(30,186){\footnotesize{${e}=\nabla u-P$ {\ elastic distortion}}}
\put(30,182){\footnotesize{$\varepsilon_p=\sym P$  {\ micro-strain}}}
\put(30,178){\footnotesize{ $\alpha=-\Curl P$ {\ micro-dislocation}}}
\put(26.5,192){\vector(-1,0){35}}
\put(3,189){\footnotesize{``relax"}}
\put(3,185){\footnotesize{$e\rightsquigarrow\sym e$}}
\put(3,181){\footnotesize{$\sigma\rightsquigarrow\sym \sigma$}}
\put(50,175){\vector(0,-1){25}}
\put(33,170){\footnotesize{``constrain"}}
\put(31,166){\footnotesize{$P\rightsquigarrow\zeta{\cdp} \id+A$}}
\put(35,162){\footnotesize{$A\in\so(3)$\&}}
\put(35,158){\footnotesize{orthogonal}}
\put(34,154){\footnotesize{ projections}}
\put(54,127){\oval(56,46)}
\put(30,146){\footnotesize{\bf Microstretch model in }}
\put(30,142){\footnotesize{\bf dislocation format}}
\put(30,138){\footnotesize{7 dof $(u,A,\zeta)$, well-posed$^*$}}
\put(30,134){\footnotesize{$\sigma$ non-symmetric,\quad $\sigma=\widehat{\mathbb{C}}.\, e$}}
\put(30,130){\footnotesize{isotropic: 4+3 parameters}}
\put(30,126){\footnotesize{no coupling: 3+3 parameters}}
\put(30,122){\footnotesize{\it constitutive variables:}}
\put(30,118){\footnotesize{${e}=\nabla u-(\zeta{\cdp}\id +A)$ {elastic distortion}}}
\put(30,114){\footnotesize{${\varepsilon}_\zeta=  \zeta{\cdp}\id$  { micro-strain}}}
\put(30,110){\footnotesize{$\nabla {\varepsilon}_\zeta= \nabla (\zeta{\cdp}\id)$  {gradient of micro-strain }}}
\put(30,106){\footnotesize{ ${\alpha}=-\Curl A$ { micro-dislocation}}}
\put(50,104){\vector(0,-1){25.2}}
\put(34,94){\footnotesize{``constrain"}}
\put(30,90){\footnotesize{$P\rightsquigarrow A\in\so(3)$}}
\put(50,103.5){\vector(-3,-1){44.5}}
\put(15,99){\footnotesize{``constrain"}}
\put(11,95){\footnotesize{$P\rightsquigarrow \zeta{\cdp} \id$}}
\put(57,61){\oval(47,35)}
\put(37,74){\footnotesize{\bf Linear Cosserat model}}
\put(37,70){\footnotesize{6 dof $(u,A)$, well-posed}}
\put(37,66){\footnotesize{$\sigma$ non-symmetric,\quad $\sigma=\widehat{\mathbb{C}}.\, e$}}
\put(37,62){\footnotesize{isotropic: 3+3 parameters}}
\put(37,58){\footnotesize{no coupling: 2+3 parameters}}
\put(37,54){\footnotesize{\it constitutive variables:}}
\put(37,50){\footnotesize{$e=\nabla u-A$\  elastic distortion}}
\put(37,46) {\footnotesize{ $\alpha=-\Curl A$\  micro-dislocation}}
\put(-52,129){\oval(53,41)}
\put(-75,144){\footnotesize{\bf Symmetric model of Teisseyre}}
\put(-75,140){\footnotesize{9 dof $(u,\sym P)$, not well-posed$^*$}}
\put(-75,136){\footnotesize{$\sigma$ symmetric\quad $\sigma=\mathbb{C}.\, \varepsilon_e$}}
\put(-75,132){\footnotesize{isotropic: 6+1 parameters}}
\put(-75,128){\footnotesize{no coupling: 4+1 parameters}}
\put(-75,124){\footnotesize{\it constitutive variables:}}
\put(-75,120){\footnotesize{$\varepsilon_{e}=\sym(\nabla u-P)$ {\ elastic strain}}}
\put(-75,116){\footnotesize{$\varepsilon_p=\sym P$  {\ micro-strain}}}
\put(-75,112){\footnotesize{$\alpha=-\Curl P$ {\ micro-dislocation}}}
\put(-12,65){\oval(54,46)}
\put(-34,84){\footnotesize{\bf Microvoid model in }}
\put(-34,80){\footnotesize{\bf dislocation format}}
\put(-34,76){\footnotesize{4 dof $(u,\zeta)$, well-posed$^*$}}
\put(-34,72){\footnotesize{$\sigma$ symmetric\quad $\sigma={\mathbb{C}}.\, \varepsilon_e$}}
\put(-34,68){\footnotesize{isotropic: 3+1 parameters}}
\put(-34,64){\footnotesize{no coupling: 2+1 parameters}}
\put(-34,60){\footnotesize{\it constitutive variables:}}
\put(-34,56){\footnotesize{$\varepsilon_{e}=\sym(\nabla u-\zeta {\cdp} \id)$ {\ elastic strain}}}
\put(-34,52){\footnotesize{${\varepsilon}_p=  \sym(\zeta {\cdp} \id)=\zeta{\cdp} \id$  {\ micro-strain}}}
\put(-34,48){\footnotesize{$\nabla {\varepsilon}_\zeta= \nabla (\zeta {\cdp} \id)$  {\ gradient of}}}
\put(-10,44){\footnotesize {micro-strain }}
\thicklines
\put(-12,17){\oval(54,40)}
\put(-34,34){\footnotesize{\bf Classical microvoid model}}
\put(-34,30){\footnotesize{4 dof $(u,\zeta)$, well-posed}}
\put(-34,26){\footnotesize{$\sigma$ symmetric\quad $\sigma={\mathbb{C}}.\, \varepsilon_e$}}
\put(-34,22){\footnotesize{isotropic: 4+1 parameters}}
\put(-34,18){\footnotesize{no coupling: 3+1 parameters}}
\put(-34,14){\footnotesize{\it constitutive variables:}}
\put(-34,10){\footnotesize{$\varepsilon=\sym \nabla u$ {total strain}}}
\put(-34,6){\footnotesize{${\varepsilon}_p=  \sym(\zeta{\cdp} \id)=\zeta{\cdp} \id$  {\ micro-strain}}}
\put(-34,2){\footnotesize{$\nabla {\varepsilon}_\zeta= \nabla (\zeta{\cdp} \id)$  {\ gradient of}}}
\put(-10,-2){\footnotesize {\ micro-strain }}
\thicklines
\put(56,17){\oval(58,42)}
\put(30,34){\footnotesize{\bf Classical microstretch model}}
\put(30,30){\footnotesize{7 dof $(u,A,\zeta)$, well-posed}}
\put(30,26){\footnotesize{$\sigma$ non-symmetric,\quad $\sigma=\widehat{\mathbb{C}}.\, e$}}
\put(30,22){\footnotesize{isotropic: 6+4 parameters}}
\put(30,18){\footnotesize{no coupling: 3+4 parameters}}
\put(30,14){\footnotesize{\it constitutive variables:}}
\put(30,10){\footnotesize{${e}=\nabla u-(\zeta{\cdp} \id +A)$ {elastic distortion}}}
\put(30,6){\footnotesize{${\varepsilon}_\zeta=  \zeta{\cdp} \id$  {\ micro-strain}}}
\put(30,2){\footnotesize{$\nabla {\varepsilon}_\zeta= \nabla (\zeta{\cdp} \id)$  {\ gradient of micro-strain }}}
\put(30,-2){\footnotesize{ ${\alpha}=-\Curl A$ {\ micro-dislocation}}}
\end{picture}
\end{center}
\caption{Relaxed micromorphic,  microstretch model, Cosserat model, microstrain model and microvoid models in
 dislocation format}\label{RMV}
\end{figure}

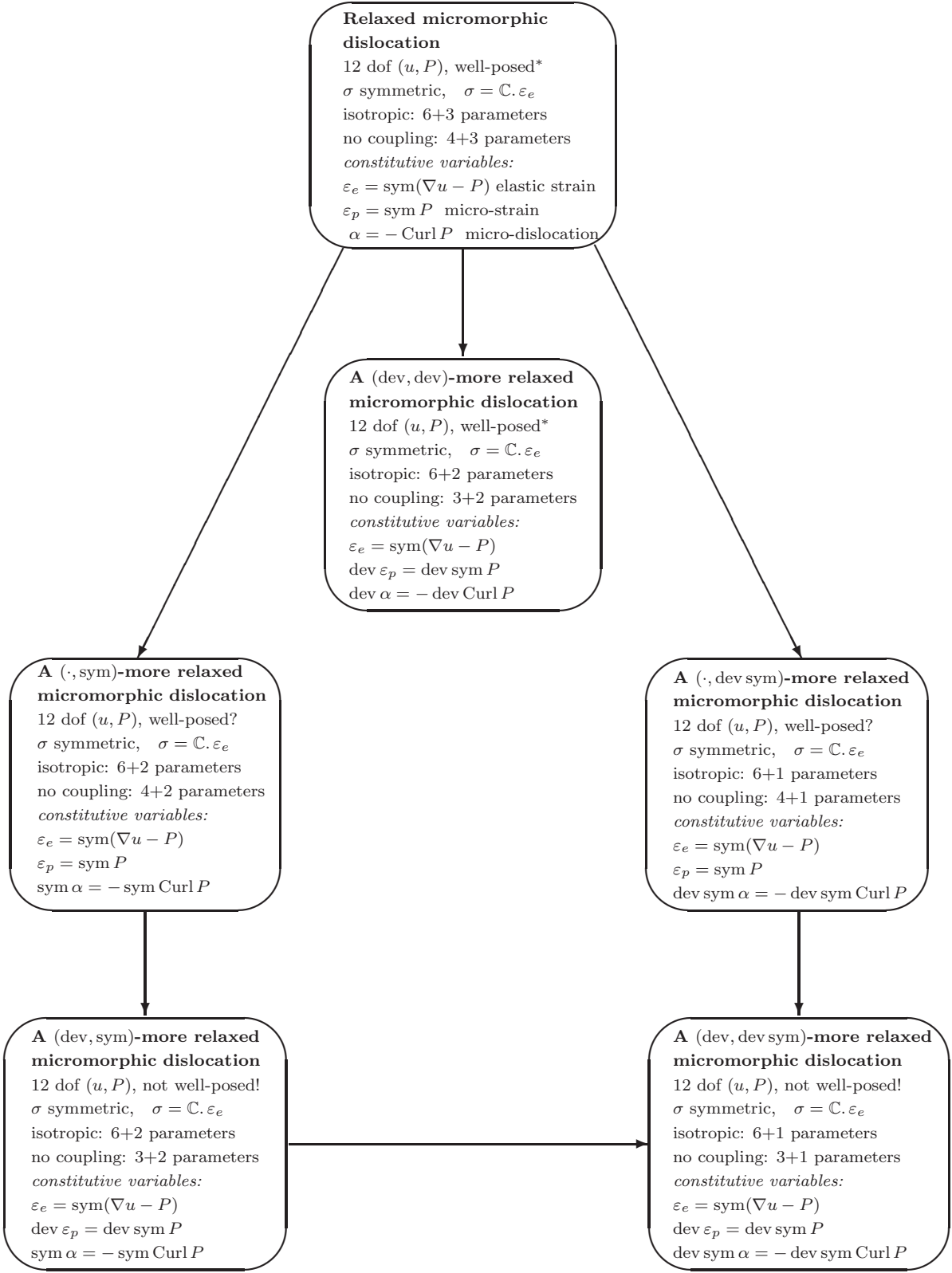
\begin{figure}
\setlength{\unitlength}{1mm}
\begin{center}
\begin{picture}(10,220)
\thicklines
\put(-40,225){\footnotesize{\bf The family of relaxed micromorphic dislocation models }}
\put(0,198){\oval(51,41)}
\put(-20,215){\footnotesize{\bf Relaxed micromorphic }}
\put(-20,211){\footnotesize{\bf dislocation}}
\put(-20,207){\footnotesize{12 dof $(u,P)$, well-posed$^*$}}
\put(-20,203){\footnotesize{$\sigma$ symmetric,\quad $\sigma={\mathbb{C}}.\, \varepsilon_e$}}
\put(-20,199){\footnotesize{isotropic: 6+3 parameters}}
\put(-20,195){\footnotesize{no coupling: 4+3 parameters}}
\put(-20,191){\footnotesize{\it constitutive variables:}}
\put(-20,187){\footnotesize{$\varepsilon_e=\sym(\nabla u-P)$ {elastic strain}}}
\put(-20,183){\footnotesize{$\varepsilon_p=\sym P$  { micro-strain}}}
\put(-20,179){\footnotesize{ $\alpha=-\Curl P$ { micro-dislocation}}}
\put(0,177.5){\vector(0,-1){18}}
\put(-20,177.5){\vector(-1,-2){34.25}}
\put(22,178){\vector(1,-2){34.5}}
\put(0,138){\oval(46,42)}
\put(-19,155){\footnotesize{\bf A $(\dev,\dev)$-more relaxed }}
\put(-19,151){\footnotesize{\bf micromorphic dislocation}}
\put(-19,147){\footnotesize{12 dof $(u,P)$, well-posed$^*$}}
\put(-19,143){\footnotesize{$\sigma$ symmetric,\quad $\sigma={\mathbb{C}}.\, \varepsilon_e$}}
\put(-19,139){\footnotesize{isotropic: 6+2 parameters}}
\put(-19,135){\footnotesize{no coupling: 3+2 parameters}}
\put(-19,131){\footnotesize{\it constitutive variables:}}
\put(-19,127){\footnotesize{$\varepsilon_e=\sym(\nabla u-P)$ }}
\put(-19,123){\footnotesize{$\dev\varepsilon_p=\dev\sym P$  }}
\put(-19,119){\footnotesize{$\dev\alpha=-\dev\Curl P$ }}
\thicklines
\put(-53,88){\oval(45,42)}
\put(-71,106){\footnotesize{\bf A $(\cdot,\sym)$-more relaxed }}
\put(-71,102){\footnotesize{\bf micromorphic dislocation}}
\put(-71,98){\footnotesize{12 dof $(u,P)$, well-posed?}}
\put(-71,94){\footnotesize{$\sigma$ symmetric,\quad $\sigma={\mathbb{C}}.\, \varepsilon_e$}}
\put(-71,90){\footnotesize{isotropic: 6+2 parameters}}
\put(-71,86){\footnotesize{no coupling: 4+2 parameters}}
\put(-71,82){\footnotesize{\it constitutive variables:}}
\put(-71,78){\footnotesize{$\varepsilon_e=\sym(\nabla u-P)$ }}
\put(-71,74){\footnotesize{$\varepsilon_p=\sym P$  }}
\put(-71,70) {\footnotesize{$\sym\alpha=-\sym\Curl P$ }}
\put(54,88){\oval(47,42)}
\put(35,105){\footnotesize{\bf A $(\cdot,\dev\sym)$-more relaxed }}
\put(35,101){\footnotesize{\bf micromorphic dislocation}}
\put(35,97){\footnotesize{12 dof $(u,P)$, well-posed?}}
\put(35,93){\footnotesize{$\sigma$ symmetric,\quad $\sigma={\mathbb{C}}.\, \varepsilon_e$}}
\put(35,89){\footnotesize{isotropic: 6+1 parameters}}
\put(35,85){\footnotesize{no coupling: 4+1 parameters}}
\put(35,81){\footnotesize{\it constitutive variables:}}
\put(35,77){\footnotesize{$\varepsilon_e=\sym(\nabla u-P)$ }}
\put(35,73){\footnotesize{$\varepsilon_p=\sym P$  }}
\put(35,69) {\footnotesize{$\dev\sym\alpha=-\dev\sym\Curl P$ }}
\put(-53,67){\vector(0,-1){17.5}}
\put(-53,28){\oval(47,42)}
\put(-72,45){\footnotesize{\bf A $(\dev,\sym)$-more relaxed }}
\put(-72,41){\footnotesize{\bf micromorphic dislocation}}
\put(-72,37){\footnotesize{12 dof $(u,P)$, not well-posed!}}
\put(-72,33){\footnotesize{$\sigma$ symmetric,\quad $\sigma={\mathbb{C}}.\, \varepsilon_e$}}
\put(-72,29){\footnotesize{isotropic: 6+2 parameters}}
\put(-72,25){\footnotesize{no coupling: 3+2 parameters}}
\put(-72,21){\footnotesize{\it constitutive variables:}}
\put(-72,17){\footnotesize{$\varepsilon_e=\sym(\nabla u-P)$ }}
\put(-72,13){\footnotesize{$\dev\varepsilon_p=\dev\sym P$  }}
\put(-72,9) {\footnotesize{$\sym\alpha=-\sym\Curl P$ }}
\put(56,67){\vector(0,-1){17.5}}
\put(56,28){\oval(50,42)}
\put(35,45){\footnotesize{\bf A $(\dev,\dev\sym)$-more relaxed }}
\put(35,41){\footnotesize{\bf micromorphic dislocation}}
\put(35,37){\footnotesize{12 dof $(u,P)$, not well-posed!}}
\put(35,33){\footnotesize{$\sigma$ symmetric,\quad $\sigma={\mathbb{C}}.\, \varepsilon_e$}}
\put(35,29){\footnotesize{isotropic: 6+1 parameters}}
\put(35,25){\footnotesize{no coupling: 3+1 parameters}}
\put(35,21){\footnotesize{\it constitutive variables:}}
\put(35,17){\footnotesize{$\varepsilon_e=\sym(\nabla u-P)$ }}
\put(35,13){\footnotesize{$\dev\varepsilon_p=\dev\sym P$  }}
\put(35,9) {\footnotesize{$\dev\sym\alpha=-\dev\sym\Curl P$ }}
      \put(-29,28){\vector(1,0){60}}
\end{picture}
\end{center}
\caption{Relation between possible relaxed micromorphic models. The non-well-posedness follows from the results in \cite{BNPS1,BNPS2,BNPS3}. }\label{morerelaxfig}
\end{figure}

The diagram from Figure \ref{morerelaxfig} gives some new possible relaxed micromorphic models and, in view of   the status of the mathematical background, we indicate the well-posedness of the dynamic and static problem.

\section{Outlook}

The new concept of metamaterials is attracting more and more the interest
of physicists and mechanicians. It is described and studied in many
works: we refer here for instance to \cite{Engheta} or \cite{Zouhdi}.

Metamaterials are obtained by suitably assembling multiple individual
elements but usually arranged in (quasi-)periodic substructures in
order to show very peculiar and especially designed mechanical properties.
Indeed, the particular shape, geometry, size, orientation and arrangement
of their constituting elements can affect e.g. the propagation of
waves of light or sound in a not-already-observed manner, creating
material properties which cannot be found in conventional materials.
Particularly promising in the design of metamaterials are those micro-structures
which present high-contrast in microscopic properties: these micro-structures,
once homogenized, may produce generalized continuum
models (see e.g. \cite{Boutin,AlibertSeppFdI,PideriSeppecher,Forest06,Forest,Misra}). The micro-structures of such metamaterials, although remaining
quasi-periodical, are conceived so that some of the physical micro-properties
characterizing their behavior diverge when the size of the REV tends
to zero, while simultaneously some other properties are vanishing
in the same limit.

In the present paper we have mathematically studied a large class of evolution
equations which are governing the propagation of linear waves in micromorphic
or generalized continua (see e.g. \cite{FdIPlacidiMadeo,PlacidiRosiMadeo,RosiMadeoGuyader}).
The mathematical existence, uniqueness and continuous dependence theorems
which we have obtained in \cite{GhibaNeffExistence} are the logical basis of the studies which will
be developed in further investigations, where the manifold variety
of propagating mechanical waves which may exist in micromorphic continua
may unfold unexpected applications in the design of  particularly
tailored metamaterials, showing very useful and up-to-now unimagined
features. Indeed, as already remarked, second gradient materials can
be seen as a particular limit case of the micromorphic media introduced
in this paper. Such materials can be obtained from micromorphic ones
constraining the micromorphic strain tensor $\sym{P}$ to be equal to
the classical strain tensor.  More precisely, the elastic distortion $\nabla u-P$ is considered to be zero. In this sense, the study of wave propagation
in micromorphic media intrinsically contains all the results which
are valid for second gradient media. Previous results on wave propagation
in second gradient elastic media have shown a wide variety of exotic
phenomena basically related to screening or transmitting properties
of interfaces embedded in such media. It has been shown that (see
e.g. \cite{FdIPlacidiMadeo,PlacidiRosiMadeo,RosiMadeoGuyader}) for
waves at frequencies which are sufficiently high to interact with
the underlying microstructure, the screening or transmitting properties
of the interface can be sensibly enhanced. It is clear that materials
which are able to show such exotic properties with respect to wave
propagation could lead to the design of technologically relevant devices
for example in the field of stealth technology or vibration and acoustic
passive control. Some preliminary results on the study of wave propagation
in micromorphic media indicate that, for particular sets of the constitutive
parameters suggested by our mathematical analysis, propagation of some types of waves can be inhibited or
waves which propagate without carrying energy can  also be observed.
Such frequency-dependent exotic properties are already observable
in the bulk of the considered micromorphic medium without considering
more complicated reflection and transmission phenomena at surfaces
of discontinuity of material properties. This means that well-conceived
micromorphic materials could be used as exotic waveguides which allows
to filter and/or switch on and off some typical waves depending on
the envisaged use. Recently the theory of material symmetry for the Cosserat continuum was extended
in \cite{Eremeyev3}. In \cite{Eremeyev3} it is mentioned that some relaxed Cosserat models can be
interpreted as micropolar liquid crystals. Although the theory of material
symmetry for the relaxed micromorhic models similar to \cite{Eremeyev3} is not elaborated into
details, such relaxed micromorphic model can be also interpreted as liquid crystal
in the sense of the material symmetry group.

We  remark that the theorems obtained in \cite{GhibaNeffExistence}
can also be used to give a better grounded basis to many results which
are already available in the literature (see e.g. \cite{FdIRosa,FdIRosa1}).

\bigskip

{\bf Acknowledgements.}  I.D. Ghiba acknowledges support from the Romanian National Authority for Scientific Research (CNCS-UEFISCDI), Project No. PN-II-ID-PCE-2011-3-0521. I.D. Ghiba  would like to thank P. Neff  for his kind hospitality during his visit at the Faculty of Mathematics, Universit\"{a}t Duisburg-Essen, Campus Essen. P. Neff is grateful to F. dell'Isola for making his visit to CISTERNA di LATINA (M\&MoCS), in spring 2013, a wonderful scientific experience. A. Madeo thanks INSA-Lyon for the financial support assigned to the project BQR
2013-0054
``Mat\'{e}riaux M\'{e}so et Micro-H\'{e}t\'{e}rog\`{e}nes: Optimisation par Mod\`{e}les de Second Gradient et
Applications en Ing\'{e}nierie".

\bibliographystyle{plain} 

\appendix
\section{ Some useful identities}\label{identitiesapp}

\setcounter{equation}{0}

In this Appendix we outline some identities which could be useful for the readers:
\begin{itemize}
\item [a)] For all matrices $A\in \so(3)$ we have the Nye's formula \cite{Nye53}
\begin{align}
-\Curl A&=(\nabla \axl A)^T-\tr[(\nabla \axl A)^T]{\cdp} \id ,\notag\\\notag
\nabla(\axl A )&= -(\Curl A)^T+\frac{1}{2}\tr[(\Curl A)^T]{\cdp} \id\,\quad\quad  \text{ ``Nye's curvature tensor"}.
\end{align}
\item [b)] For all differentiable functions $\zeta:\mathbb{R}\rightarrow\mathbb{R}$ on $\Omega$ we have
$
\Curl(\zeta{\cdp} \id)=\left(
\begin{array}{ccc}
0 &\zeta_{,3} &-\zeta_{,2}\\
-\zeta_{,3}& 0& \zeta_{,1}\\
\zeta_{,2} &-\zeta_{,1} &0
\end{array}\right)\in\so(3)
$
and
$
\Curl\Curl(\zeta{\cdp} \id)=\left(
\begin{array}{ccc}
-(\zeta_{,22}+\zeta_{,33}) &\zeta_{,12} &\zeta_{,13}\\
\zeta_{,12}& -(\zeta_{,11}+\zeta_{,33}) & \zeta_{,23}\\
\zeta_{,13} &\zeta_{,23} &-(\zeta_{,11}+\zeta_{,22})
\end{array}\right)\in\Sym(3)
$
\item [c)] If $A=\left(
\begin{array}{ccc}
0 &-x_3&x_2\\
x_3& 0& -x_1\\
-x_2 & x_1 &0
\end{array}\right),$
then $\Curl A=\id \in\Sym(3)$.
\item[d)] $\tr(\Curl S)=0$\ \ for all $S\in\Sym(3)$.
\item[e)] $
\Curl[(\Curl S)^T]\in \Sym(3)\, ,
$
for all $S\in\Sym(3)$.
\item[f)] $
\Curl[(\Curl A)^T]\in \so(3)\, ,
$
for all $A\in\so(3)$.
\item[g)] In view of b),  e) and f) we have
$$
\Curl [( \Curl \sym P)^T]\in \Sym(3),\quad\quad\quad \Curl [(\Curl \skew P)^T]\in \so(3) \quad\quad \forall \  P\in\mathbb{R}^{3\times3}.
$$
\item[h)] $\tr[\Curl (\skew \Curl  A )]=0$ \ \ for all $A\in\so(3)$.
\item[i)] Saint-Venant compatibility conditions\footnote{Kr\"{o}ner's notation}: if ${\mathbf{inc}}(S):=\Curl((\Curl S)^T)=0$ and $S\in\Sym(3)$ \break then $S=\sym \nabla u$ in a simply connected domain.
\item[j)] $\nabla\axl(\skew \nabla u)=[\Curl(\sym \nabla u)]^T$ \quad but \quad  $\nabla\axl(\skew P)\neq[\Curl(\sym P)]^T$ for general $P\in\mathbb{R}^{3\times 3}$.
\item[k)] $
    \mathfrak{a}_1\|X\|^2+\mathfrak{a}_2\langle X,X^T\rangle+\mathfrak{a}_3[\tr(X)]^2=(\mathfrak{a}_1+\mathfrak{a}_2)\|\dev\sym X\|^2+(\mathfrak{a}_1-\mathfrak{a}_2)\|\skew X\|^2+\frac{\mathfrak{a}_1+\mathfrak{a}_2
    +3\mathfrak{a}_3}{3}[\tr (X)]^2
    $  for all $X\in\mathbb{R}^{3\times 3}$.
\item[l)] For all $P\in\mathbb{R}^{3\times 3}$ we have
\begin{align}
&\tr[\big(\Curl( \skew P)\big)^T]=2\,{\rm div} \axl (\skew P)\,\\\notag
&\Curl\{\tr[\big(\Curl( \skew \axl (\skew P))\big)^T]{\cdp} \id\}=-2\anti\nabla({\rm div} \axl (\skew P)).
\end{align}

\item[m)]
For all $P\in\mathbb{R}^{3\times 3}$ we have
\begin{align}
\Curl\{[\Curl (\skew P)]^T\}=\,-\anti\nabla({\rm div} \axl (\skew P))\,.
\end{align}
\item[n)]
We have the identity
\begin{align}\label{happyfor}
-2\Curl\{[\Curl (\skew P)]^T\}+\Curl\{\tr[\big(\Curl( \skew P)\big)^T]{\cdp} \id\}=0, \quad\quad \text{for all}\quad \quad P\in\mathbb{R}^{3\times3}.
\end{align}
\item[o)] If  \ \ $\alpha_1=-6\alpha_3$ and \ \ $ \alpha_2=6\alpha_3$, then
    \begin{align}\hspace{-1cm}
\Curl \{\alpha_1 \dev \sym \Curl P+ \alpha_2 \skew \Curl P+ {\alpha_3}&\ \tr(\Curl P)\!\cdot\!\id\}\notag\\&=-6\alpha_3\,\Curl\{[\Curl(\sym P)]^T\} \in \Sym(3)
\end{align}
 for all $P\in\mathbb{R}^{3\times3}$.
\item[p)] Given $P\in\Sym(3)$, then we have
\begin{align}
\Curl \{\alpha_1 \dev \sym \Curl P+ \alpha_2 \skew \Curl P+ {\alpha_3}\ \tr(\Curl P)\!\cdot\!\id\} \in \Sym(3).
\end{align}
 if and only if $\ \alpha_1=-\alpha_2$.
\item[q)] $\langle v,\axl(W)\rangle_{\mathbb{R}^3}=\frac{1}{2}\langle \anti(v),W\rangle_{\mathbb{R}^{3\times3}}$ \quad $\forall \ W\in\so(3)$. The  adjoint of the operator $\axl:\so(3)\rightarrow\mathbb{R}^3$ is the mapping $\axl^*:\mathbb{R}^3\rightarrow\so(3)$, $\axl^*(\cdot)=\frac{1}{2}\,\anti(\cdot)$.
\end{itemize}
\end{document}